\documentclass[twocolumn, times]{aastex631}

\newcommand{\mum}{\ifmmode{\rm \mu m}\else{$\mu$m}\fi}

\newcommand{\cosmology}{$\Omega _m = 0.287$ and~$H_0=69.3$km~s$^{-1}$Mpc$^{-1}$}

\usepackage{CJKutf8, upgreek}

\shorttitle{}
\graphicspath{{./}{figures/}}

\begin{document}

\title{\large AGN Selection and Demographics: A New Age with JWST/MIRI}

\author[0000-0002-6221-1829]{Jianwei Lyu (\begin{CJK}{UTF8}{gbsn}吕建伟\end{CJK})}
\affiliation{Steward Observatory, University of Arizona, 933 North Cherry Avenue, Tucson, AZ 85721, USA}

\author[0000-0002-8909-8782]{Stacey Alberts}
\affiliation{Steward Observatory, University of Arizona, 933 North Cherry Avenue, Tucson, AZ 85721, USA}

\author[0000-0003-2303-6519]{George H. Rieke}
\affiliation{Steward Observatory, University of Arizona, 933 North Cherry Avenue, Tucson, AZ 85721, USA}

\author[0000-0003-4702-7561]{Irene Shivaei} 
\affiliation{Steward Observatory, University of Arizona, 933 North Cherry Avenue, Tucson, AZ 85721, USA}
\affiliation{Centro de Astrobiolog\'{\i}a (CAB), CSIC-INTA, Ctra. de Ajalvir km 4, Torrej\'on de Ardoz, E-28850, Madrid, Spain}

\author[0000-0003-4528-5639]{Pablo G. P\'erez-Gonz\'alez}
\affiliation{Centro de Astrobiolog\'{\i}a (CAB), CSIC-INTA, Ctra. de Ajalvir km 4, Torrej\'on de Ardoz, E-28850, Madrid, Spain}

\author[0000-0002-4622-6617]{Fengwu Sun} 
\affiliation{Steward Observatory, University of Arizona, 933 North Cherry Avenue, Tucson, AZ 85721, USA}

\author[0000-0003-4565-8239]{Kevin N. Hainline}
\affiliation{Steward Observatory, University of Arizona, 933 North Cherry Avenue, Tucson, AZ 85721, USA}

\author[0000-0002-4735-8224]{Stefi Baum} 
\affiliation{Department of Physics and Astronomy, University of Manitoba, Winnipeg, MB R3T 2N2, Canada}

\author[0000-0001-8470-7094]{Nina Bonaventura}
\affiliation{Steward Observatory, University of Arizona, 933 North Cherry Avenue, Tucson, AZ 85721, USA}

\author[0000-0002-8651-9879]{Andrew J.\ Bunker} 
\affiliation{Department of Physics, University of Oxford, Denys Wilkinson Building, Keble Road, Oxford OX1 3RH, UK}

\author[0000-0003-1344-9475]{Eiichi Egami} 
\affiliation{Steward Observatory, University of Arizona, 933 North Cherry Avenue, Tucson, AZ 85721, USA}

\author[0000-0002-2929-3121]{Daniel J.\ Eisenstein}
\affiliation{Center for Astrophysics $|$ Harvard \& Smithsonian, 60 Garden St., Cambridge MA 02138 USA}

\author[0000-0001-5097-6755]{Michael Florian}
\affiliation{Steward Observatory, University of Arizona, 933 North Cherry Avenue, Tucson, AZ 85721, USA}

\author[0000-0001-7673-2257]{Zhiyuan Ji}
\affiliation{Steward Observatory, University of Arizona, 933 North Cherry Avenue, Tucson, AZ 85721, USA}

\author[0000-0002-9280-7594]{Benjamin D.\ Johnson} 
\affiliation{Center for Astrophysics $|$ Harvard \& Smithsonian, 60 Garden St., Cambridge MA 02138 USA}

\author{Jane Morrison}
\affiliation{Steward Observatory, University of Arizona, 933 North Cherry Avenue, Tucson, AZ 85721, USA}

\author[0000-0002-7893-6170]{Marcia Rieke} 
\affiliation{Steward Observatory, University of Arizona, 933 North Cherry Avenue, Tucson, AZ 85721, USA}

\author[0000-0002-4271-0364]{Brant Robertson} 
\affiliation{Department of Astronomy and Astrophysics University of California, Santa Cruz, 1156 High Street, Santa Cruz CA 96054 USA}

\author[0000-0002-0303-499X]{Wiphu Rujopakarn}
\affiliation{Department of Physics, Faculty of Science, Chulalongkorn University, 254 Phayathai Road, Pathumwan, Bangkok 10330, Thailand}
\affiliation{National Astronomical Research Institute of Thailand, 260 Moo 4, T. Donkaew, A. Maerim, Chiangmai 50180, Thailand}

\author[0000-0002-8224-4505]{Sandro Tacchella}
\affiliation{Kavli Institute for Cosmology, University of Cambridge, Madingley Road, Cambridge CB3 0HA, UK} 
\affiliation{Cavendish Laboratory, University of Cambridge, 19 JJ Thomson Avenue, Cambridge CB3 0HE, UK}

\author{Jan Scholtz} 
\affiliation{Kavli Institute for Cosmology, University of Cambridge, Madingley Road, Cambridge CB3 0HA, UK}
\affiliation{Cavendish Laboratory, University of Cambridge, 19 JJ Thomson Avenue, Cambridge CB3 0HE, UK}

\author[0000-0001-9262-9997]{Christopher N.\ A.\ Willmer} 
\affiliation{Steward Observatory, University of Arizona, 933 North Cherry Avenue, Tucson, AZ 85721, USA}

\received{Oct 18, 2023}
\revised{Feb 05, 2024}
\accepted{Mar 07, 2024}
\submitjournal{ApJ}

\begin{abstract}
Understanding the co-evolution of supermassive black holes (SMBHs) and their host systems requires a comprehensive census of active galactic nuclei (AGN) behavior  across a wide range of redshift, luminosity, 
obscuration level and galaxy properties. We report significant progress with JWST towards this goal from the Systematic Mid-infrared Instrument Legacy Extragalactic Survey (SMILES). Based on comprehensive SED analysis of 3273 MIRI-detected sources, we identify 217 AGN
candidates over a survey area of $\sim$34 arcmin$^2$, including a primary sample of 111 AGNs in normal massive galaxies ($M_{*}>10^{9.5}~M_\sun$) at $z\sim$0--4, an extended sample of 86 AGN {\it candidates} 
in low-mass galaxies ($M_{*}<10^{9.5}~M_\sun$) and a high-$z$ sample of 20 AGN {\it candidates} at $z\sim$4--8.4. Notably, 
about 80\% of our MIRI-selected AGN candidates are new discoveries despite the extensive pre-JWST AGN searches. Even among the massive galaxies where the previous AGN search is believed to be thorough, 34\% of the MIRI AGN identifications are new, highlighting the impact of obscuration on previous selections. By combining
our results with the efforts at other wavelengths, we build the most complete AGN sample to date and 
examine the relative performance of different selection techniques. We find the obscured AGN fraction increases from $L_{\rm AGN, bol}\sim10^{10}~L_\odot$  to $10^{11}~L_\odot$ and then drops towards higher luminosity. Additionally, the obscured AGN fraction 
gradually increases from $z\sim0$ to $z\sim4$ with most high-$z$ AGNs obscured. We discuss how AGN obscuration, intrinsic SED variations, 
galaxy contamination, survey depth and selection techniques complicate the construction of a complete AGN sample.
\end{abstract}


\section{Introduction}

Supermassive black holes (SMBHs) are found at the centers of most, if not all, galaxies and are thought to be intimately 
linked to the evolution of galaxies across cosmic time
\citep[see reviews by e.g.,][]{Alexander2012, Kormendy2013}.
The formation and growth of SMBHs are regulated by the supply of accreting material 
with the efficiency influenced by various host galaxy properties at a wide range of physical scales \citep[e.g.,][]{Hopkins2008, Hopkins2010, Inayoshi2020}. As SMBHs accrete matter, they release
enormous amounts of energy in the form of radiation and powerful outflows that shape the host
systems through feedback mechanisms \citep[e.g.,][]{Fabian2012, Heckman2014}. How to unravel the nature of this SMBH-galaxy connection has been
a major quest in extragalactic astronomy. The growing phase of SMBHs, 
commonly known as Active Galactic Nuclei (AGNs), is particularly interesting, as they represent the most active phase of the SMBH-galaxy interaction and have distinctive observational characteristics to separate them out from other objects.

Since the original identifications through optical spectroscopy by \cite{Seyfert1943}, extensive ground-based optical surveys have successfully yielded statistical samples of AGN over wide ranges of galaxy properties and cosmic time  \citep[e.g.,][]{Adams1977, Filippenko1985, Ho1997, Richards2002, Kauffmann2003, Hao2005}{ , particularly with the development of various emission line diagnostics \citep[e.g.,][]{Baldwin1981, Kewley2001, Kauffmann2003}}. 
Meanwhile, a large number of AGNs missed 
in the optical have been revealed by new techniques developed at other wavelengths such as radio { \cite[e.g.,][]{Edge1959, Heckman2014, Padovani2016, Tadhunter2016}, X-rays \cite[e.g.,][]{Gursky1971, Hickox2018} 
and ultraviolet excess \cite[e.g.,][]{Markarian1977, Schmidt1983, Hainline2011}}. Nevertheless, all these methods have limitations --- only 
a minority of AGNs are prominent in the radio and the other techniques are significantly hampered by absorption by the dust and gas surrounding 
the AGN. Luckily, most of the luminosity of obscured AGNs emerges in the mid-IR, producing  SEDs dominated by emission by warm dust. To 
mitigate the deficiencies of AGN selections at other wavelengths, various AGN infrared selection techniques, 
particularly mid-IR { broad-band} color diagnostics with {\it Spitzer} and {\it WISE} data have been developed { \cite[e.g.,][]{Lacy2004, Stern2005,Alonso2006, Stern2012, Assef2013, Assef2018, Mateos2012, Padovani2017, Hickox2018, Hviding2018}}, however, these methods are mostly effective on power law SEDs \citep{Donley2012}, i.e., on lightly 
obscured sources, and they leave the heavily obscured population only partially explored (see reviews by \citealt{Hickox2018, LyuRieke2022}).
In addition, the high-$z$ obscured and faint AGN population (e.g., $z\gtrsim2$) is even less explored
given the very limited wavelength coverage and sensitivity beyond $\sim$8~$\mu$m.
Consequently, there is wide disagreement in estimates of the fraction of the AGN population that has 
been overlooked due to strong obscuration, from 10\% \citep{Mendez2013} to 30\% \citep{Delmoro2016} to 
much more \citep{Ananna2019}, and how it evolves with AGN and/or galaxy properties. { As pointed by the 
Astro2020 Decadal Survey \citep[][Appendix D-Q3c.]{astro2020}: ``the cosmic census of AGNs is currently 
patchy, which limits our understanding of the co-evolution of galaxies and their SMBHs. The highest redshifts 
remain largely unprobed, our knowledge of the most heavily obscured AGNs is incomplete, even at the lowest redshifts, 
and nuclear activity in the lowest-mass galaxies is poorly constrained. '' }

With the successful launch and operation of JWST, we have entered a new era. Notably, from NIRSpec and NIRCam spectral
observations at $\lambda<5\mum$, people have reported a number of high-$z$ AGNs or AGN candidates through rest-frame UV-optical emission line properties \citep[e.g.,][]{Harikane2023, Kocevski2023, Larson2023, Ubler2023}  with the highest records at $z\sim10$ \citep{Maiolino2023, Goulding2023}. Since
most of these searches target preferentially the broad-line systems that do not represent the whole AGN population, it is unclear how many AGNs have been missed. 
Previous systematic searches of obscured AGNs in the IR have been handicapped because of the very limited number of spectral bands 
with relevant data past $\sim$ 8 $\mu$m, leaving the SED models under-constrained. This obstacle has been lifted by the 
Mid-Infrared Instrument (MIRI; \citealt{Wright2023}) on JWST, which provides nine photometric bands from 5 to 26 $\mu$m that 
continuously sample the AGN hot dust emission features ($\lambda_{\rm peak}\ge 3\mu$m) up to redshift $z\sim8$. Compared to the
{\it Spitzer} mission, the MIRI sensivities are about 10--100 times higher with about eight times better spatial resolution at
similar wavelengths, allowing the search and characterization of the AGN population in unprecedented precision,
depth and redshift range \citep[e.g.,][]{Yang2023}

\begin{figure*}[htp]
    \begin{center}
  \includegraphics[width=0.397\hsize]{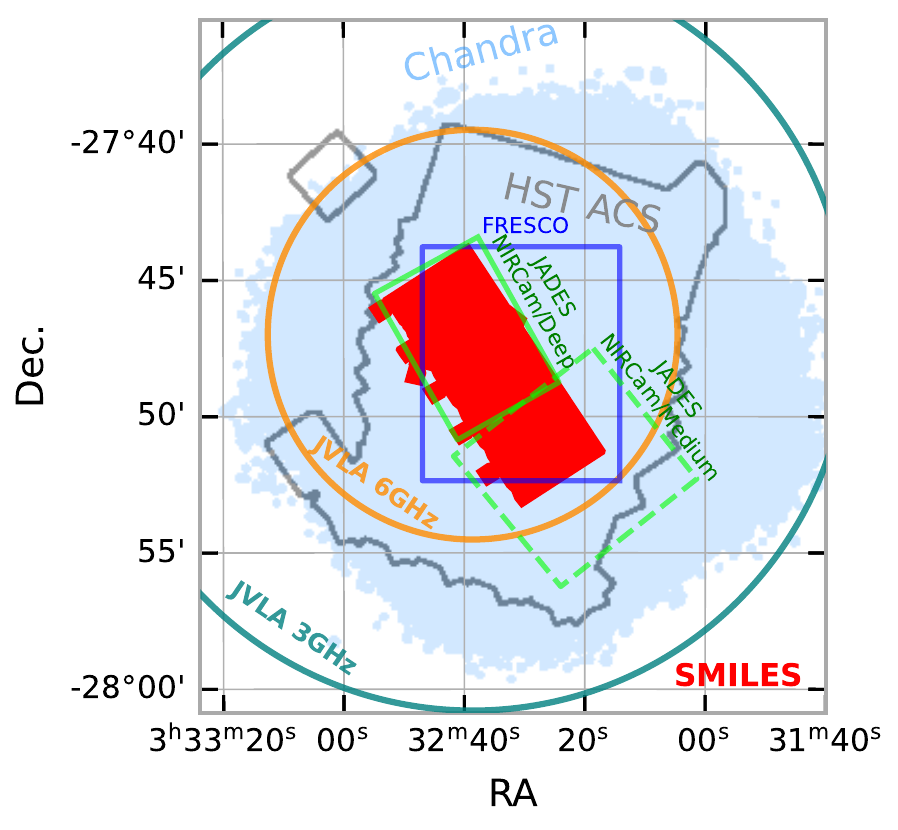}
  \includegraphics[width=0.592\hsize]{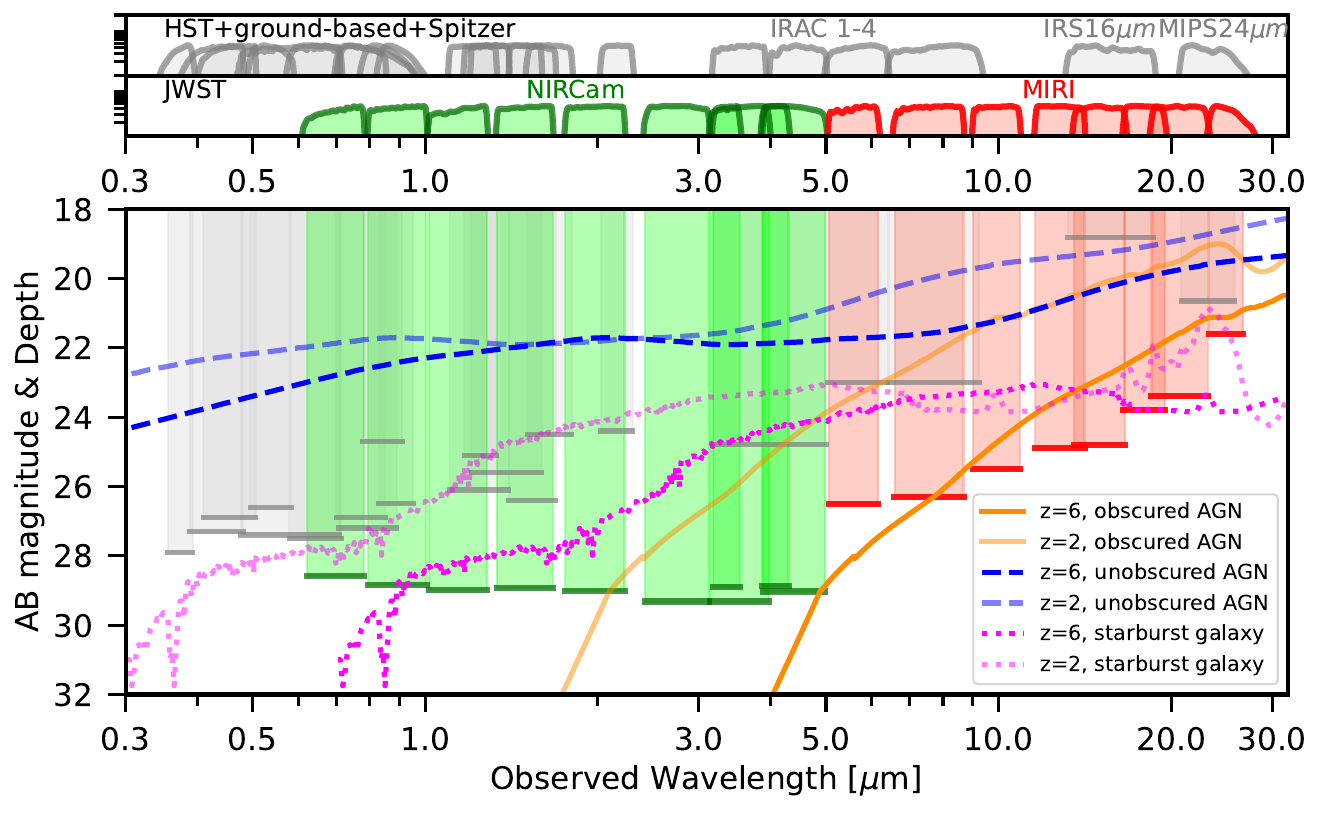}  
    \caption{Survey layout from X-ray to radio (left) and the 5$\sigma$ flux limits of photometric bands 
    at 0.3 -- 26 $\mu$m (right) used for SED fittings in GOODS-S/HUDF. On the left, the footprint of our MIRI 
    survey is shown as the red-shaded region with JADES NIRCam deep (solid green line) and medium (dashed green line) 
    observations, FRESCO NIRCam/grism coverage (solid blue line), HST ACS GOODS-S coverage (thick gray line), {\it Chandra}
    X-ray coverage (light blue shaded region) and JVLA radio observations at 3 GHz (dark green lines) and 6 GHz (dark orange lines). 
    On the right, we denote the pre-JWST filters in gray, JWST/NIRCam filters in green and JWST/MIRI filters in red. Besides 
    the flux limits of these filters as a function of wavelength, we also show some representative SEDs of obscured 
    AGNs (orange), unobscured AGNs (blue) and starburst galaxies (magenta) at $z$=2 and $z$=4, { where JWST observations are expected to bring 
    major advances to identify and characterize these objects}. 
    }
  \label{fig:footprint-sens}
    \end{center}
\end{figure*}

In this paper, we present the first results of AGN selection and demographics using the 
Systematic Mid-infrared Instrument Legacy Extragalactic Survey (SMILES; \citealt{Rieke-1207}) led
by the US MIRI Science Team. This survey is composed of 3$\times$5 MIRI pointings in 8 imaging bands at $\lambda\sim$5--27~$\mum$ 
over the central region of the Great Observatories Origins Deep Survey--South (GOODS-S; \citealt{Giavalisco2004}), 
including the Hubble Ultra Deep Field (HUDF; \citealt{Beckwith2006}) 
(see Figure~\ref{fig:footprint-sens} for the survey footprint and sensitivities).
In the pre-JWST era, this field was  extensively observed across the electromagnetic spectrum, often with the deepest observations anywhere  in the whole sky, offering
the best ancillary dataset for AGN study. In the UV to the near-IR,
it has been visited multiple times by the Hubble Space Telescope (HST) as well as many ground-based facilities that include both imaging
and spectral observations. AGNs have been identified through emission line features \citep[e.g.,][]{Santini2009, Silverman2010} or time variability \citep[e.g.,][]{Pouliasis2019}. In the X-ray, 
this field is covered as part of the Chandra Deep Field South (CDF-S; \citealt{Giacconi2002}) with 
a total exposure time about 7 Ms at 0.5--7 keV, enabling the identification of many X-ray detected AGNs 
\citep{Luo2017}. In the radio-band, super deep data at 3 GHz and 6 GHz with the Jansky Very Large Array (JVLA) 
have also been obtained and used in AGN searches \citep{Alberts2020}.  In the infrared, very deep 
IRAC and confusion-limited MIPS 24 $\mu$m images from {\it Spitzer} are available and have been utilized 
to identify AGNs by mid-IR colors \citep[e.g.,][]{Donley2008} and through SED fitting \citep{Alberts2020,Lyu2022}.
These multi-wavelength datasets and the corresponding results provide the best benchmark to show
the advances brought by JWST and to characterize the relative performance of the MIRI AGN selection.

Besides the pre-JWST multi-wavelength dataset, our MIRI survey also has almost full overlap with the deep NIRCam imaging observations from the JWST 
Advanced Deep Extragalactic Survey (JADES; \citealt{Eisenstein2023}) and a large overlap with spectroscopic data from the 
First Reionization Epoch Spectrosopic COmplete Survey (FRESCO; \citealt{Oesch2023}), which can help to reveal the nature of these MIRI sources and
characterize their properties. Compared to other MIRI imaging surveys in JWST cycles 1 and 2 (e.g., CEERS, COSMOS-Web), SMILES has larger sky coverage and/or more complete wavelength coverage. Combined with very deep JWST/NIRCam and HST data, this MIRI
dataset provides a unique opportunity for AGN study over a wide range of redshifts and properties.


Taking advantage of the deep JWST (and HST) photometric data that continuously covers from 0.4 to 26 $\mu$m, we
will conduct comprehensive SED analyses to identify AGNs with a wide range of luminosity and obscuration levels from $z\sim0$ to $z\sim8$. 
We will demonstrate the new discovery space offered by MIRI and present a major leap in AGN selection compared to all the previous IR missions. 
Combined with the super deep X-ray and radio data in the same field, we are able to build the most complete
AGN sample in terms of bolometric luminosity and discuss the limits of different wavelengths and selection techniques. 
These results can establish the foundation of a number of follow-up investigations on AGN-related science 
such as AGN obscuration, luminosity functions, AGN duty cycles, host galaxy properties and AGN-galaxy relations.

This paper is structured as follows. Section~\ref{sec:data} describes the observations, data reduction and photometry measurements.
In Section~\ref{sec:analysis}, we introduce the AGN selection methods and present the results, which include 
SED identifications and other techniques with and without using the JWST data. In
Section~\ref{sec:result}, we report the AGN number densities from MIRI, compare them to the AGN selected at other wavelengths
and characterize the relative performance of different selection techniques across the electromagnetic spectrum. Section~\ref{sec:obscured_agn}
characterizes the obscured AGN fraction from SED analysis and studies its possible evolution. The nature of AGNs missed at different wavelengths is also explored. In Section~\ref{sec:discussion}, we discuss various issues that can complicate the construction of a complete AGN sample. 
A summary is provided in Section~\ref{sec:summary}.

Throughout this work, we assume a flat $\Lambda$CDM cosmology with \cosmology \citep{WMAP9}.

\section{Data and Measurements}\label{sec:data}

\subsection{JWST/MIRI Observations and Image Reduction}

Our JWST MIRI imaging observations (JWST ID 1207; PI: George Rieke) comprise 15 separate pointings that produce a 3$\times$5 mosaic 
at the central 34 square arcminutes of the GOODS-S/HUDF sky area. 
To improve the PSF sampling, and mitigate cosmic rays and
detector artifacts, we used the 4-Point-Sets Dither optimized for point sources with the SMALL pattern. 
The full MIRI field was used with the FASTR1 readout pattern. 
For each pointing, the total science exposure was 
about 2.17 hours with the time spread over 8 MIRI filters: F560W, F770W, 
F1000W, F1280W, F1500W, F1800W, F2100W and F2550W with the shortest integrations lasting 10.7 min (F1000W) and the longest 36.4 min (F2100W).  The exposure time per band 
was aimed at optimizing detection of a typical AGN or star-forming galaxy at $z\sim2$, except for F2550W which has a shorter exposure time aimed at $z\sim1$ galaxies.
Most of these observations were successfully carried out during the scheduled Dec 7--14, 2022 time window with the same position angle (PA), but 2 pointings 
failed due to telescope guiding failures and the relevant data was re-taken on Jan 1 and 28, 2023. The different observing  visits 
introduce position angle rotations of 26.6 and 48.2 degrees for these two pointings relative to the others and cause 
some irregularity in the final MIRI mosaic layout.

The data reduction used the JWST Calibration Pipeline v1.10.0 \citep{Bushouse2022} and JWST Calibration Reference System (CRDS) context jwst\_1084.pmap.  
Starting from the raw (``uncal") data files obtained from MAST\footnote{\url{https://mast.stsci.edu/portal/Mashup/Clients/Mast/Portal.html}}, 
the pipeline module \texttt{calwebb\_detector1} was used with default parameters to apply detector-level corrections, including linearity, dark current, 
first and last frame effects, and jump detection.  For the latter, we implemented newly introduced algorithms with default parameters to correct 
for cosmic ray showers, i.e., large cosmic ray events that can have an impact over a large number of pixels. 

Next, the default step \texttt{calwebb\_image2} was run to apply instrumental corrections and calibrations to produce ``cal" files.  Given that MAST data 
products can still suffer from strong background noise structures, at this point we apply a custom, external background subtraction from
the Rainbow Database \citep[][]{PerezGonzalez2005, PerezGonzalez2008, Barro2011a, Barro2011b} JWST pipeline. 
 Briefly, as explained in more detail in \citet[][]{Alvarez2023}, we use an iterative process that progressively masks sources and median filters out large gradients 
and striping along detector columns and rows to produce ``clean" cal files.  For each (unfiltered) cal file, a ``super background" is then 
constructed by median combining and scaling all other clean cal files.  We found that better results are achieved by using all (relatively 
contemporaneous) cal files to construct a super background as opposed to constructing a background per visit. The latter can result in over-subtraction 
around extended sources due to inadequate availability of background pixels at the source location in detector space. The final super backgrounds were then subtracted from each cal file and an additional 265x265 pix$^2$ box median subtraction is applied to remove any remaining varying background, mainly caused by cosmic ray showers.

The background-subtracted cal files were then astrometrically-corrected on a per visit and per filter basis using tweakreg\footnote{\url{https://jwst-pipeline.readthedocs.io/en/latest/jwst/tweakreg/README.html}} with custom routines external to the JWST pipeline.  The F560W image is first registered to the JADES image mosaic (which is registered to GAIA DR3) by creating matched catalogs of high S/N F444W and F560W sources, which will have similar morphologies and therefore centroids. The corrected F560W positions were then used to correct the next longer wavelength image and so on, so as to minimize the effects of morphological changes\footnote{Due to the decreasing number of high S/N sources, the astrometrically-corrected F1500W catalog was used to correct the F1800W, F2100W, and F2550W images.}. The astrometric accuracy achieved by this process is 0.01--0.02$^{\prime\prime}$ ($1\sigma$) for all filters except F2550W, which has a low number of high S/N sources and an astrometric accuracy of $\sim0.04^{\prime\prime}$. Finally, the astrometry-corrected and background-subtracted cal files were processed through \texttt{calwebb\_image3}, which produced the final mosaics with a pixel scale of 0.06$^{\prime\prime}$.  The final 5$\sigma$ point source sensitivities are 0.21, 0.21, 0.44, 0.60, 0.68, 1.7, 2.8, and 15 $\mu$Jy in F560W, F770W, F1000W, F1280W, F1500W, F1800W, F2100W, and F2550W respectively, for apertures encompassing $65\%$ of the encircled energy of the PSFs (see below).  Compared to the JWST ETC predictions, our measured detection limits have been improved by a factor of 1.6--2 for F770W-F1800W.

\begin{figure*}[ht]
    \begin{center}
  \includegraphics[width=1.0\hsize]{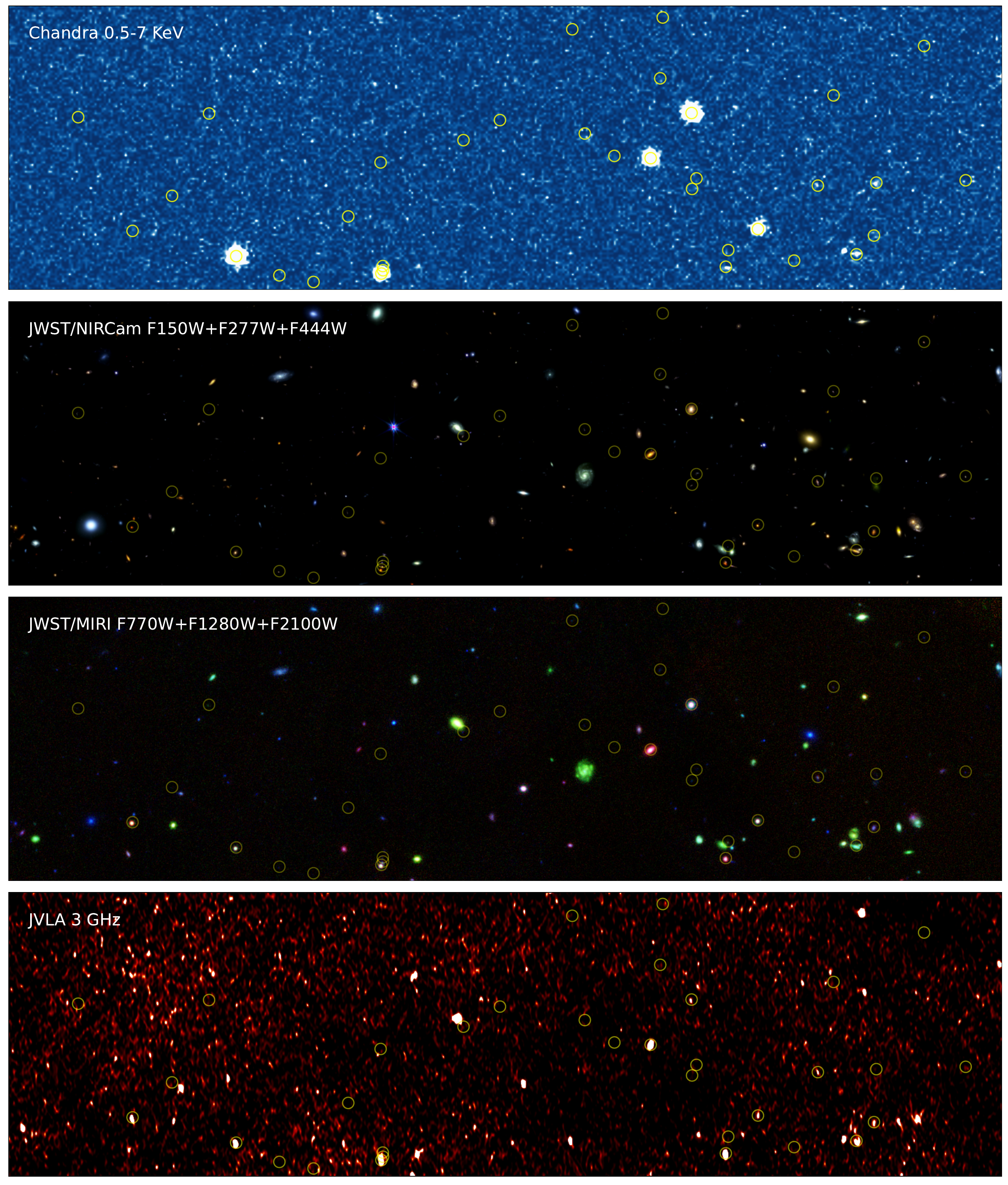}
    \caption{Showcase of multi-wavelength data used for AGN searches in SMILES with a field of view 3.5 arcmin$\times$1.0 arcmin. The yellow circles indicate the AGNs or AGN candidates identified from this work. The JWST NIRCam and MIRI data are shown as three-color images with the long wavelength in red, intermediate wavelength in green and short wavelength in blue. }
  \label{fig:image_example}
    \end{center}
\end{figure*}

Figure~\ref{fig:image_example} shows a representative cutout of our JWST/MIRI images together with JWST/NIRCam, 
Chandra X-ray, and JVLA radio data in the same region. Almost all of the X-ray and radio sources have been detected in 
the well-resolved MIRI images (see Section~\ref{sec:support-data}). The MIRI image resolutions are at 0.2--0.8$\arcsec$ 
(PSF FWHM) while {\it Chandra} is at $\sim$0.5$\arcsec$ (field center), JVLA at $\sim$0.3--1$\arcsec$ and JWST/NIRCam at 0.02--0.16$\arcsec$, 
allowing reliable associations and accurate photometry across the full wavelength range.

\subsection{MIRI Source Identification and Photometry Measurements}

Source detection and photometric catalogs were produced using a modified version of the JADES photometric pipeline (\citealt{Rieke2023}).  The object detection uses stacked MIRI F560W and F770W high S/N images and a series of iterative steps, 
starting with the creation of a low S/N threshold blended segmentation map which is then processed to remove spurious noise 
detections and deblend sources (see \citealt{Rieke2023} for details).  The final 
segmentation map is used to define source centroids and 2.5x scaled Kron apertures, which are then used to measure photometry in all filters.  
Aperture corrections are applied by determining the fraction of flux outside a given Kron aperture using a model PSF for each filter.  Model PSFs 
are created using WebbPSF \citep{Perrin2014}\footnote{\url{https://github.com/spacetelescope/webbpsf}} for the F1000W-F2550W filters.  The F560W 
and F770W PSFs include an extended cross-like imaging artifact known as a ``cruciform'' \citep{Gaspar2021} which is not yet adequately modeled by 
WebbPSF.  As such, we constructed empirical PSFs using high dynamic range imaging of stars taken during commissioning (Gaspar, private communication). 
Photometric uncertainties are determined by placing random apertures across the mosaics to account for correlated pixel noise 
\citep[e.g.,][]{Whitaker2011, Rieke2023}.

The nature of our source detection does not impose a hard S/N threshold and our resulting catalog contains both real and spurious detections down to faint limits.  
We conduct visual inspections with the aid of deep NIRCam and HST images to throw out spurious sources. Finally, we have 3273 MIRI
sources with 3-$\sigma$ detection in at least one band. 

By comparing with previous {\it Spitzer} photometry of the same sources, we found that our flux measurements in the F560W and F770W bands are systematically higher
and concluded the origin is likely due to an underestimation of the cruciform during the flux calibration step.
To mitigate the problem, we derived the correction factors as follows. We fitted the ASTRODEEP photometry up to IRAC band-2\footnote{The IRAC 
bands 3 and 4 are also affected by the same ``cruciform'' problem because of the same detector architecture as in  MIRI, so they are not used 
in the calibration.} of 25 isolated stars with single stellar population models from the Flexible Stellar Population Synthesis (FSPS) 
code\footnote{Although designed for fitting galaxies, at the wavelengths of interest these models are dominated by the output of red giants, 
and thus represent a typical average over the expected behavior of the stars.} and computed the synthetic MIRI fluxes in F560W and F770W.  
The median offsets between the observed (aperture-corrected) Kron flux and the model flux were then computed; the values are 1.26 for F560W 
and 1.04 for F770W. Finally, we divide the F560W and F770W fluxes by these factors during the SED fitting in Section~\ref{sec:analysis}. The photometry is archived in the v0.4.2 MIRI photometry catalog.

\subsection{JWST/NIRCam, HST/ACS, HST/WFC3 Photometry and SED Constructions}

Most of the SMILES survey area overlaps with the NIRCam image footprint from JADES \citep{Eisenstein2023}. This includes the JADES GOODS-S NIRCam/Deep (ID: 1180; PI: Daniel Eisenstein), partly taken from September 29 to
October 10, 2022 and already publicly released \citep{Rieke2023}, and the JADES GOODS-S NIRCam/Medium (ID 1210; PI: Nora Luetzgendorf), taken from October 20 to 24, 2023. All
these datasets include deep F090W, F115W, F150W, F200W, F277W, F335M, F356W, F410M, F444W images, offering critical insights into the nature and properties of the MIRI sources. { In addition, two additional NIRCAM pointings in the HUDF were conducted with the JWST Extragalactic Medium-band Survey (JEMS; \citealt{Williams2023}), offering F182M, F210M, F430M, F460M, F480M coverage over 15.6 arcmin$^2$. In contrast to the previous multi-band photometric analysis with data collected by different missions/instruments across the years,} these NIRCam observations were carried out no more than 3 months earlier than the MIRI data, which together with the time dilation (typical redshifts are $z$=1--3), greatly mitigates the possible influence from AGN variability on the source SEDs.\footnote{ For a typical AGN with a BH mass $\sim10^7~M_\odot$, the light crossing time for the UV/optical disk is 0.06--6 days; the observed relative continuum time lags in the rest-frame UV/optical band are a few days \citep[see review by e.g.,][]{Cackett2021}. Meanwhile, the rest-frame near-IR emission of the AGN torus is typically time lagged to the optical emission by $\tau\sim100\times(L_\odot/10^{11}~L_\odot)^{0.5}$ days with the variability signals quickly going away at longer wavelengths \citep[see review by][]{Lyu2022c}. Given the fact that most AGNs are obscured and the time lag is further diluted by redshift with a factor of $(1+z)$, AGN variability should not impact the JADES+SMILES photometry for SED analysis in general.}

Although the previous HST data is not as deep as JADES NIRCam data at the same wavelength, it is still indispensable 
as HST provides measurements at $\lambda<0.8~\mum$ that are missed by NIRCam. The addition of HST
bands also improves the SED analysis such as constraining the photo-$z$. The HST/ACS F435W, F606W, F775W, F814W, F850LP 
and HST/WFC3 F105W, F125W, F140W, F160W images of this field have been reprocessed by the JADES team, including registering 
the astrometry against JWST/NIRCam images and measuring the HST and NIRCam fluxes in a consistent way. 

For the NIRCam and HST photometry in the JADES field, we have adopted the PSF-convolved KRON flux from the JADES v0.8.1 catalog. We searched for the NIRCam counterparts for all the MIRI sources within a radius of 0.3 arcsec. For MIRI sources outside the current JADES footprint, we searched for 
the nearest CANDELS source within a radius of 0.6 arcsec and adopted the corresponding ASTRODEEP \citep{Merlin2021} photometry up
to IRAC band 2 (4.5~$\mu$m).

We have verified the quality of the JADES+MIRI photometry measurements with synthetic JWST photometry based on the SED fittings of 
the same sources with the ASTRODEEP+{\it Spitzer} IRS 16~$\mum$ and MIPS 24~$\mum$ data. On a statistical level, they agree with each other within 1-4\%.

\subsection{Redshift Constraints for MIRI Sources}\label{sec:data-redshift}

\begin{figure}[htp]
    \begin{center}
  \includegraphics[width=1.0\hsize]{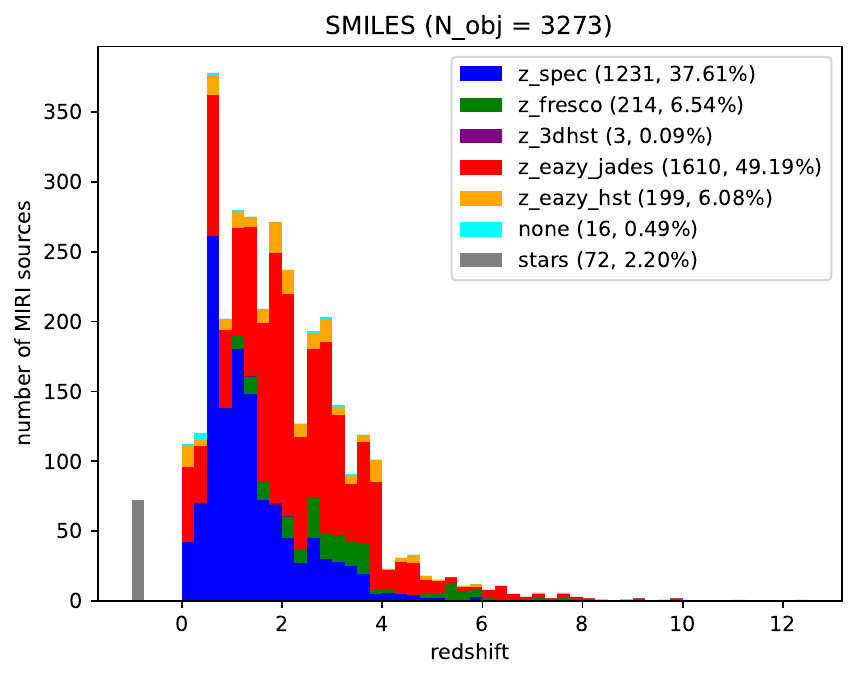}
    \caption{Redshift distribution of MIRI sources. We highlight the sources of the final adopted redshifts with different colors.}
  \label{fig:redshift-dist}
    \end{center}
\end{figure}

To aid the SED analysis, redshifts of the MIRI sources are collected from various sources that can be grouped into three categories:

\begin{itemize}
    \item spectroscopic redshifts: These include high confidence spectroscopic redshift measurements
   published in the literature over last twenty years (\citealt{Kodra2023}; H. Hathi, private communication), the redshift catalog 
   of the MUSE HUDF surveys published this year \citep{Bacon2023},  and also the 
   recently released redshifts from JADES NIRSpec data release 1 \citep{Bunker2023}.
    \item grism redshifts: Most of the MIRI field has also been covered by FRESCO grism spectral observations with NIRCam/F444W (3.8--5.0 $\mu$m) at R$\sim$1600 \citep{Oesch2023}. Since
    many objects only show a single emission line, we need to guide the redshift measurements with photo-$z$ from JADES NIRCam as priors. Typically, the confidence 
    level of the prism redshift is the same as the spectroscopic ones above given the spectral resolution and depth of the NIRCam grism observations. When  a  
    FRESCO grism redshift is not available, we search for the 3D-HST grism redshift from WFSC3/G141 (1.1--1.65 $\mum$) at R$\sim$130. 
    Given the shorter wavelength coverage and poorer resolution, the latter is mostly only useful for objects at low redshifts.
    \item photometric redshifts: Thanks to the extensive multi-band HST and JWST/NIRCam image coverage of the JADES, reasonably accurate
    photo-z measurements are possible for objects without spectroscopic observations. The JADES team has updated EAZY \citep{Brammer2008} configurations and provided the photo-z measurements based on fitting HST and JWST/NIRCam photometry \citep[see details in][]{Hainline2023}. We adopt this product by default. For objects outside the JADES/NIRCam coverage, we adopt the optimized 
     photometric redshifts for CANDELS published in \citet{Kodra2023}.
\end{itemize} 
Figure~\ref{fig:redshift-dist} gives a summary of the redshift distribution of the MIRI sources. Despite the rather short exposure times of these MIRI bands, we detected over 30 objects at $z>6.0$. About 45\% of the sample have spectroscopic or grism redshifts. The EAZY-based photometric redshifts based on NIRCam data alone have 5\% outliers and a scatter (excluding outliers) of $\left< z_{spec} - z_{phot} \right>$ of 0.024, 1$\sigma$ \citep{Rieke2023}. With the combination of MIRI photometry at longer wavelength, we
expect the updated photo-$z$ computed from our SED fitting with Prospector (see Section~\ref{sec:sed-model}) to be improved further with even less outliers.

\subsection{Other Supporting Datasets and Catalogs}\label{sec:support-data}

Following the strategies in \cite{Lyu2022},
we cross-matched our MIRI sources with the CDF-S and JVLA catalogs to look for AGN signatures within a search radius 
of 3.0$\arcsec$. As the astrometry of the {\it Chandra} data was registered in the pre-Gaia era, we investigated the coordinate
differences between the closest-matched MIRI and X-ray sources and applied a systematic
shift of $\Delta$RA=0.128$\arcsec$ and $\Delta$Dec=$-$0.288$\arcsec$ to all the CDF-S sources to remove the 
relative astrometric offsets.
There are 202 X-ray sources within the MIRI footprint but four of them do not have MIRI counterparts. Upon visual inspection, 
their X-ray detections are below 3-$\sigma$. For the radio sources, there are 183 objects but one does not have a MIRI counterpart. The nature of 
this radio source will be explored in the future.

In Section~\ref{sec:other-selection}, we will discuss AGN selections for sources detected in X-ray or radio.

\section{AGN Identifications}\label{sec:analysis}

With the data described above, we  conducted SED analysis to search for AGNs. Section~\ref{sec:sed-model} introduces 
the semi-empirical SED models used for AGN identifications. In Section~\ref{sec:sed-fitting-test}, we
conduct the initial round of SED analysis to get an overview of the sample properties of all the MIRI sources and
check the performance of our models. Section~\ref{sec:sed-selection} describes how the AGNs are identified
from SED analysis with the removal of possible false positives. In Section~\ref{sec:other-selection}, we summarize the
AGN samples selected by other methods in the same field, including those utilizing the JWST data and those in the
pre-JWST era.

\subsection{SED Models and Fitting Setup}\label{sec:sed-model}

\begin{figure*}[htp]
    \begin{center}
  \includegraphics[width=1.0\hsize]{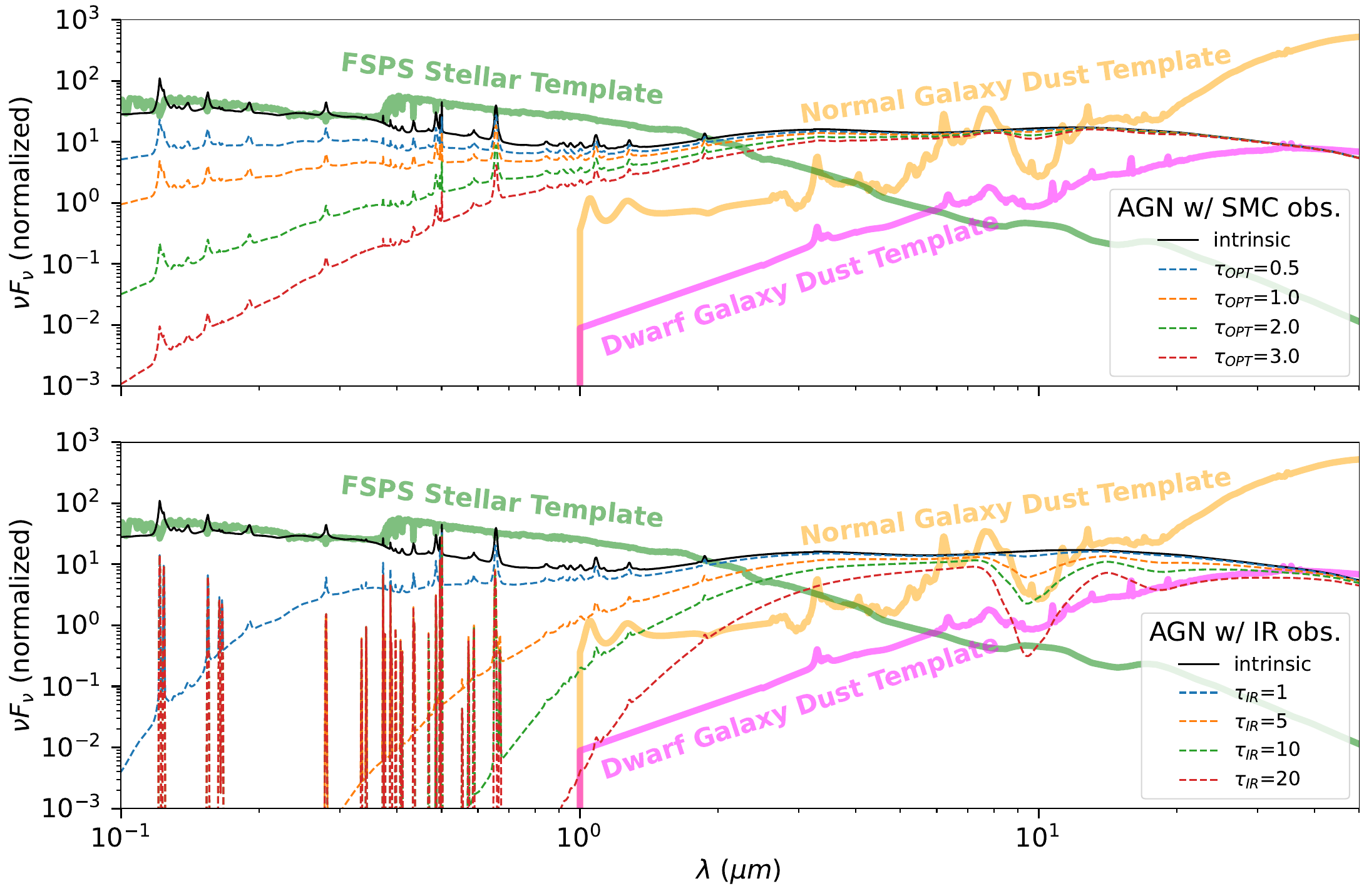}
    \caption{Illustration of how the AGN SED shape is changed by different levels of attenuation as well as the templates for the galaxy stellar and dust components in our model. The upper panel shows the AGN SED obscured by the SMC
    extinction curve, mainly used for UV-optical reddened AGNs. The lower panel presents the AGN SEDs 
    modified by the empirical IR attenuation law.  We show
    a galaxy stellar SED with a stellar age of 500 Myr (light green thick line), and the pure dust emission templates 
    for normal star-forming galaxies (light orange line) and low-metallicity dwarf galaxies (light magenta line) to compare with the AGN SEDs in both panels. See text for details.}
  \label{fig:agn-model}
    \end{center}
\end{figure*}

As discussed in detail in \citet{LyuRieke2022}, reliable AGN identification through SED fitting
requires careful minimization of the free parameters of the models while achieving accurate fits to the data. To meet
this requirement, we introduced a fitting package for AGN SED selection and analysis in \citet{Lyu2022} that features 
a modified version of the Prospector code \citep{Johnson2021} with the Flexible Stellar Population Synthesis 
(FSPS; \citealt{Conroy2009, Conroy2010}) model for the stellar component; semi-empirical SED models for the AGN 
component \citep{Lyu2017a, Lyu2017b, Lyu2018, LyuRieke2022}; and a well-proven template for dust emission from 
the star-forming galaxies \citep{Rieke2009, Lyu2022}. The same tool is also used for this work with a few updates 
described below. In Figure~\ref{fig:agn-model}, we show the SEDs of different components and 
illustrate some of their variations.

The configuration of the stellar component is very similar to \cite{Lyu2022} --- we assumed the same Kroupa initial mass function and delayed-tau 
star-formation history but adopted the Calzetti attenuation curve with a flexible slope as in \citet{Kriek2013}. Since there are objects at very high redshifts, IGM absorption is also included. In total, 
there are seven free parameters to describe the stellar component: (1) stellar mass formed ($mass$); (2) stellar metallicity ($logzsol$); 
(3) stellar age ($tage$); (4) $e$-folding time of the star formation history ($tau$); (5) the attenuation level at 5500~\AA~($dust2$); 
(6) the power law index of the attenuation ($dust\_index$); (7) the IGM absorption factor ($igm\_factor$). For stellar populations older than $\sim$ 10 Myr, regardless
of the metallicity and star formation history, the near-to-mid-IR infrared photospheric emission is dominated by red 
giants and supergiants with very similar  spectra, so any degeneracy of the parameters (2)--(4) will not influence the AGN identification.

For the star forming galaxy (SFG) dust emission component, by default we also adopt the template for $log(L_{\rm IR}/L_\odot)=11.25$, which is the best description of the 
average properties of normal SFGs at a wide range 
of redshifts that constitute our {\it primary} sample (see Appendix A in \citealt{Lyu2022} for details). Furthermore, as shown later, our MIRI data 
is deep enough to probe the dwarf galaxy regime, where the systems can be low-metallicity and
their dust emission SEDs likely behave differently from those of more luminous SFGs. These galaxies constitute our {\it extended} sample. We therefore adopt the Haro 11 SED template
developed in \citet{Lyu2016} as the choice to match the dust emission SED for dwarf galaxies. This object has been selected
from the Dwarf Galaxy Survey \citep{Remy-Ruyer2013} to represent the extreme SED behavior for low-metallicity star-forming galaxies 
and it features a much higher dust temperature ($T$=46.5~K) than normal
dust SFGs ($T\sim$20--30K) with a warmer mid-IR SED and weak PAH features. \citet{DeRossi2018} have also shown that the Haro 11 SED template 
is the best choice among other options for young, presumably sub-solar metallicity, galaxies  undergoing extremely  vigorous episodes of star formation. In Section~\ref{sec:sed-fitting-test},
we will test the performance of these two sets of galaxy dust emission templates with MIRI data and explore the situation where the Haro 11 template is preferred for  our SED fits.  Once the template is decided, there is only one free parameter $L\_IR\_obs$
to scale it to match the data together with other components.

Compared to \citet{Lyu2022}, several improvements have been made for the AGN component: 
\begin{itemize}
\item Based on the SDSS DR7 optical spectral archive \citep{Abazajian2009} and near-IR AGN templates in the 
literature \citep{Glikman2006, Hernan-Caballero2016}, we increased the resolution of the AGN template from the UV to
the near-IR and added AGN emission lines to the model. These emission lines are further decomposed into narrow
and broad components with the separation at FWHM=1200 km/s and these two groups of lines can be modified in different ways (e.g., 
obscuration at different physical scales).
\item We introduced a hybrid extinction configuration for the AGN component that considers the possibly different dust 
grain properties at different scales or angular directions. This setup features a Small Magellanic Cloud (SMC) curve 
for the commonly seen UV-optical extinction in type-1 AGNs and an empirical attenuation law constructed for the AGN IR 
obscuration. The former extinction law is associated with small dust grains at large physical scales far away
from the central engine and the latter is for the circumnuclear obscuration, such as the dusty torus, which is likely to be 
dominated by larger dust grains because of the harsh environment \citep{Baskin2018}. 
\end{itemize}
Figure~\ref{fig:agn-model} presents this updated AGN model with various emission lines and shows how the SED can be modified by the different
levels of extinction characterized by the two attenuation laws parameterized by $tau\_opt$ and $tau\_ir$. 
In our fitting, the AGN continuum and broad emission line component are bound together and can be obscured by 
both attenuation laws (i.e., $tau\_opt$ and $tau\_ir$) while the relative strength of the narrow line component can 
be only changed by the SMC extinction ($tau\_ir$). We adopt a uniform prior for $tau\_opt$ = [0, 3.0] and 
$tau\_ir$ = [0, 20.0] as these ranges cover the majority of AGN obscuration without too much redundancy. 
For the purpose of AGN selections, we also adopt the normal AGN template following \citet{Lyu2022, Lyu2022c}. In total, 
there are three parameters for the AGN component: (1) AGN luminosity ($L\_AGN$), (2) AGN SMC extinction level ($tau\_opt$) 
and (3) AGN IR extinction level ($tau\_ir$).

During the SED fittings, we fix the redshift if the object has a spectroscopic or grism redshift (from FRESCO or 3D-HST). For objects that 
only have JADES or 3D-HST photo-z, we will re-compute the photo-z with the new SED fittings that include the MIRI 
data points. Accordingly, we have up to 12 (11 if the redshift is fixed) free parameters for the SED fitting. We use 
Dynamic Nested Sampling to find the best-fitting results and assess the model degeneracies. Readers are highly encouraged 
to read the relevant sections in \citet{Lyu2022} for details. { A summary of the fitting parameters is given in Table~\ref{tab:model-setup}.} 

\begin{deluxetable*}{@{\extracolsep{4pt}}ccccc}
    \tabletypesize{\footnotesize}
    \tablewidth{1.0\hsize}
    \tablecolumns{5}
    \tablecaption{Parameter Setup of the Modified {\it Prospector} Code \label{tab:model-setup}
    }
    \tablehead{
  \colhead{Parameter} &
  \colhead{Free?} & 
  \colhead{Prior/Value} &
  \colhead{Unit} &
  \colhead{Comment}
}
\startdata
\multicolumn{5}{c}{Stellar Component (FSPS)} \\
\hline
  mass        & True  &  [1e6, 1e12], LogUniform      &  $M_\odot$   &   Solar masses formed   \\
  logzsol     & True  &  [-2, 0.19], TopHat           &   $\log (Z/Z_\odot)$  & Stellar metallicity \\
  sfh         & N/A   &  4  &  N/A   &  Delay-tau SFH is selected \\
  tage        & True  &  [0.001, 13.8], TopHat        &  Gyr   & stellar age \\
  tau         & True  &  [0.1, 30], LogUniform        &  Gyr$^{-1}$  & E-folding time of the SFH\\
  dust\_type  & N/A   &  4  &  N/A   & \citet{Kriek2013} attenuation law is selected \\
  dust2       & True  &  [0.0, 4.0], TopHat           &  dimensionless & stellar optical depth at 5500\AA   \\
  dust\_index & True  &  [-0.6, 0.3], Tophat          &  dimensionless &  power-law multiplication of Calzetti law \\
  \hline
\multicolumn{5}{c}{AGN Component} \\
\hline
  L\_AGN      & True  &  [1e-4, 1e6], LogUniform    &  $10^{10}L_\odot$  & AGN bolometric luminosity \\
  f\_hd       & False &  1.0, fixed      &  dimensionless  & relative strength of the AGN hot-dust component \\
  f\_wd       & False &  1.0, fixed      &  dimensionless  & relative strength of the AGN warm-dust component \\
  f\_pol      & False &  0.0, fixed      &  dimensionless  & relative strength of the AGN polar-dust component \\
  tau\_ir    & True  &  [0, 20], TopHat      &  dimensionless &  AGN optical depth at 5500\AA for the IR obscuration \\
  tau\_opt    & True  &  [0, 3], TopHat      &  dimensionless &  AGN optical depth at 5500\AA for the UV-optical obscuration \\
  \hline
\multicolumn{5}{c}{SFG Dust Component} \\
\hline
  L\_IR\_obs  & True  & [1e-4, 1e6], LogUniform & dimensionless &  Scaling factor of the SFG IR component \\
  L\_IR\_temp & False & 11.25, fixed &  $\log (L_{\rm IR, SF}/L_\odot)$  &  SFG template IR luminosity (8--1000~$\mu$m)\\
  f\_pah      & False &  1.0, fixed   & dimensionless & relative strength of the PAH component\\ 
\enddata
\end{deluxetable*}

Our approach has a number of advantages. First, our models for the AGN component and the SFG dust component are derived 
from empirical observations and have been extensively tested, as elaborated 
in detail in \citet{Lyu2022, LyuRieke2022}. This allows us to narrow the number and range of free parameters while retaining realistic fits, 
in contrast to approaches that are more strongly based on theoretical models. In addition, the SED model of our AGN component not only 
provides a good fit to the AGN observations, but can change continuously from the unobscured phase to the most heavily obscured 
phase, offering an unbiased view on the distribution of AGN obscuration properties. Moreover, compared to most other tools, our AGN model 
includes broad and narrow emission lines, a typically over-looked but now very needed feature for the much improved photometric data in 
the JWST era. Finally, although preliminary, our SED fittings have the option to use an empirical low-metallicity galaxy template to 
fit the IR SEDs of the dwarf galaxy population.

More details of this SED fitting package, including its public release, will be presented in a separate paper in the future.

\subsection{SED Fitting of MIRI Sources}\label{sec:sed-fitting-test}

We fit the SEDs for all the MIRI sources with the redshift priors described in Section~\ref{sec:data-redshift}. 
Within the current JADES footprint, the fits can utilize a maximum of 27 photometric bands that include
five HST/ACS bands at 0.44 -- 0.9~$\mu$m, fourteen JWST/NIRCam bands at 0.9 -- 4.4~$\mu$m and eight JWST/MIRI bands at 5.6 -- 25.5~$\mu$m. For 
the 89 MIRI sources outside the current JADES NIRCam catalog footprint, besides the MIRI photometry, we have 23 bands 
from the ASTRODEEP catalog at 0.3 -- 4.5~$\mu$m, which includes HST ACS and WFC3, ground-based deep optical and near-IR observations and {\it Spitzer}/IRAC bands 1 and 2. In the first round, we adopted the normal luminous SFG dust template for the fitting to get
an overall picture of the sample properties and to test the performance 
of our SED models.

\subsubsection{Testing the SED Models for the Massive Galaxy Population}

\begin{figure*}[htp]
    \begin{center}
        \includegraphics[width=0.495\hsize]{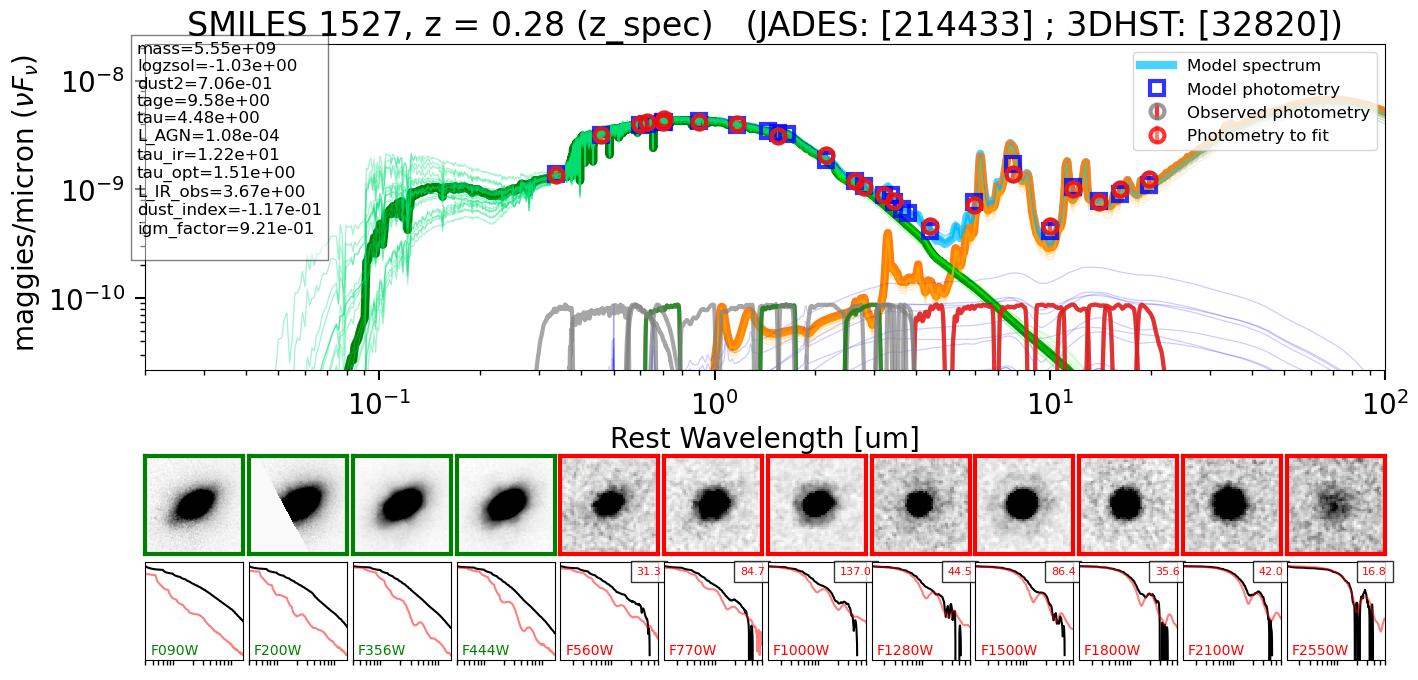}
        \includegraphics[width=0.495\hsize]{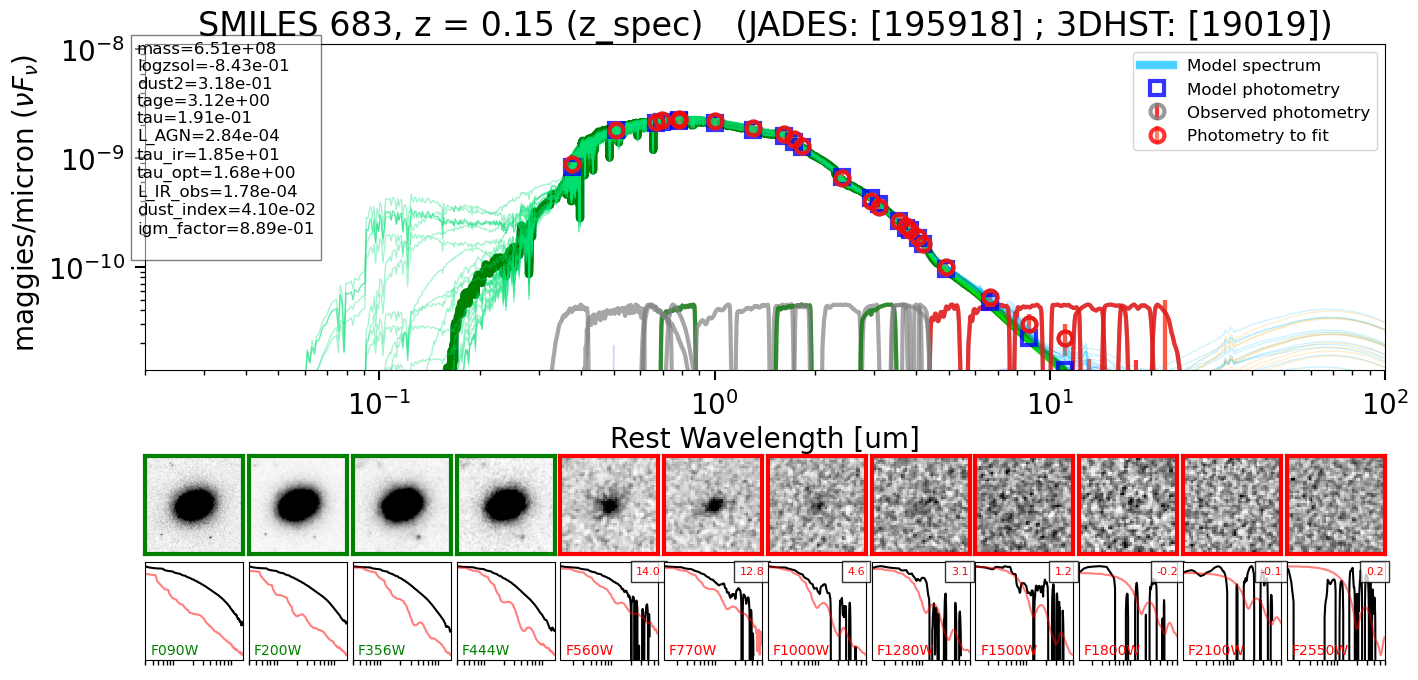}
        \includegraphics[width=0.495\hsize]{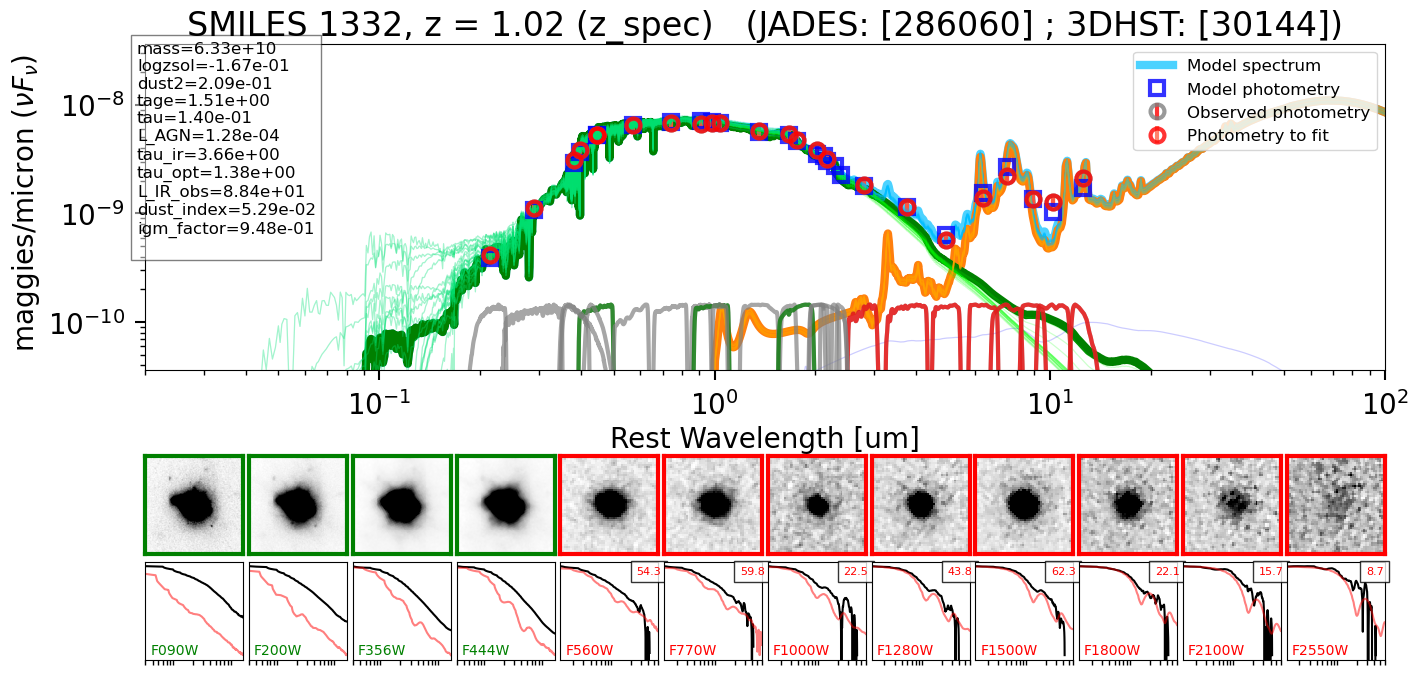}
        \includegraphics[width=0.495\hsize]{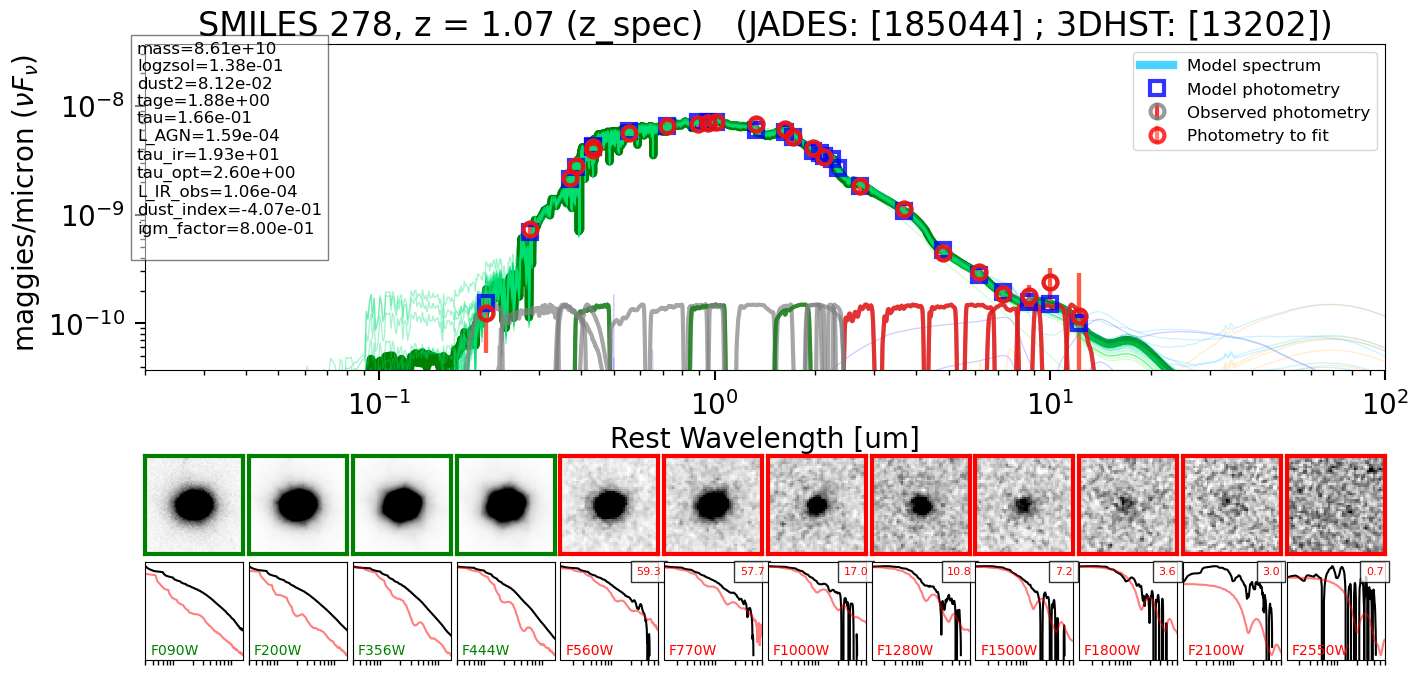}
        \includegraphics[width=0.495\hsize]{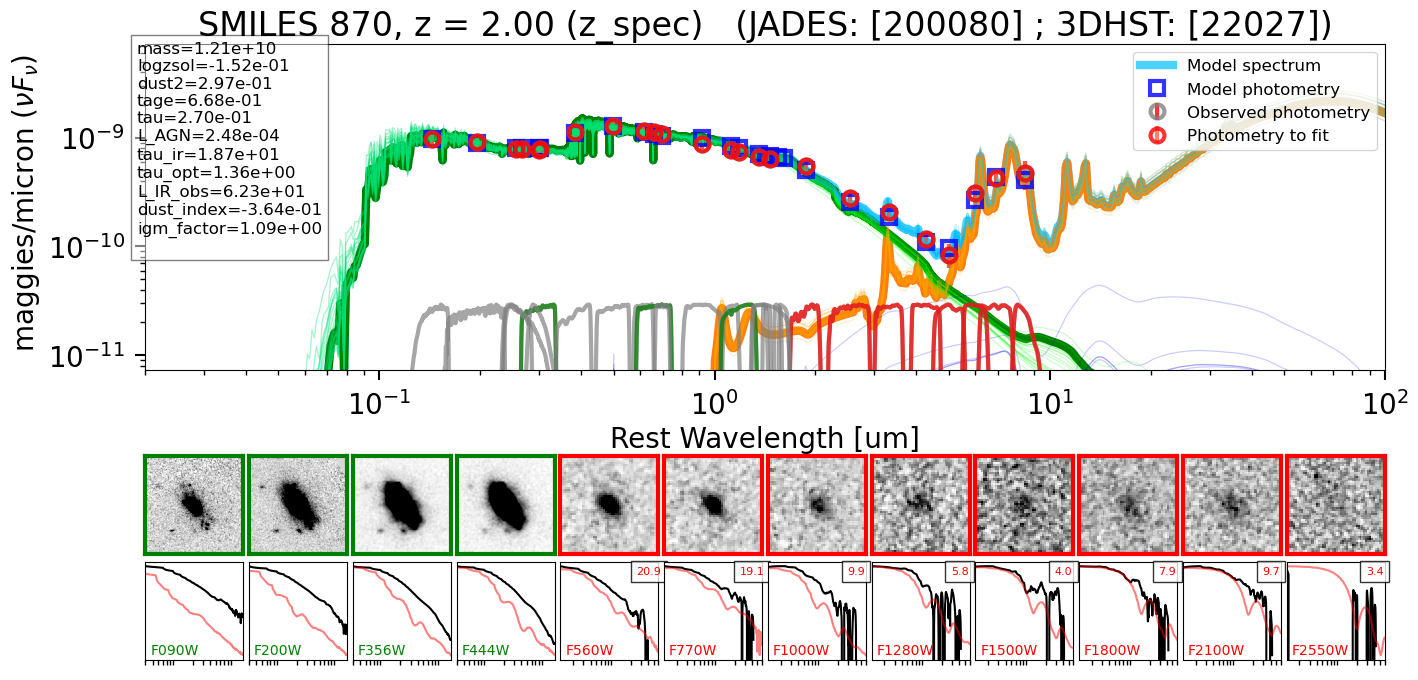}
        \includegraphics[width=0.495\hsize]{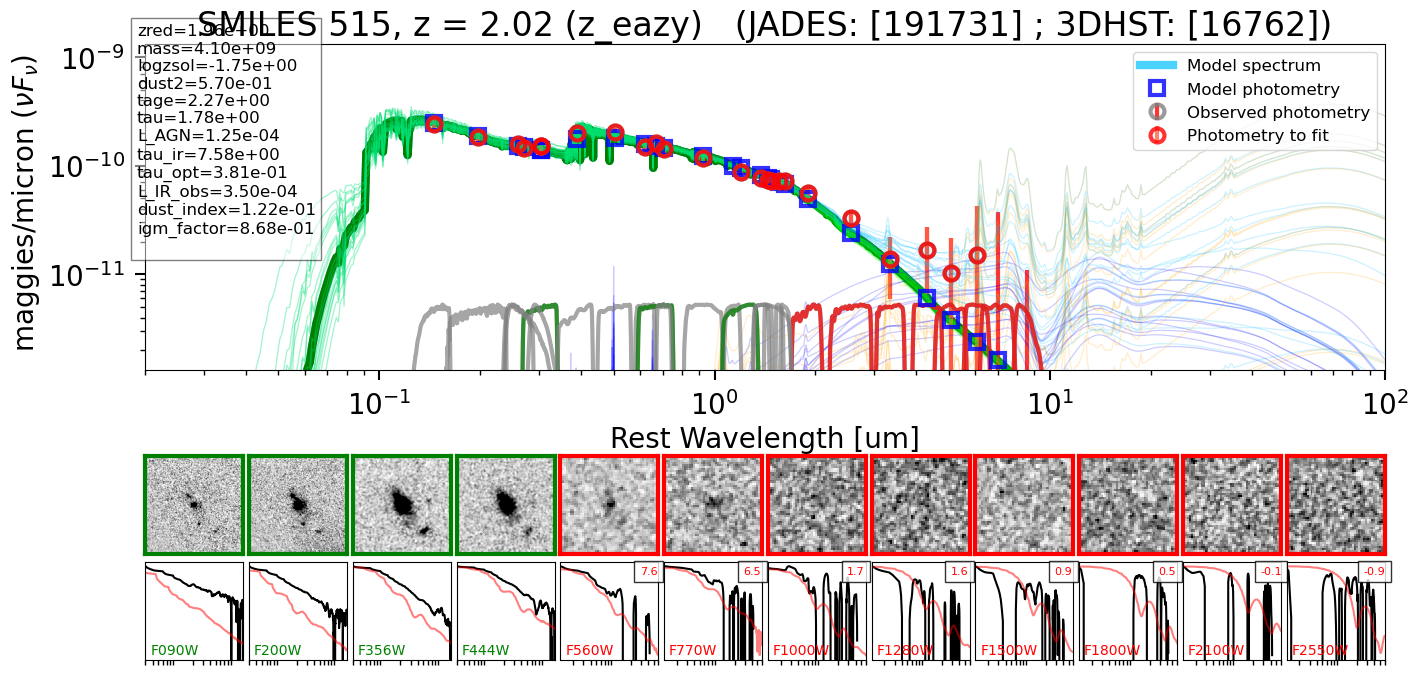}
        \includegraphics[width=0.495\hsize]{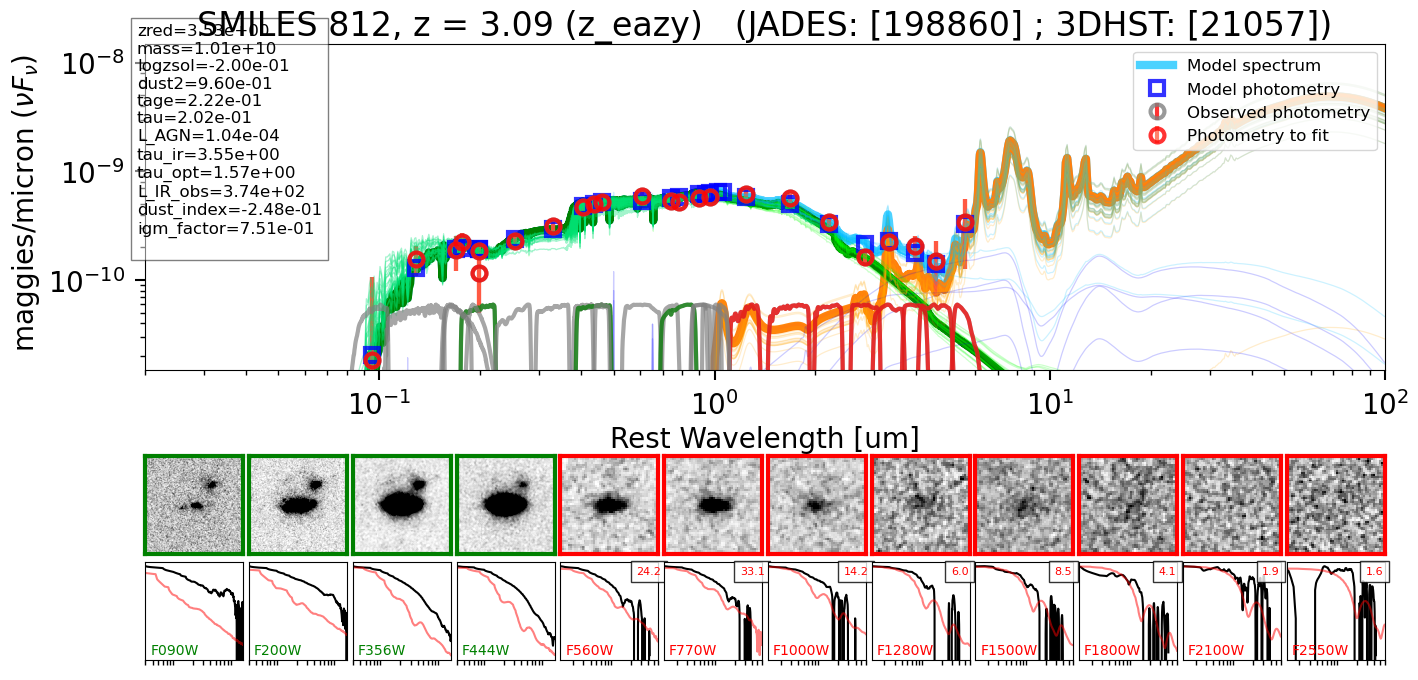}
        \includegraphics[width=0.495\hsize]{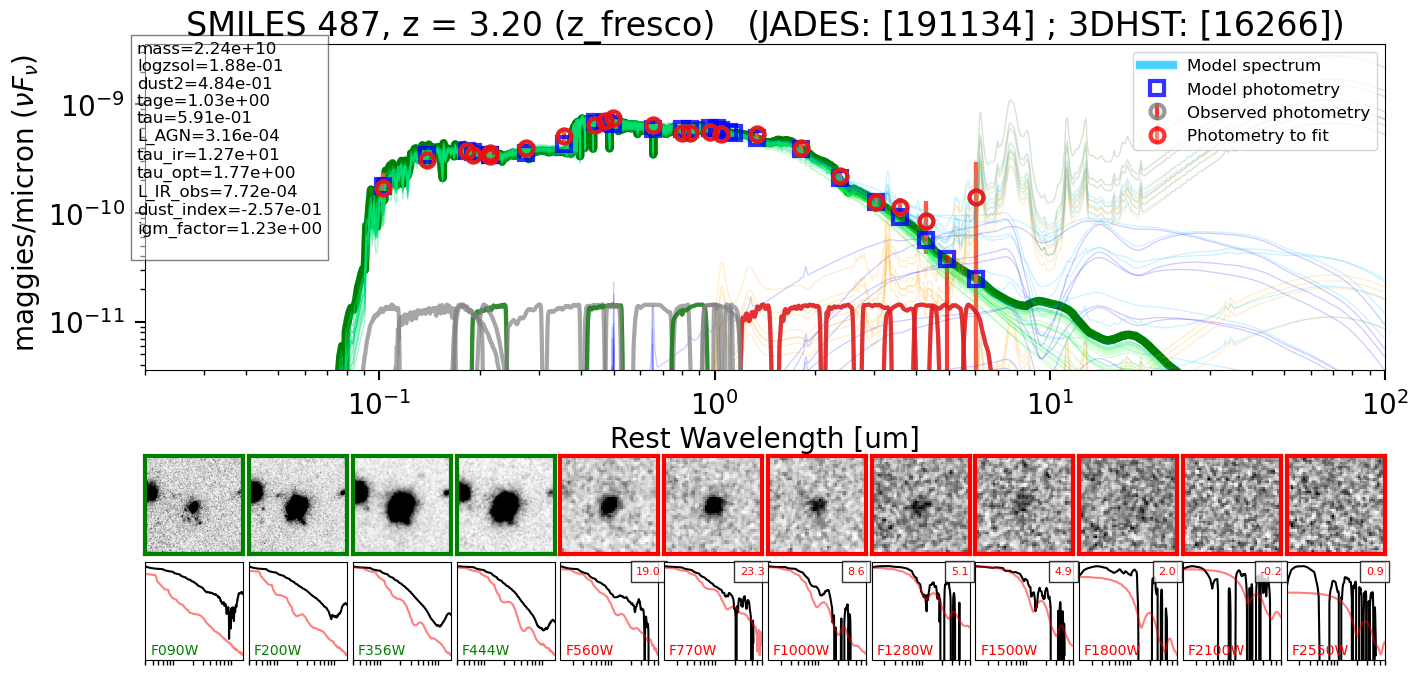}
        \includegraphics[width=0.495\hsize]{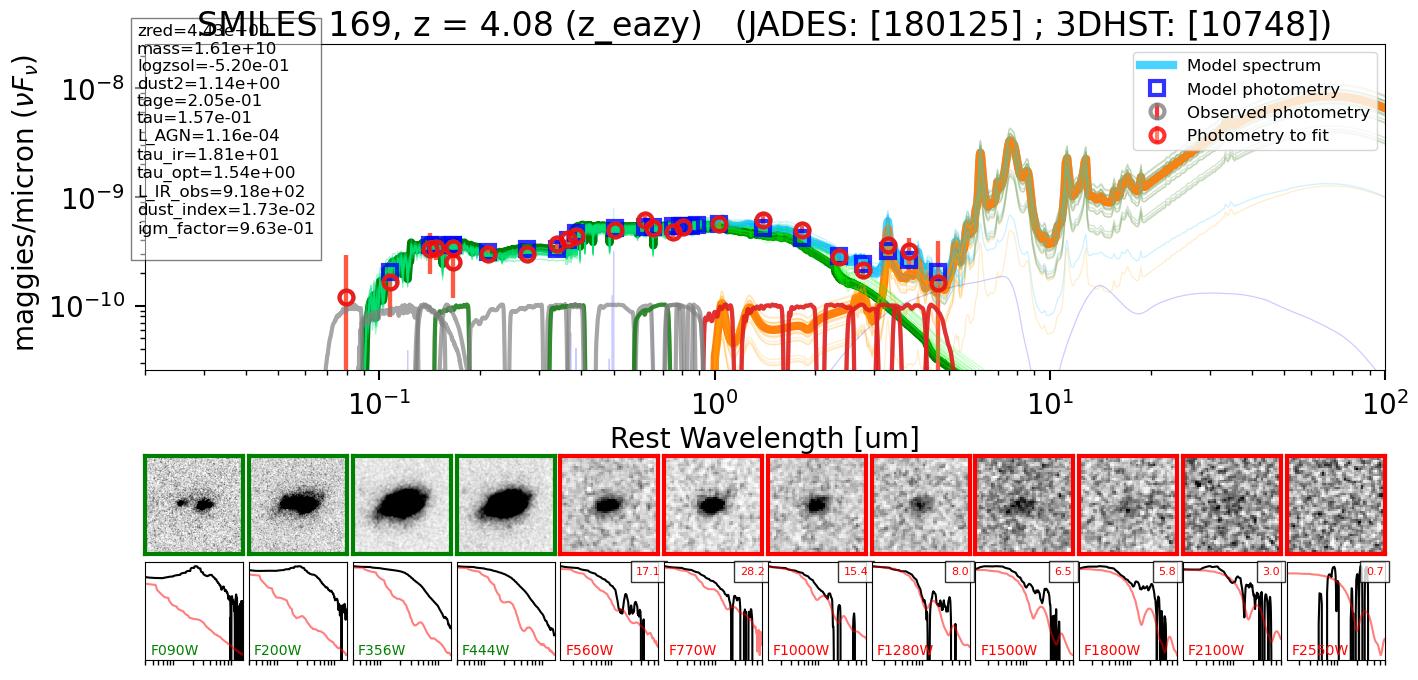} 
        \includegraphics[width=0.495\hsize]{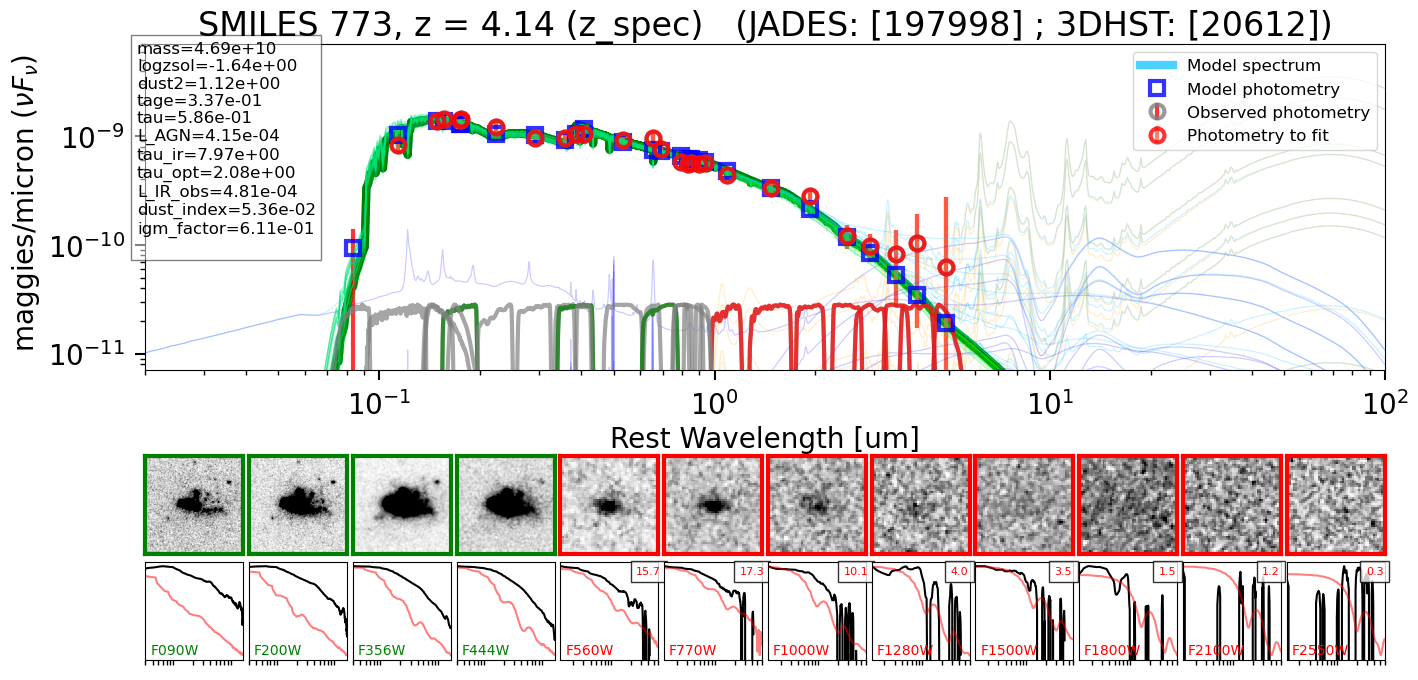}
    \caption{Example JWST NIRCam+MIRI SED fittings of dusty star-forming galaxies (left) and IR-quiescent galaxies (right). The object redshifts increase from top to bottom. Below each SED plot, we also show 3\arcsec$\times$3\arcsec cutouts of the source NIRCam and MIRI images and comparisons of the { normalized} source light profile (black line) and the corresponding ePSF (red line). {( The x-axis of the light profile panels spans from 0 to 6.0 arcsec.)}}
  \label{fig:sed-galaxy-examples}
    \end{center}
\end{figure*}

To demonstrate the robustness of our SED models, we present some representative SED fittings of dusty star-forming galaxies (DSFGs) and quiescent galaxies (QGs) from $z\sim0.2$--4 in Figure~\ref{fig:sed-galaxy-examples}. As we require good MIRI coverage to test our SED templates, most of these galaxies are  relatively massive for their redshifts, by selection (i.e., $M_*\gtrsim10^{9.5 - 10}~M_\odot$) and they form our primary sample. For DSFGs, various 
 SED features are successfully matched by our templates, showing that
the MIRI bands can be used to trace the various PAH features at e.g., 3.3, 6.2, 8.6, 11.3~$\mu$m. For the QG population, the 
MIRI photometry is deep enough to constrain the stellar Rayleigh-Jeans tail at rest-frame 5~$\mu$m even for galaxies at $z\sim4$, as shown in the bottom-right panels of Figure~\ref{fig:sed-galaxy-examples}.

The high-quality MIRI photometry as well as the satisfactory performance\footnote{ On average, the predicted photometry from our SED fittings agrees with the observations within 10\% and the median value of the reduced chi-square of the fittings is 1.6.} of our SED models for massive DSFGs and QGs establish a
solid foundation for identifying AGN signatures through SED analysis in such systems.

\subsubsection{Testing the SED Models for the AGN Population}

\begin{figure*}[htp]
    \begin{center}
        \includegraphics[width=0.495\hsize]{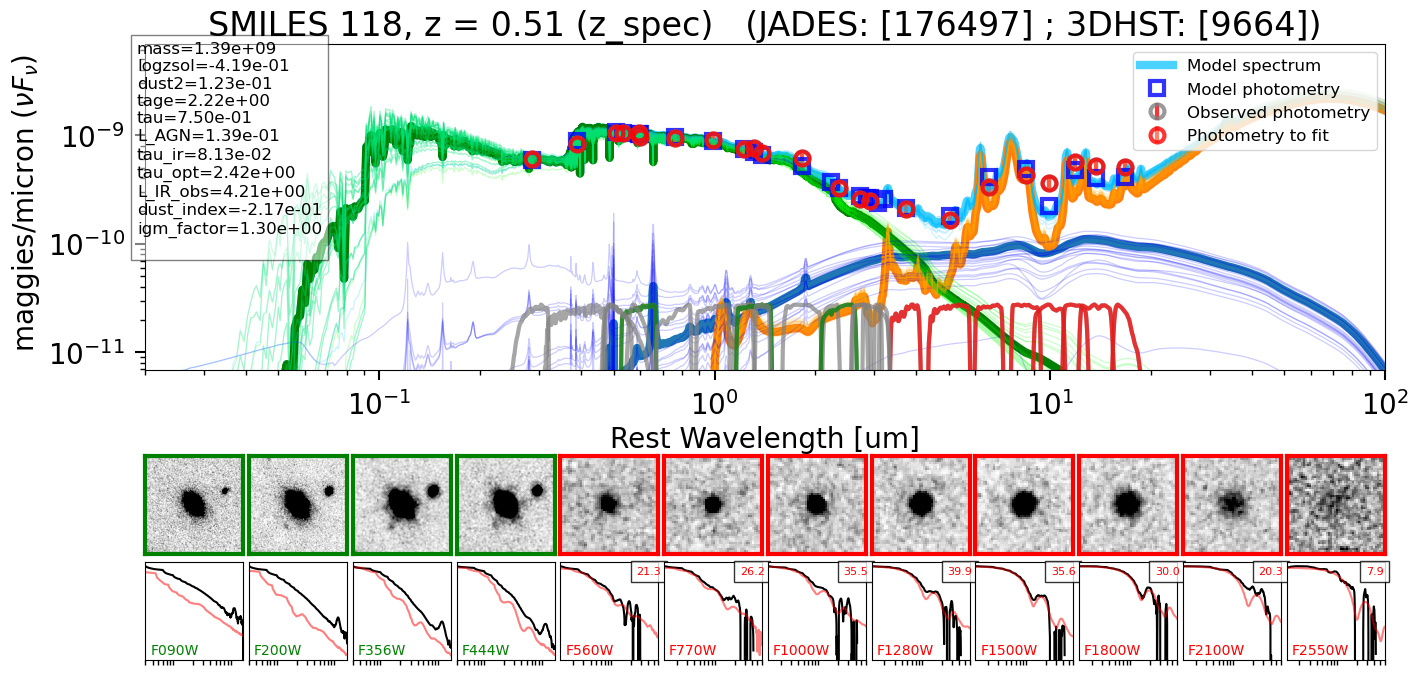}
        \includegraphics[width=0.495\hsize]{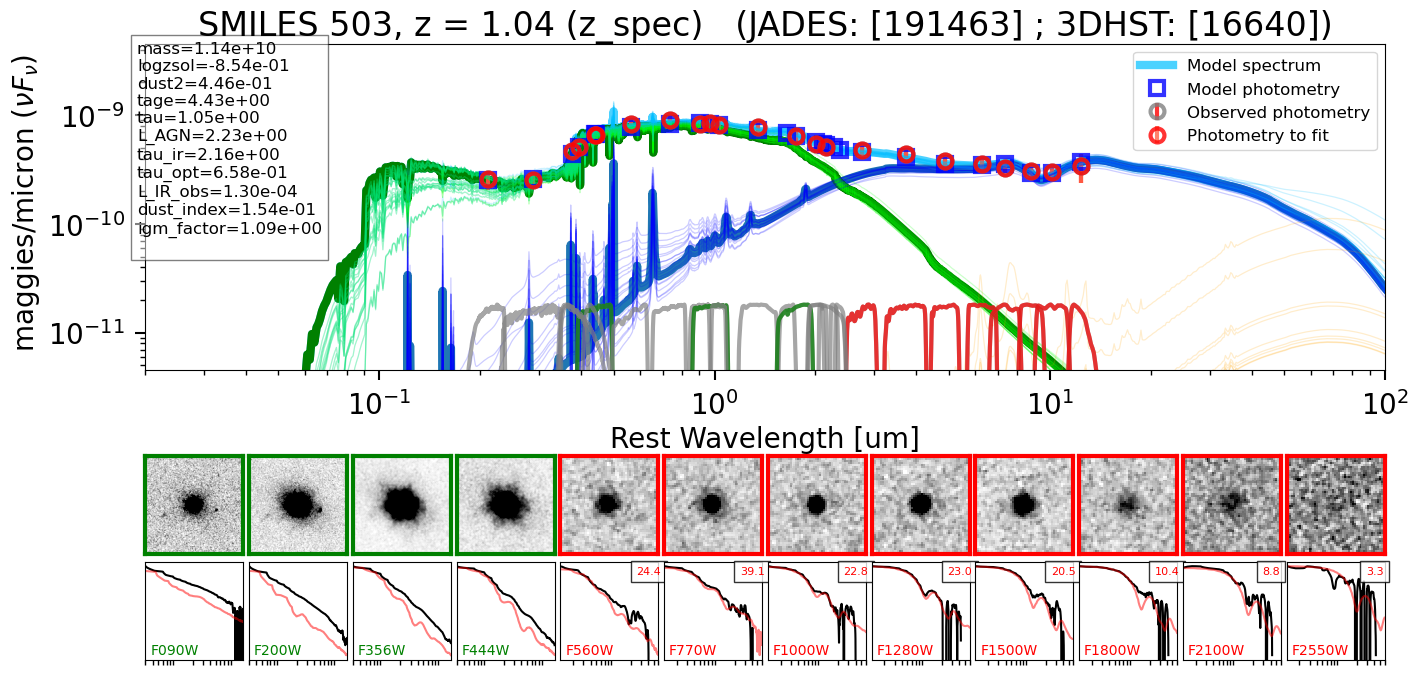}
        \includegraphics[width=0.495\hsize]{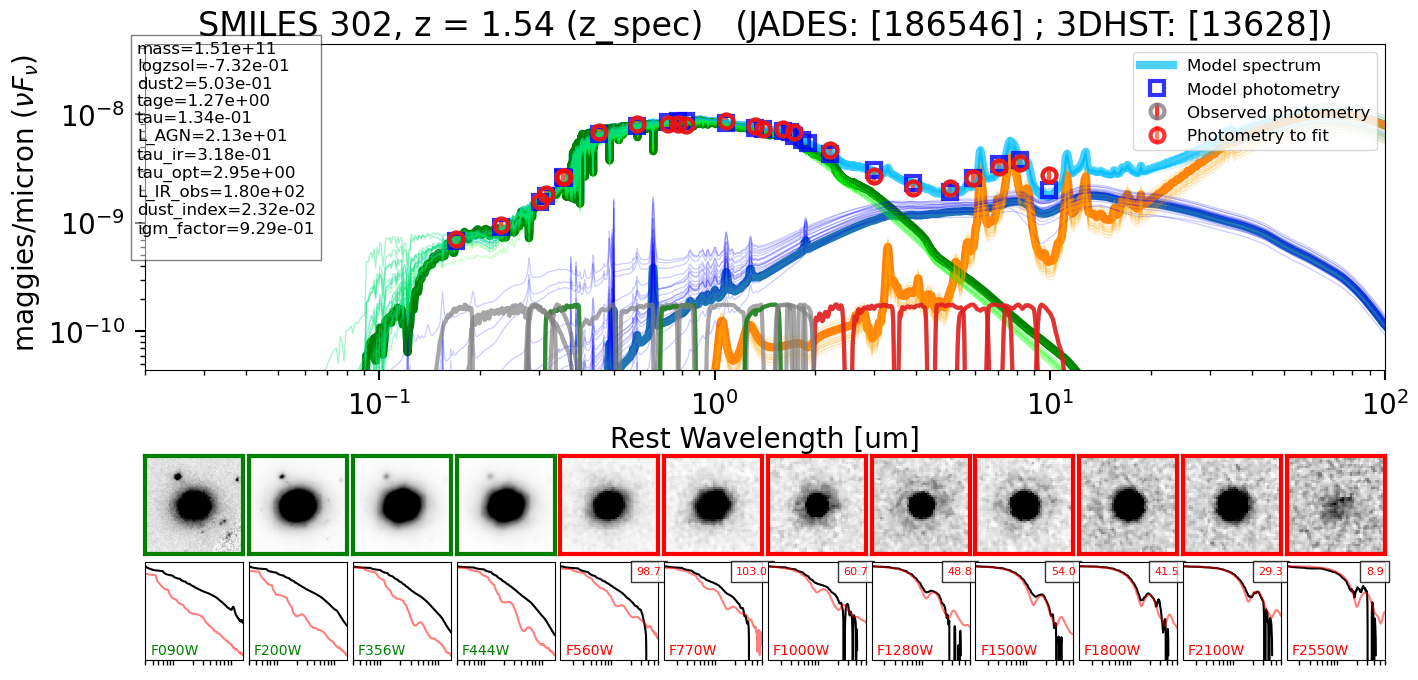}
        \includegraphics[width=0.495\hsize]{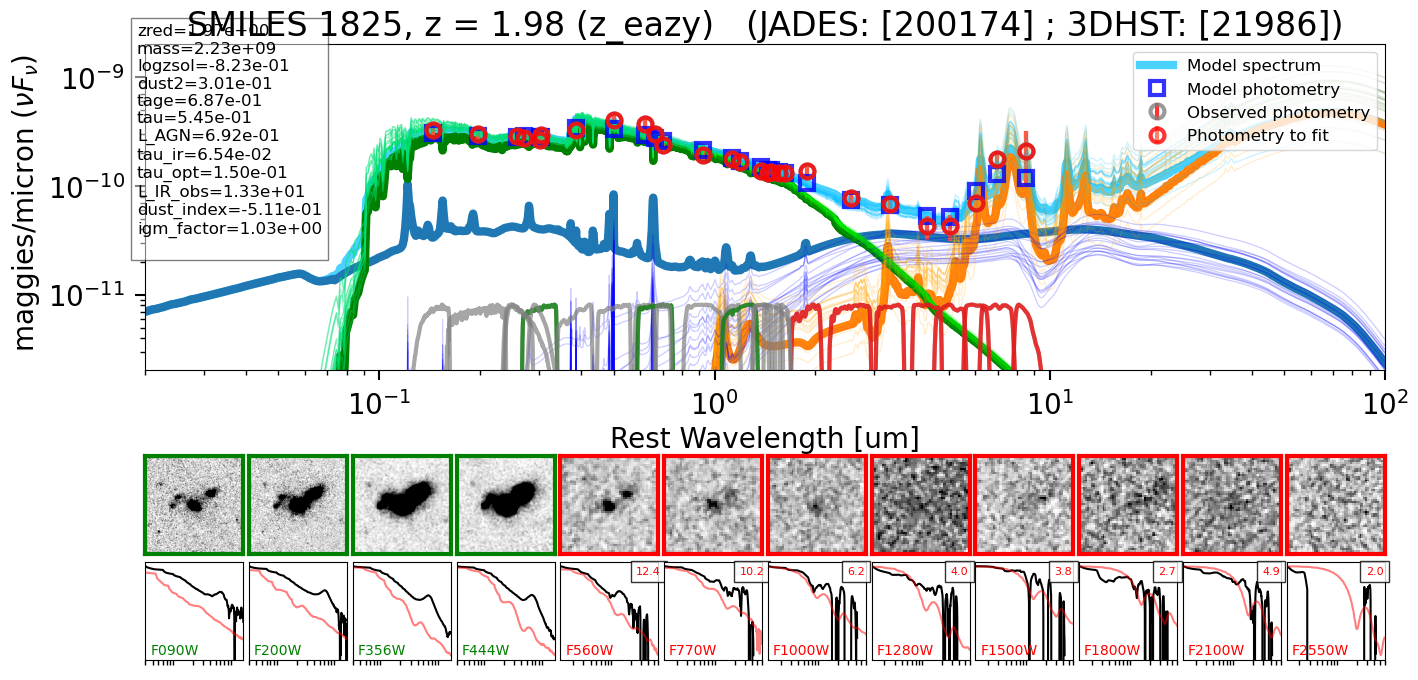}
        \includegraphics[width=0.495\hsize]{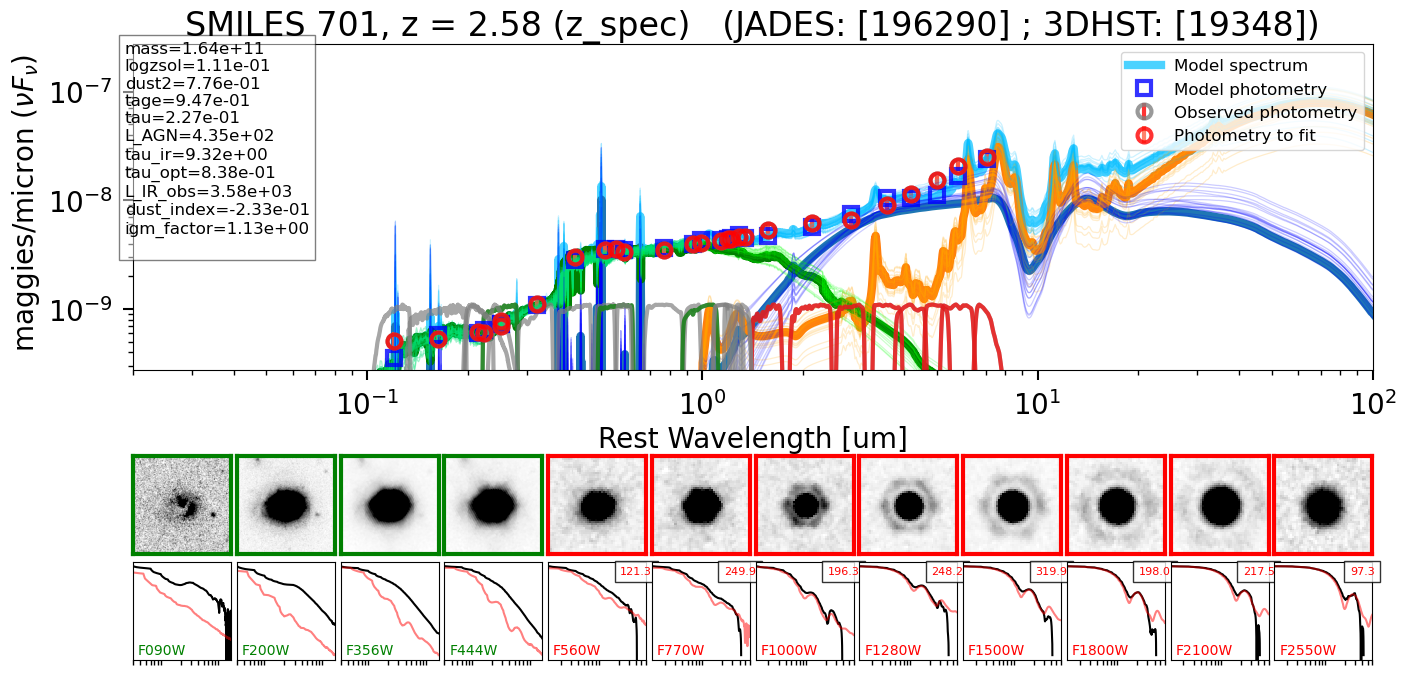}
        \includegraphics[width=0.495\hsize]{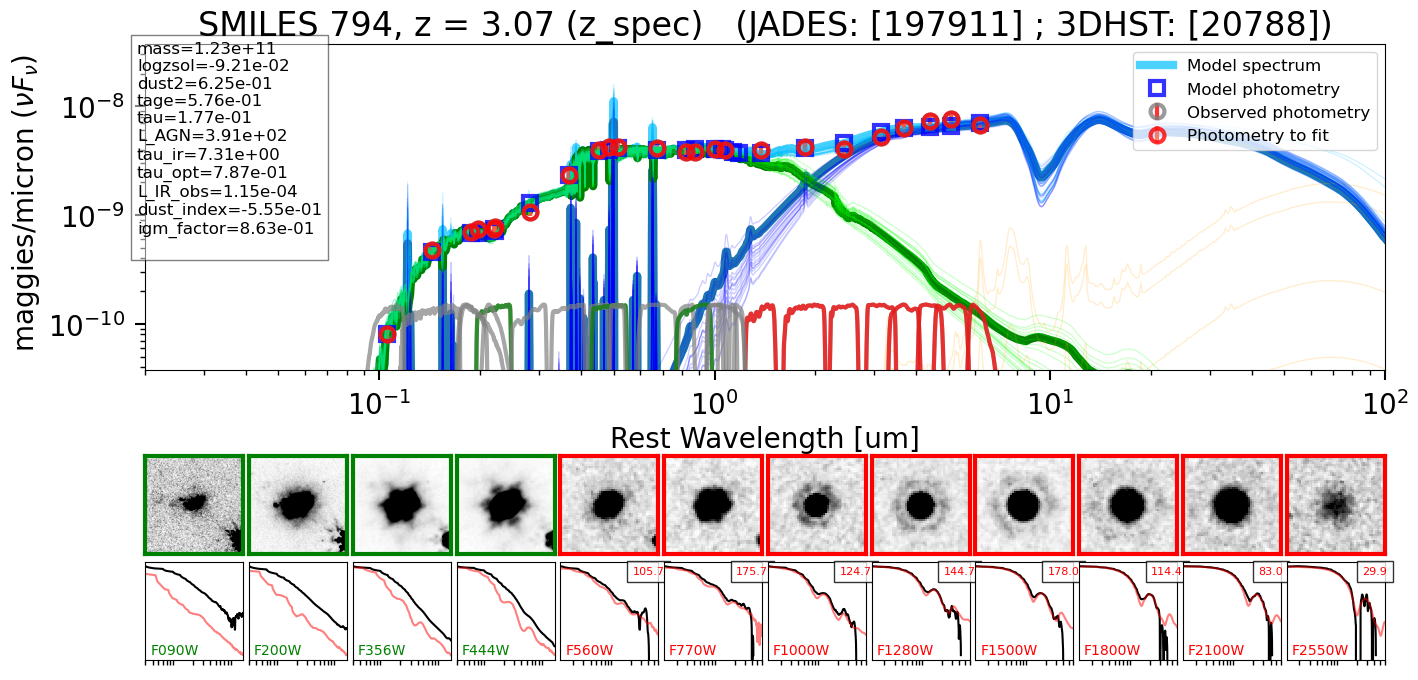}
        \includegraphics[width=0.495\hsize]{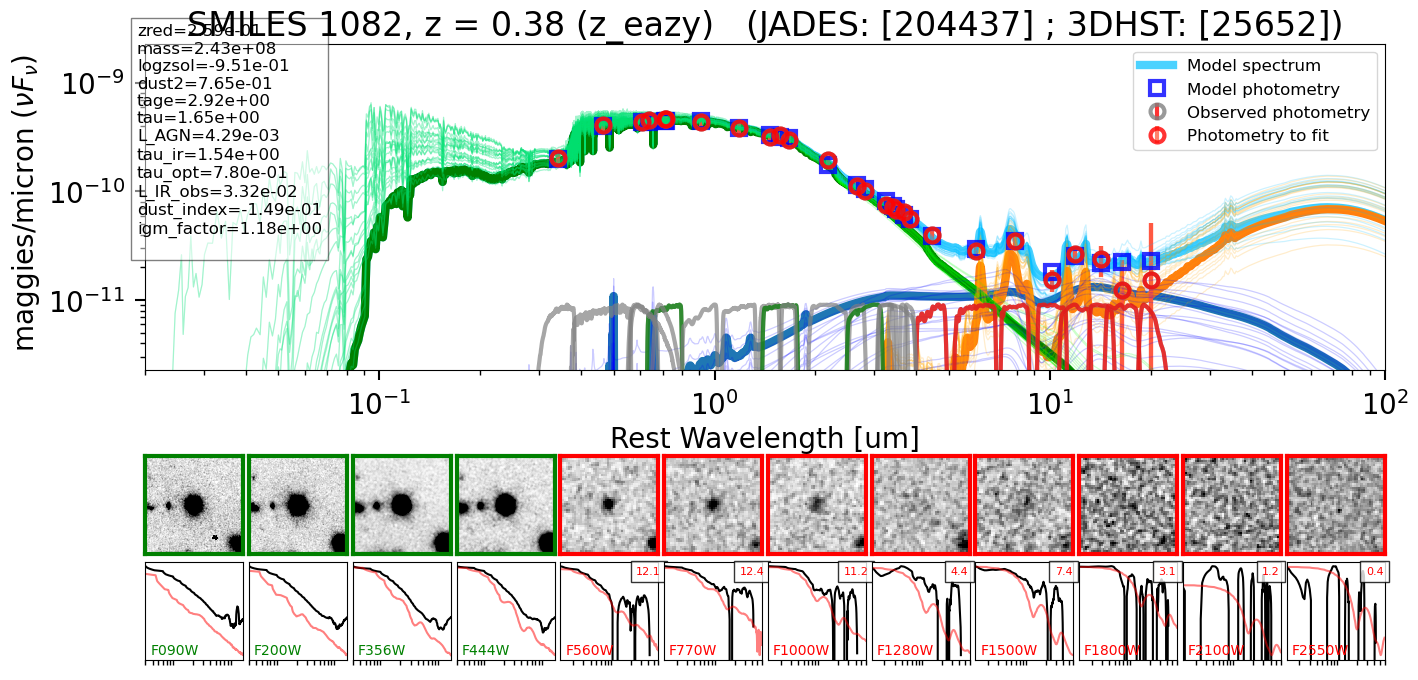}
        \includegraphics[width=0.495\hsize]{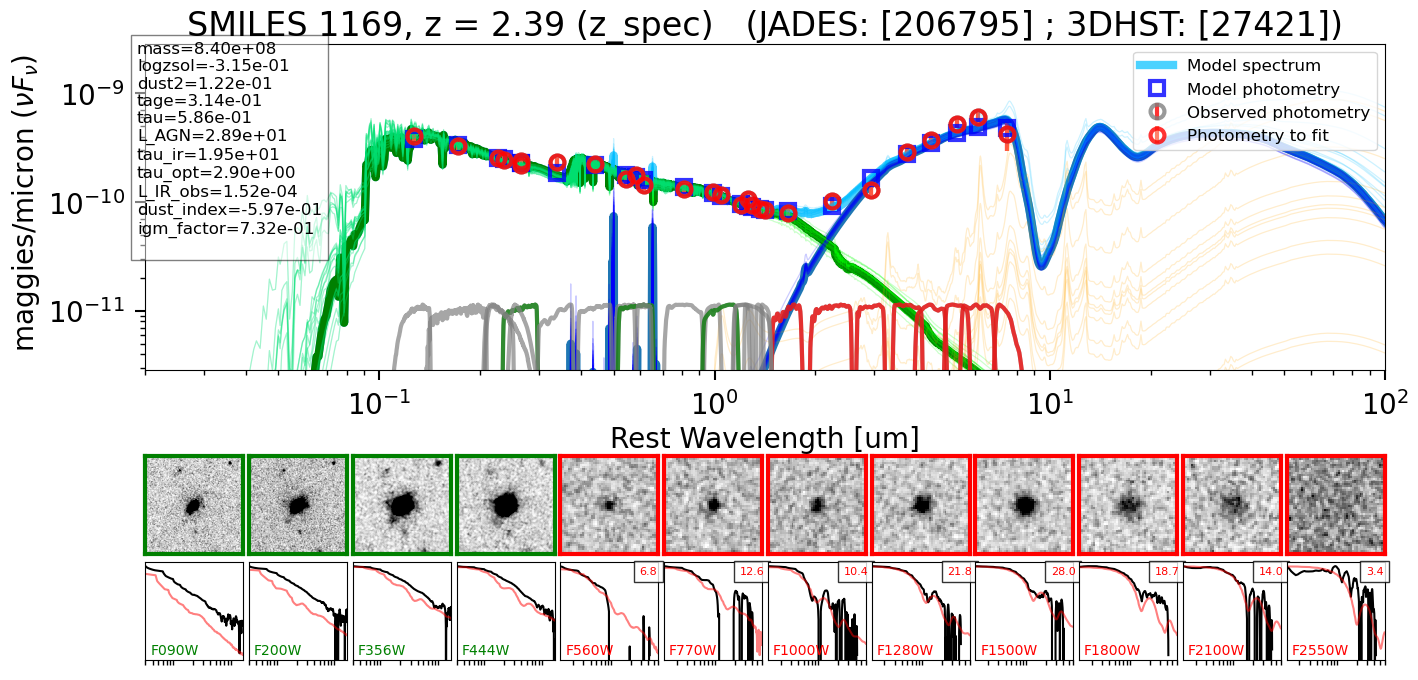}
        \includegraphics[width=0.495\hsize]{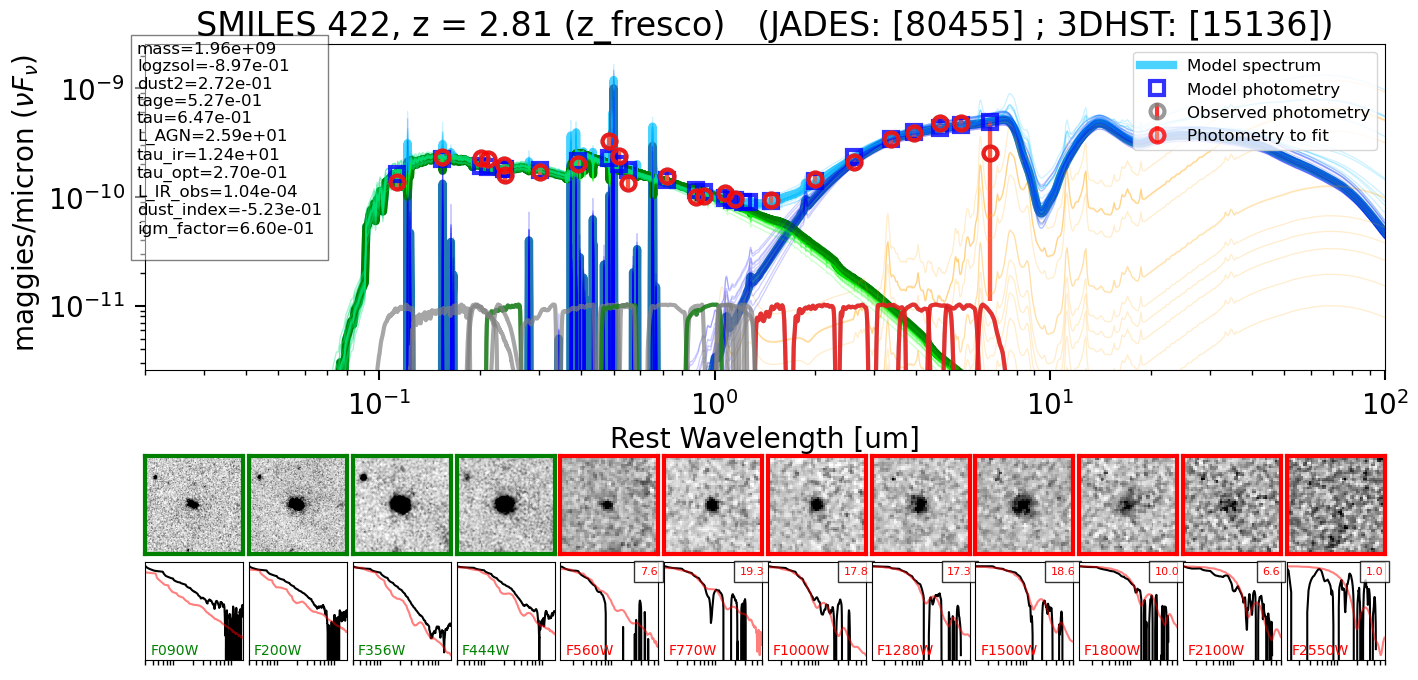} 
        \includegraphics[width=0.495\hsize]{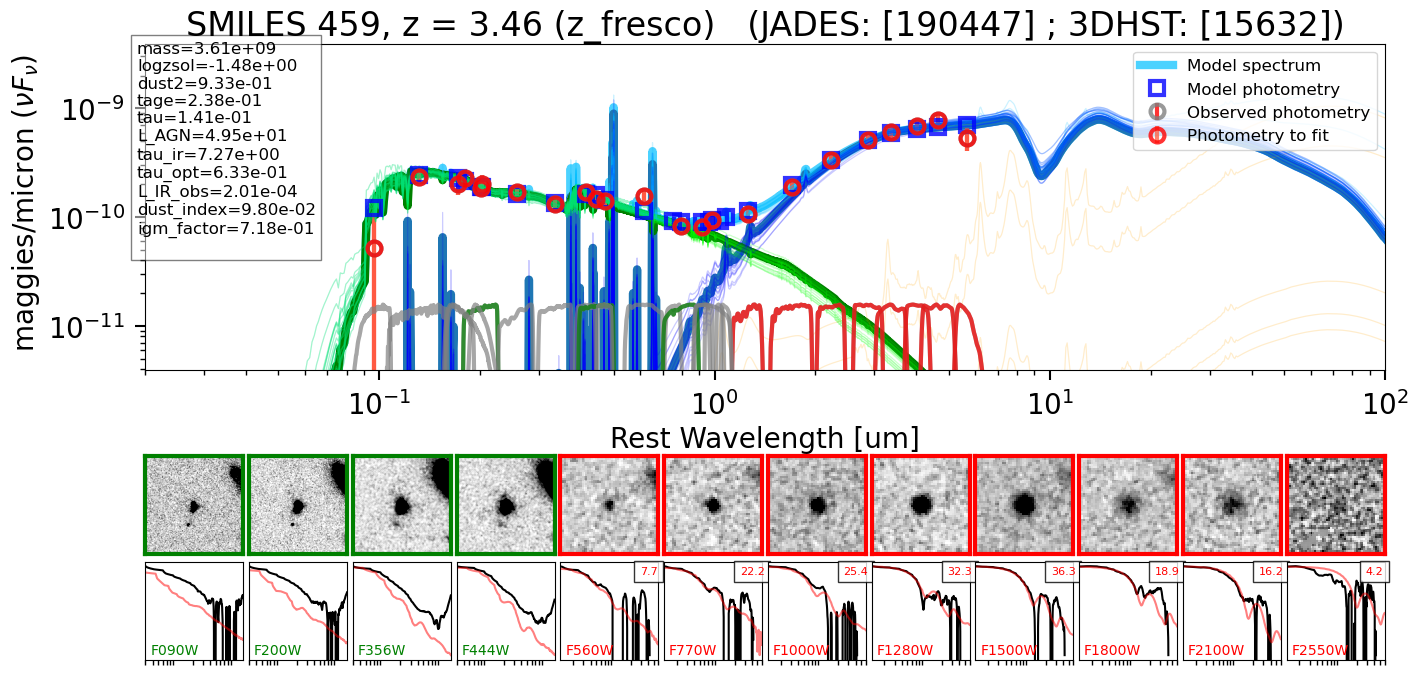}
    \caption{Similar to Figure~\ref{fig:sed-galaxy-examples} but for AGNs. In the top six panels, we show 
    normal AGNs with a range of nuclear obscuration and host galaxy contamination at $z<4$. The four panels in the bottom row are example AGNs in low-mass galaxies.}
  \label{fig:sed-agn-examples}
    \end{center}
\end{figure*}

Figure~\ref{fig:sed-agn-examples} shows some representative examples of MIRI sources with AGN evidence revealed
by SED fittings. Compared to galaxies without AGNs shown in Figure~\ref{fig:sed-galaxy-examples}, the SEDs of 
these objects present much larger variations that 
can be explained by the different level of AGN-galaxy contrast and the SED variations of the AGN component itself. Our examples
include AGNs with a wide range of redshift, obscuration level and galaxy contamination as revealed by the very successful SED 
decompositions.

\subsubsection{The Complicated Situation of the Dwarf Galaxy Population}\label{sec:dwarf-galaxy-sed}

\begin{figure*}[htp]
    \begin{center}
  \includegraphics[width=0.48\hsize]{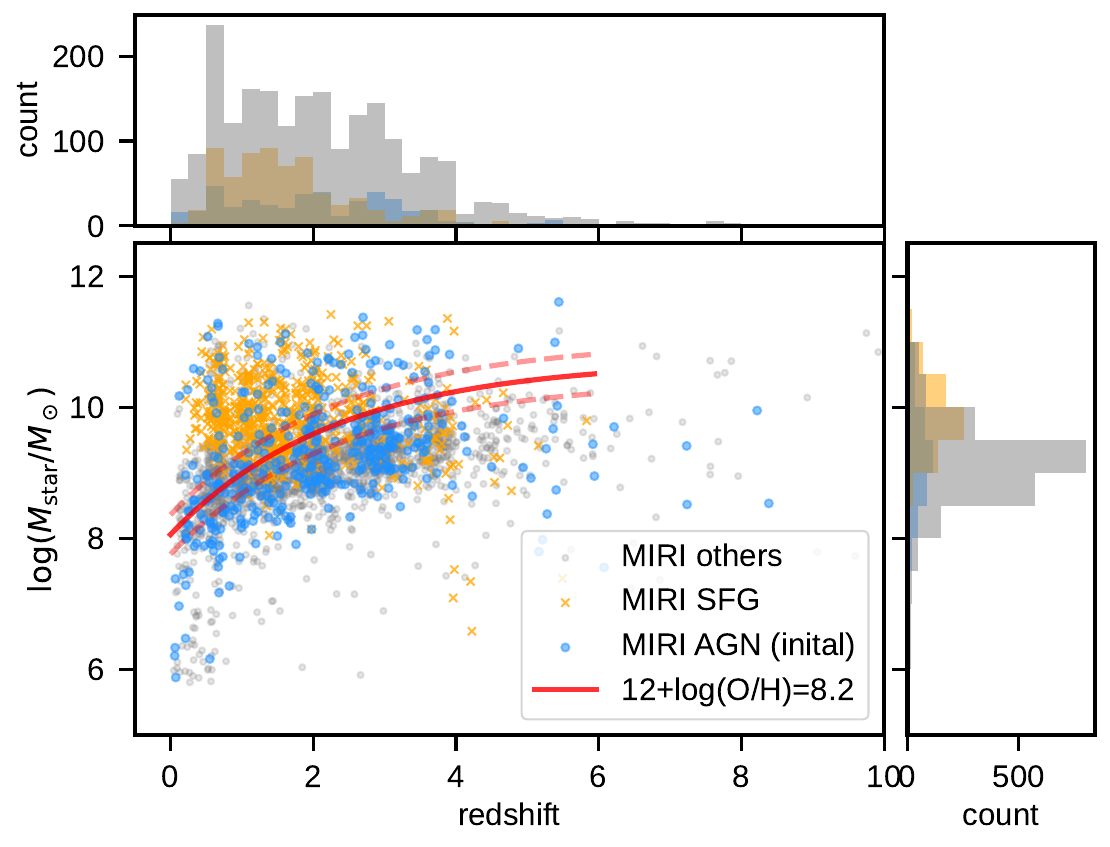}
  \includegraphics[width=0.48\hsize]{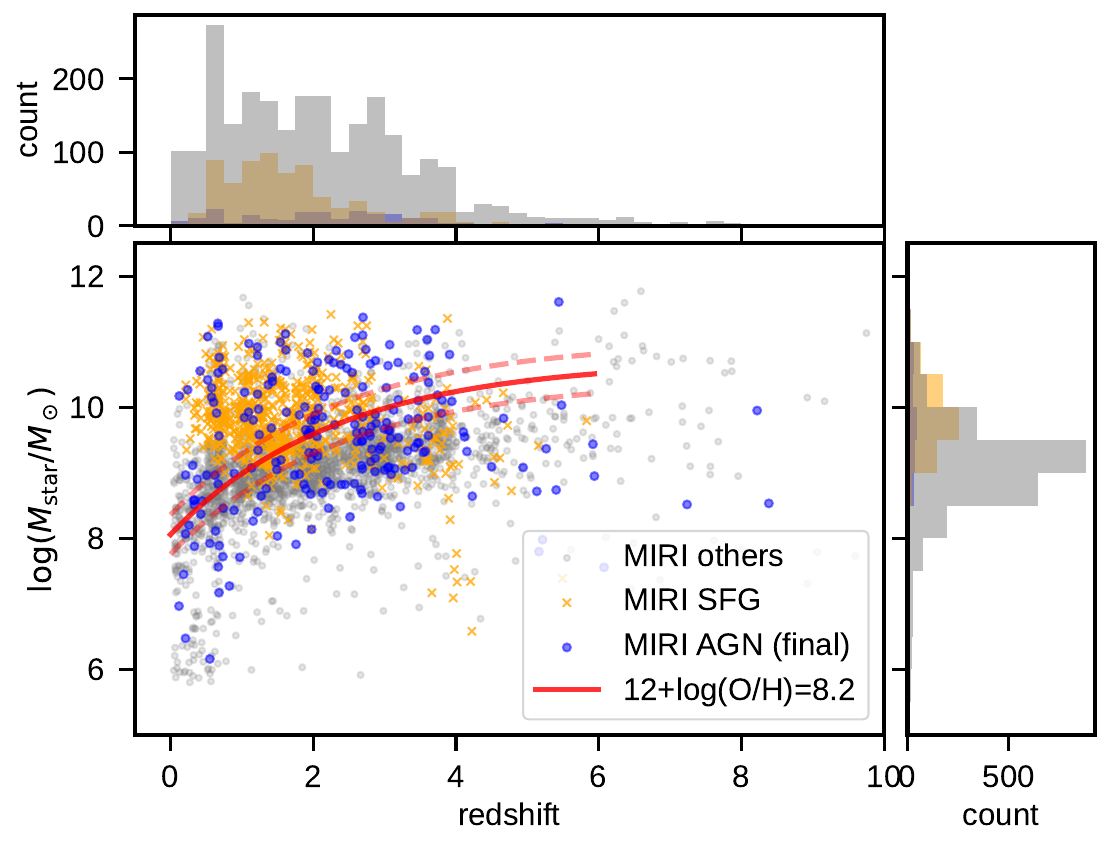}
    \caption{The stellar mass as a function of redshift for MIRI sources. We highlight all AGNs identified with the normal SFG template regardless
    of the object redshift and stellar mass in light blue on the left, and the refined AGN sample after adopting the Haro 11 dwarf SFG template
    to reject possible false positives among low-mass galaxies in dark blue on the right. In both two panels, we show the normal SFG galaxies 
    as orange crosses and other galaxies as
    gray dots. The red solid line represents the stellar mass threshold for 12+log(O/H)$\sim$8.2 with the $\pm$0.3 dex variation range as red dashed line. { If we adopt the same normal galaxy template for all stellar masses (left panel), there would be 500 AGN candidates selected. In contrast, we end up with 217 AGN candidates after considering the dwarf galaxy complication (right panel), as described in Sections~\ref{sec:dwarf-galaxy-sed} and \ref{sec:extended-selection}}}
  \label{fig:mstar-z}
    \end{center}
\end{figure*}

Besides reproducing the photometric SEDs, our fittings provide measurements of the galaxy stellar mass. As shown in Figure~\ref{fig:mstar-z}, this 
MIRI sample includes a considerable number of dwarf galaxies ($M_*<10^{9.5}~M_\sun$), where our templates for normal SFGs may not work. To determine where this may become an issue,  
we visually inspected the fittings of all the MIRI sources and picked out those pure SFGs where the updated \citet{Rieke2009} template
gives a good match (orange crosses). These ``normal'' SFGs have a stellar mass distribution peaked between $10^{9.5}$ and $10^{10}~M_\odot$, which is
about 0.5 dex above the median stellar mass of the whole MIRI galaxy population. However, we found that the IR SED features (e.g., PAH strength) of many low-mass galaxies are not 
captured by these ``normal'' SFG templates. Compared to the normal galaxy populations, these
low-mass galaxies could have different IR SED behavior associated with low-metallicity ISMs \citep[e.g.,][]{Remy-Ruyer2013, Shivaei2020, Shivaei2022}. 
In particular, the reduced PAH emission strengths and warmer dust continuum emission associated with the dwarf galaxy component can mimic the mid-IR colors of obscured AGNs to some level \citep[e.g.,][]{Hainline2016}, making the AGN identification with classical SED models not convincing. 
Thus, we need to introduce a method to remove most, if not all, false positives of AGN candidates in this extended sample.

Based on empirical observations, a correlation between PAH strength and metallicity has been revealed from the low-$z$ Universe \citep[e.g.,][]{Engelbracht2005, Engelbracht2008, Madden2006} to cosmic noon at $z\sim2.0$ \citep{Shivaei2017}. While the PAH intensity and relative strength of different bands stay almost 
the same at higher metallicity, below $\sim$ 12 + log(O/H) = 8.2 the PAH strength drops suddenly \citep{Engelbracht2005, Engelbracht2008, Marble2010, Aniano2020}.\footnote{The behavior at z $\sim$ 2 is not clear. \citet{Shivaei2017} indicate a drop below log(O/H)$\sim$8.3 for O3N2 metallicities, but a drop at a higher level for N2 metallicities. The physical basis for a large change with metallicity at redshift $\sim$ 2 is not clear, so we assume the local (and O3N2) behavior.} Given the correlation between the galaxy stellar mass and metallicity observed
over a wide range of stellar mass and redshift \citep[e.g.,][]{Tremonti2004, Lee2006, Zahid2013}, we  use  the stellar
mass from our SED fittings to approximate the gas metallicity 12 + log(O/H) following the empirical fittings 
in \citet{Ma2016}, and adopt the dwarf galaxy dust template (see Section~\ref{sec:sed-model}) when the metallicity (or stellar mass) threshold is not met. According to fittings in \citet{Ma2016},
the corresponding mass threshold $M_{\rm *, lim}$ as a function of redshift $z$ for 12 + log(O/H) $\sim$8.2 is 
\begin{equation}\label{equ:mass_lim}
\log(M_{\rm *, lim}/M_\odot) = 10.714 + 2.657\exp(-0.43z)
\end{equation}
This curve is shown as the red solid line in Figure~\ref{fig:mstar-z}. To account for the uncertainties
of e.g., stellar mass measurements and the dispersion of the mass-metallicity relation, we will compare the  normal and dwarf galaxy dust template fittings for all objects within $\pm$0.3 dex of $M_{\rm *, lim}$ (dashed red lines), and use the normal galaxy dust model results for galaxies above +0.3 dex of this mass threshold and use the dwarf model below  -0.3 dex of it. In addition, this threshold is not a hard cut as 
we will compare the fittings with the normal SFG and Haro 11 templates for objects within $\pm$0.3 dex of the mass cut, 
corresponding to a 12+(O/H) range from 8.3 to 8.1.

Despite these steps, it is possible that the AGN selections in our primary sample are mildly contaminated by cases with stellar-heated hot dust. In the case of our extended sample, Haro 11 is one of the most extreme cases with significant 
warm dust emission and weak PAH strength among the Dwarf Galaxy Survey \citep{Remy-Ruyer2013}. The usage of this template therefore may have biased our results against AGN 
identifications in this sample, as discussed in Section~\ref{sec:extended-selection}.

\subsection{AGN Identifications through SED Analysis}\label{sec:sed-selection}

We now describe our AGN identifications through SED analysis for the three sub-samples of our full object list. Following the same strategies in \citet{Lyu2022}, we picked out AGN candidates by analyzing the quality of the SED fittings, including inspections of any degenerate models from the posterior distributions of the fitted parameters. Typically, the AGN solution will be accepted if the distribution
of the fitted AGN luminosity parameter \textit{L\_AGN} has one single isolated peak and the AGN-dominated bands are not all noisy (i.e., S/N$<$3.0).  In addition, the high spatial resolution provided by NIRCam and MIRI allows us to further check if the AGN evidence inferred from the SED 
analysis is supported by the galaxy having a point-like core. Depending
on galaxy stellar mass and redshift, the AGN identifications have some additional considerations described below.

\subsubsection{AGNs in Massive Galaxies (Primary Sample)}
\label{sec:primary-selection}

For typical massive galaxies ($M_*\gtrsim10^{9.5}~M_\odot$) at $z=$0--4, the MIRI data has good coverage of the AGN warm dust features and our templates are well-tested.
As a result, we treat the selected objects as SED-identified AGNs. In total, we identify 111 AGNs under this category. 

In the top six panels of Figure~\ref{fig:sed-agn-examples}, we show some representative SED fittings together with some NIRCam and MIRI images at the
source location. In general, once the AGN emission dominates in a spectral band according to our SED decomposition, the corresponding radial profile of the source in that band is  
consistent with the instrument PSF, supporting the validity of our SED interpretation.

\subsubsection{AGNs in Dwarf Galaxies (Extended Sample)}
\label{sec:extended-selection}

\begin{figure}[htp]
    \begin{center}
  \includegraphics[width=1.0\hsize]{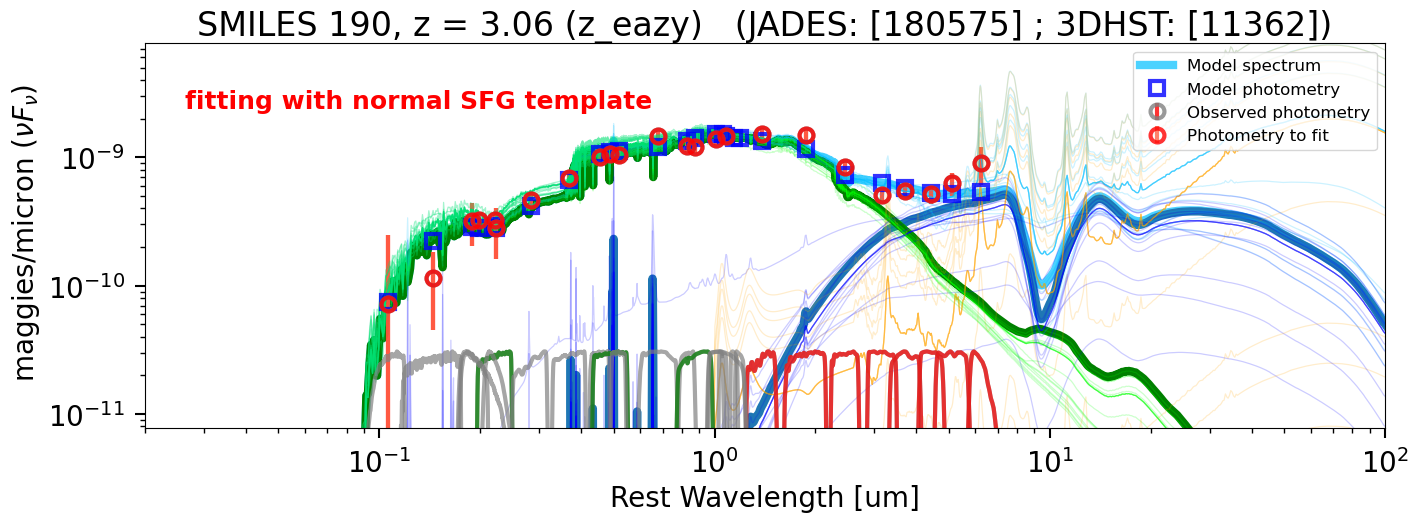}
  \includegraphics[width=1.0\hsize]{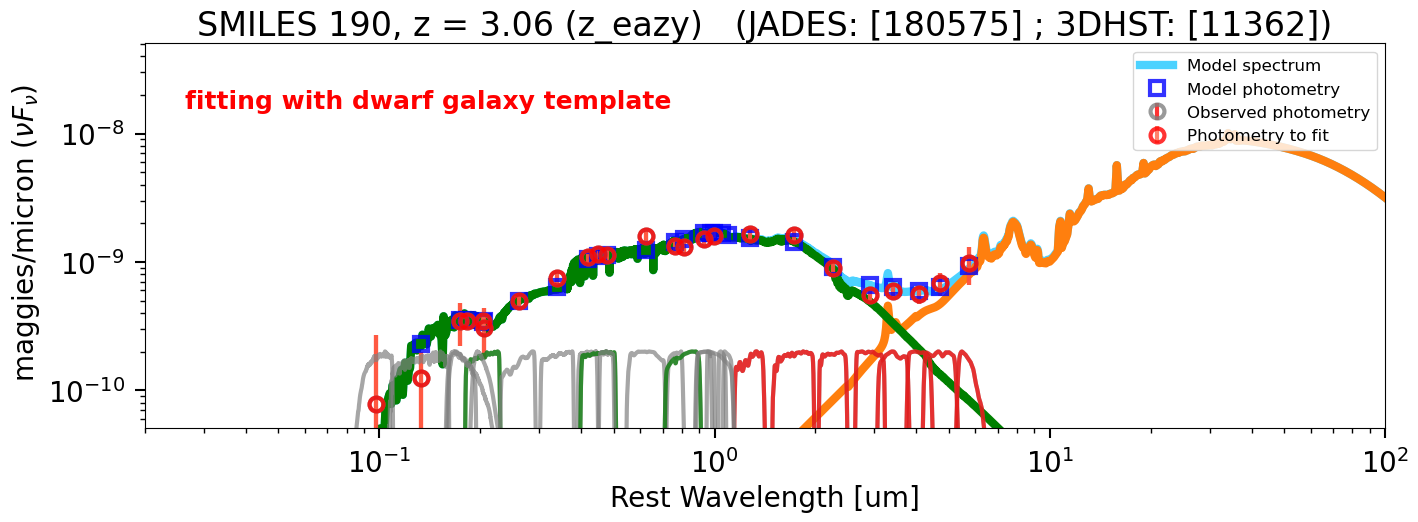}
    \caption{Example SED fittings of a low-mass galaxy. The top panel shows the object is identified as an AGN if we use the typical templates for normal SFGs and
    the bottom panels shows the fittings with the Haro 11 template, where the AGN evidence is gone.}
  \label{fig:dwarf-fitting-example}
    \end{center}
\end{figure}

As discussed in Section~\ref{sec:dwarf-galaxy-sed}, low-metallicity dwarf galaxies have IR SED shapes that can mimic the relatively hot dust emission 
features of AGN (see one example in Figure~\ref{fig:dwarf-fitting-example}). For such galaxies, we replace the massive ``normal'' SFG template with a dwarf galaxy template based upon Haro 11 to fit the SEDs and look for AGN 
candidates. An additional challenge is the reduction of MIRI coverage at longer wavelengths at the lower infrared luminosities of these objects. These two effects restrict our identification of AGNs in these galaxies to cases with SEDs significantly bluer (hotter) than that of Haro 11. Our selection is probably missing some obscured AGNs with SEDs similar to that of Haro 11.

However, using the Haro 11 template greatly reduces the chance of 
a source being mis-classified as an AGN, especially for SF-AGN composite systems. We have shifted the mass threshold to test whether the value of our adopted mass 
threshold is reasonable. If we shift it upward by 0.3 dex (corresponding to higher gas metallicity), some previous AGNs selected 
with the normal SFG template are re-classified as pure star-forming (dwarf) galaxies as expected. However, most such objects are 
either X-ray sources or show point-like morphology in the NIRCam or MIRI images, indicating that they { might be} real AGNs and the 
additional shift of the threshold is not desirable.\footnote{ We note that the X-ray detection or the point-like morphology do not justify the source is an AGN as some star-bursting dwarf galaxies do have very compact IR morphology or X-ray emission (e.g., Haro 11; \citealt{Prestwich2015, Lyu2016}). It is not clear how common these features are.} As 
a result, we think the current mass threshold in Equation~\ref{equ:mass_lim} is valid.

We adopted the definition of dwarf galaxies to be $M_*\lesssim10^{9.5}~M_\odot$, regardless of the object redshift. In the first round, 
187 AGN candidates that belong to this category are picked out. We have visually inspected all of these fits and rejected cases where the results 
are ambiguous due to low signal to noise, inconsistencies in the photometry from band to band, etc. Finally, 86 convincing AGN detections are left. 
This value is a lower limit given the limitations discussed above, particularly if AGNs with SEDs similar to that of Haro 11 have been 
overlooked. Methods to improve the AGN identifications throughout this extended sample need further development.

In Figure~\ref{fig:sed-agn-examples}, the second row from the bottom shows two example dwarf galaxies where the AGN
revealed by the MIRI SEDs is consistent with the obscured AGN template. The SED of MIRI 431 even shows evidence of strong silicate absorption,
a typical IR tracer for a heavily embedded nucleus.

\subsubsection{AGNs at $z>4$ (High-$z$ Sample)}

For galaxies at $z>$ 4, the MIRI data at the longest wavelengths are typically not deep enough to constrain
the AGN hot dust SED shape well and the other MIRI bands are probing the rest-frame near infrared (or even optical)  bands where some of the signatures of 
obscured AGNs are no longer accessible.  Therefore, the identification of high-$z$ AGNs is relatively less robust and likely only yields a subset. { Given the possibility that the IR SEDs of galaxies at $z\gtrsim4$ resemble that of Haro 11 \citep{DeRossi2018}, we adopt the Haro 11 template for the SED fittings of these high-$z$ systems to make a conservative selection.}
In total, we identify 20 AGN candidates at z $>$ 4 with the highest redshift at $z = $8.4. 

\begin{figure*}[htp]
    \begin{center}
        \includegraphics[width=0.495\hsize]{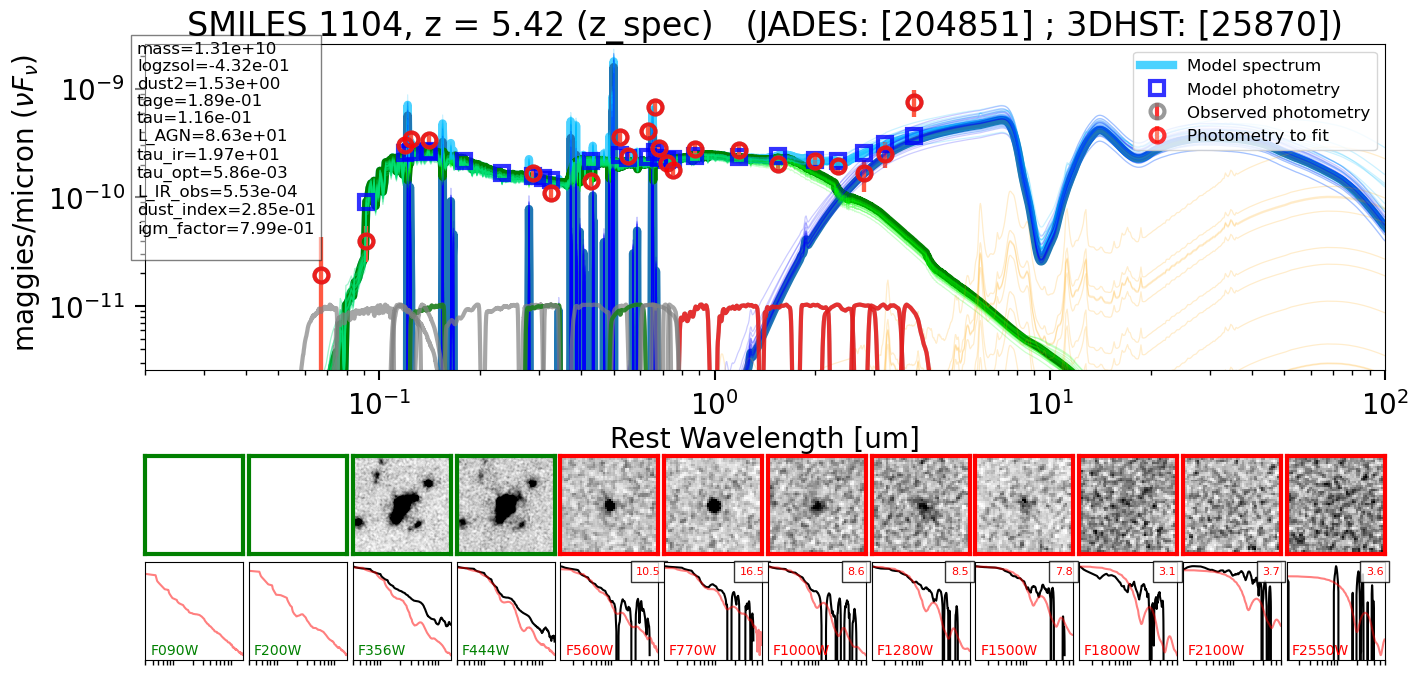}
        \includegraphics[width=0.495\hsize]{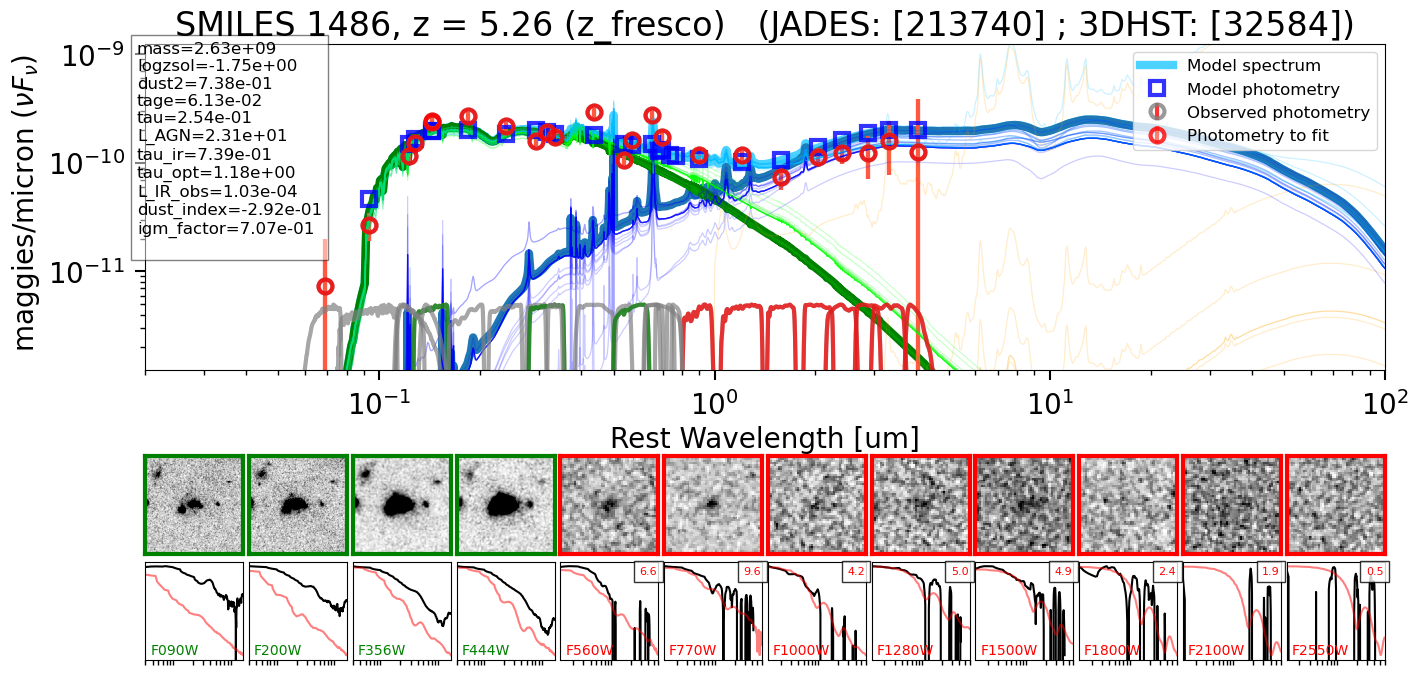}
        \includegraphics[width=0.495\hsize]{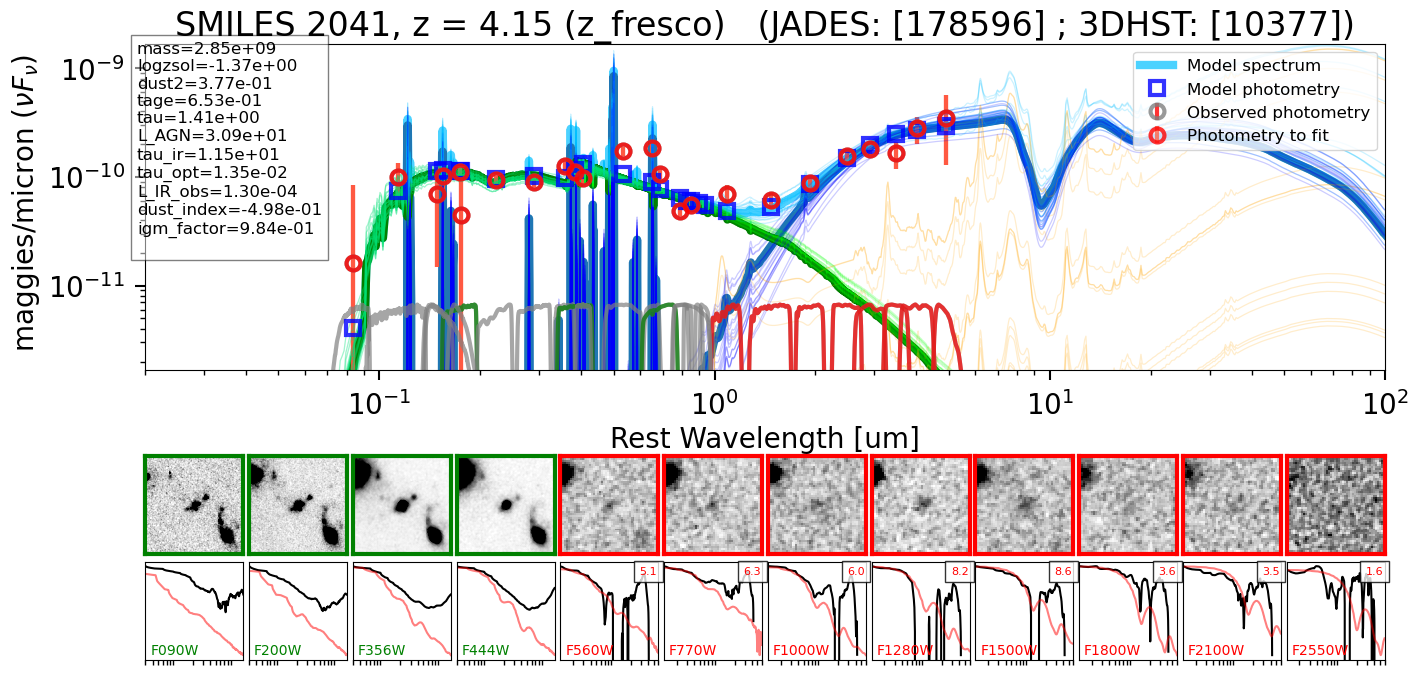}
        \includegraphics[width=0.495\hsize]{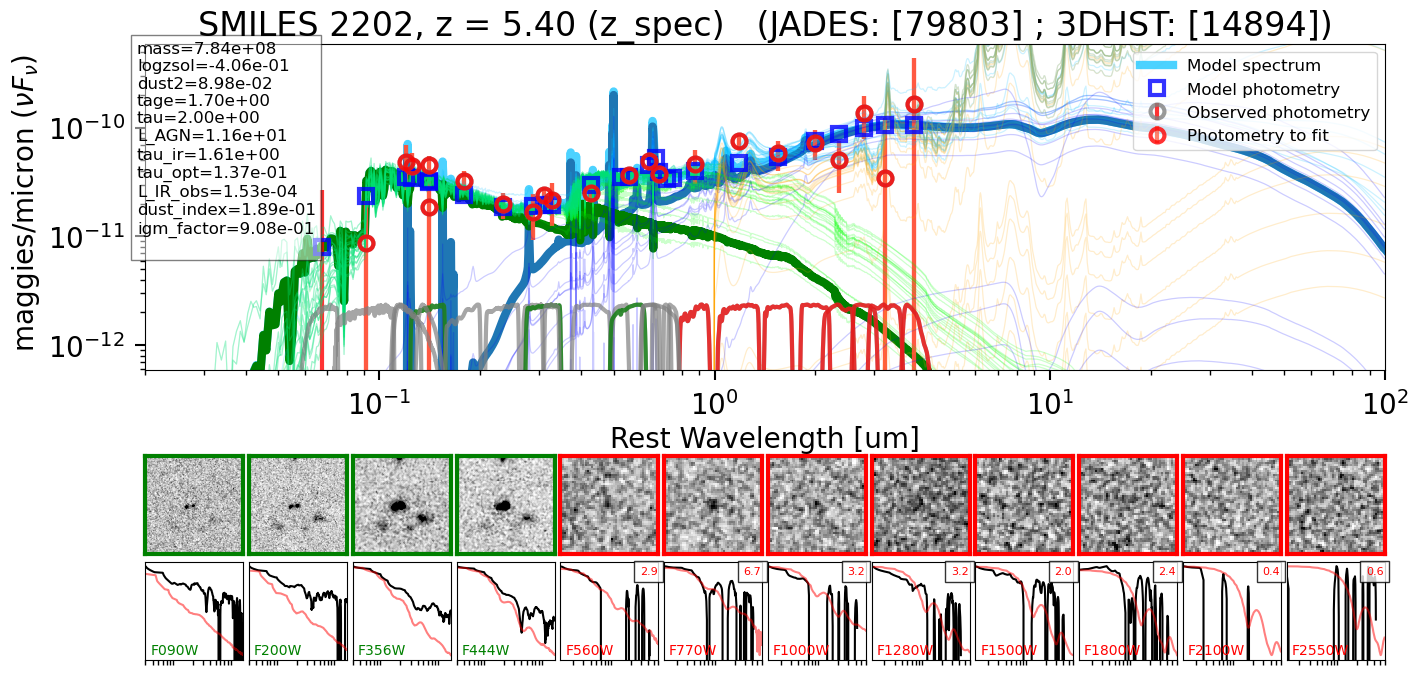}
        \includegraphics[width=0.495\hsize]{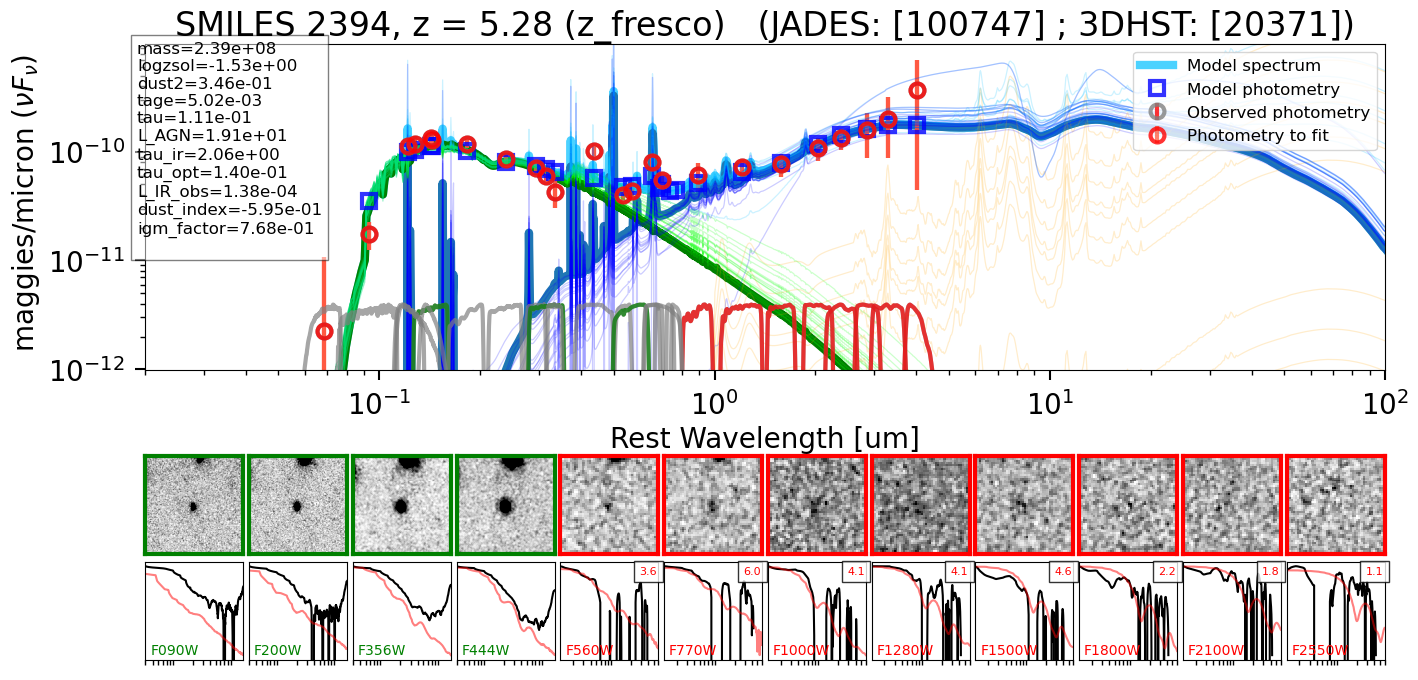}
        \includegraphics[width=0.495\hsize]{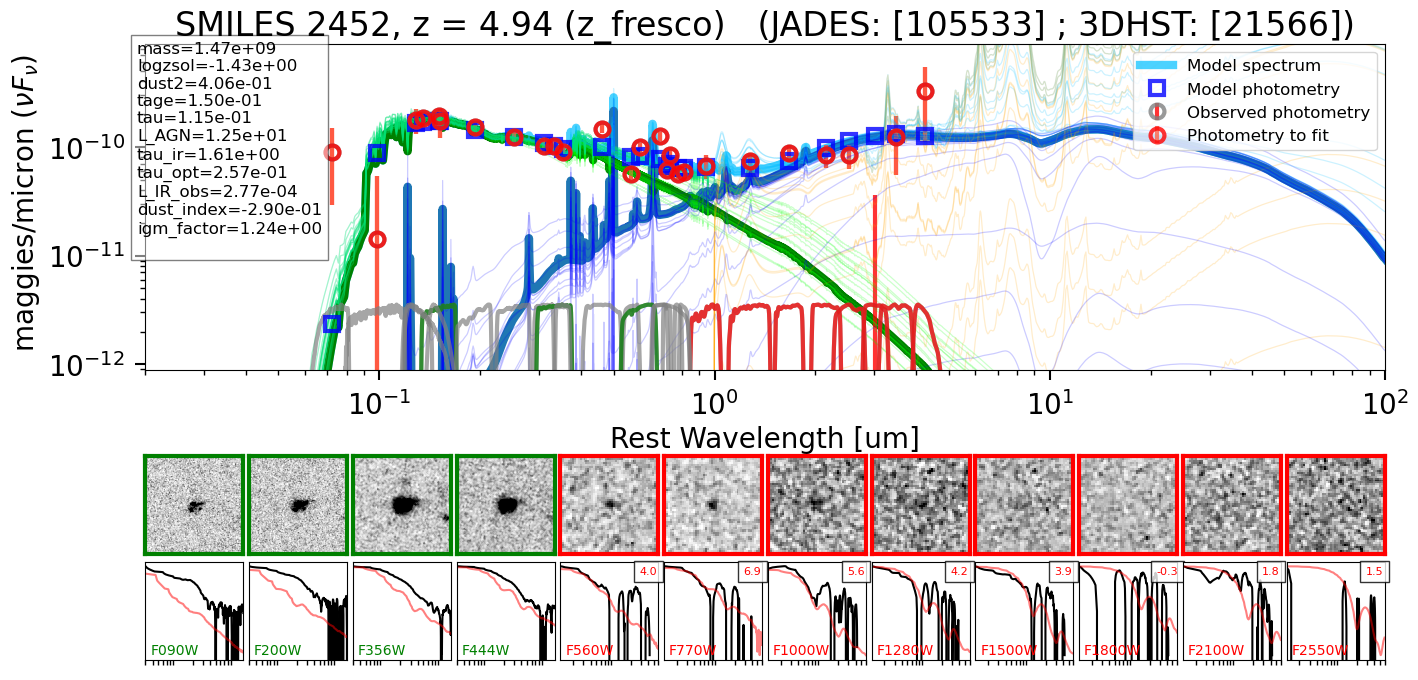}
    \caption{Similar to Figure~\ref{fig:sed-galaxy-examples} but for high-z AGN candidates. The limited rest frame infrared coverage still allows identification of obscured AGNs, although not as completely as for lower redshifts. }
  \label{fig:sed-hzagn-examples}
    \end{center}
\end{figure*}

Figure~\ref{fig:sed-hzagn-examples}  presents six examples of AGNs identified at $z >$ 4. All of these galaxies have spectroscopic redshifts and the AGN nature of MIRI 1104 is confirmed by 
the broad H$\alpha$ emission revealed by the FRESCO spectrum \citep{Matthee2023}. The infrared excesses in all of these cases extend to wavelengths significantly shorter than the excesses of star-forming galaxies. These examples demonstrate that MIRI observations can find obscured AGNs at $z > 4$, although the samples will be incomplete.

\begin{center}
  $\ast$~ ~~~ $\ast$~ ~~~ $\ast$
\end{center}

\begin{figure*}[htp]
    \begin{center}
  \includegraphics[width=1.0\hsize]{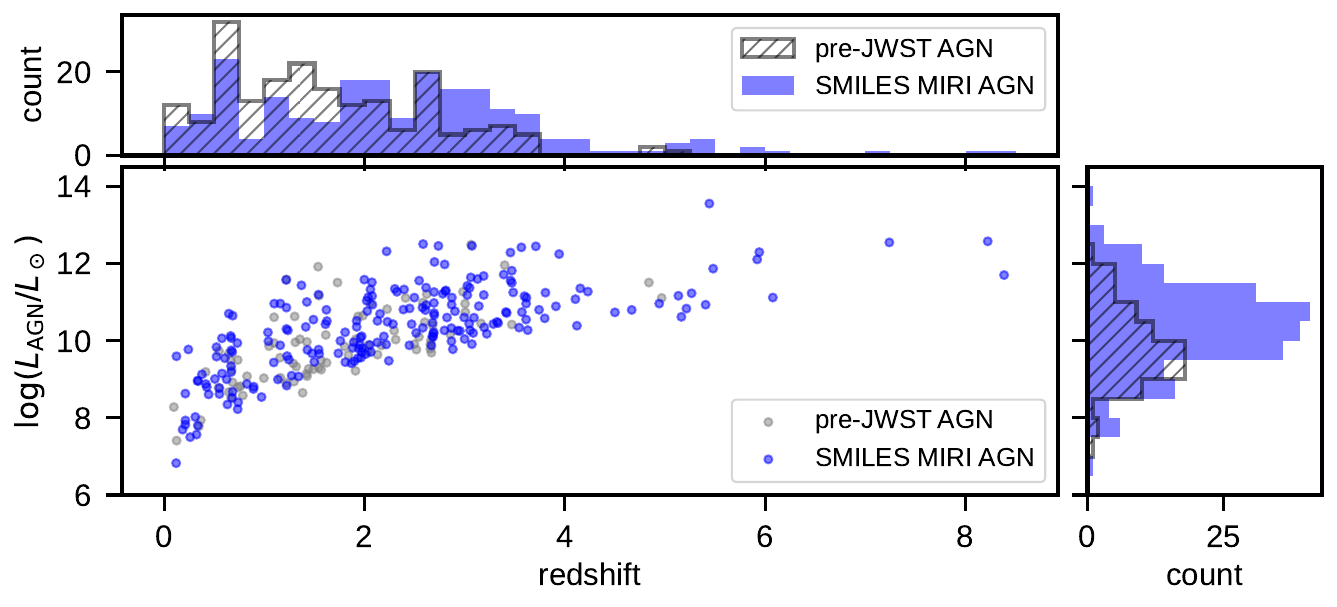}
    \caption{AGN luminosity as a function of redshift. The MIRI AGNs are shown in blue and the pre-JWST AGNs are shown in gray.}
  \label{fig:agn-lum-z}
    \end{center}
\end{figure*}

Combining the results above, we found 217 AGNs (or candidates) from SED analysis in total as listed in Table~\ref{tab:agn_sample}.
In Figure~\ref{fig:agn-lum-z}, we present the AGN luminosity distribution as a function of redshift. 
Compared to the AGN sample selected in the pre-JWST era \citep{Lyu2022}, the new sample contains 
more objects with lower bolometric luminosities and/or higher redshifts.

\begin{deluxetable*}{ccccccccc}
\tablecaption{List of MIRI-selected AGN in the SMILES Footprint}
\tablehead{
\colhead{MIRI ID}&
\colhead{R.A.} &
\colhead{Decl} &
\colhead{$z$} &
\colhead{z\_type} &
\colhead{subsample} &
\colhead{$\log(L_{\rm AGN, bol}/L_\odot)$} &
\colhead{$\log(M_{\rm star}/M_\odot)$} &
\colhead{selection code$^\tablenotemark{(a)}$} 
}
\startdata
      6  &   53.1243250   &  -27.882700    &  2.08     &     z\_eazy\_hst   &            norm   &   9.44   &  10.27   &  oxxxxxxx   \\
     50  &   53.1097673   &  -27.872172    &  0.12     &     z\_eazy\_hst   &           dwarf   &   5.11   &   6.97   &  oxxxxxxx   \\
     67  &   53.1378320   &  -27.870122    &  2.07     &         z\_eazy    &            norm   &   9.80   &   9.98   &  oxxxxxxx   \\
     68  &   53.1015140   &  -27.869963    &  3.46     &         z\_eazy    &            norm   &  10.57   &  11.18   &  oxxxxxxx   \\
     79  &   53.1278054   &  -27.869105    &  0.44     &         z\_spec    &           dwarf   &   6.89   &   8.57   &  oxxxxxxx   \\
     88  &   53.1321226   &  -27.868305    &  3.03     &       z\_fresco    &            norm   &   9.47   &   9.77   &  oxxxxxxx   \\
     $\cdots$  &   $\cdots$   &  $\cdots$    &  $\cdots$     &     $\cdots$     &       $\cdots$        &   $\cdots$   &   $\cdots$   &   $\cdots$  \\
\enddata
\tablenotetext{(a)}{AGN classification code for different selection methods. `o' is given if the object is classified as AGN by this method and 'x' means not. From the left to the right, the methods are [jwst\_sed], [jwst\_rl], [xtype\_lumcut], [xtype\_x2r], [rtype\_rl], [rtype\_fss], [otype\_sp], [var].}
\tablecomments{
Only a portion of this table is shown here to demonstrate its form and content. 
A machine-readable version of the full table is available.}
\label{tab:agn_sample}
\end{deluxetable*}

As mentioned above, there are several caveats in the AGN search among the extended and high-$z$ samples and our selections are conservative to reduce the number of false positives. In the future, we will improve these selections with more comprehensive analysis and additional data.

\subsection{Other Selections in the SMILES Footprint}\label{sec:other-selection}

Besides SED fitting of MIRI sources, there are many other methods to identify AGNs. Now we describe the identification
of radio-loud AGNs, broad-line AGNs and mid-IR variable AGNs from JWST data with the combination of datasets at other wavelengths when necessary,
and then summarize AGN samples identified without the usage of JWST data within the same footprint.

\subsubsection{Radio-Loud AGNs with MIRI and JVLA}

\begin{figure}[htp]
    \begin{center}
  \includegraphics[width=1.0\hsize]{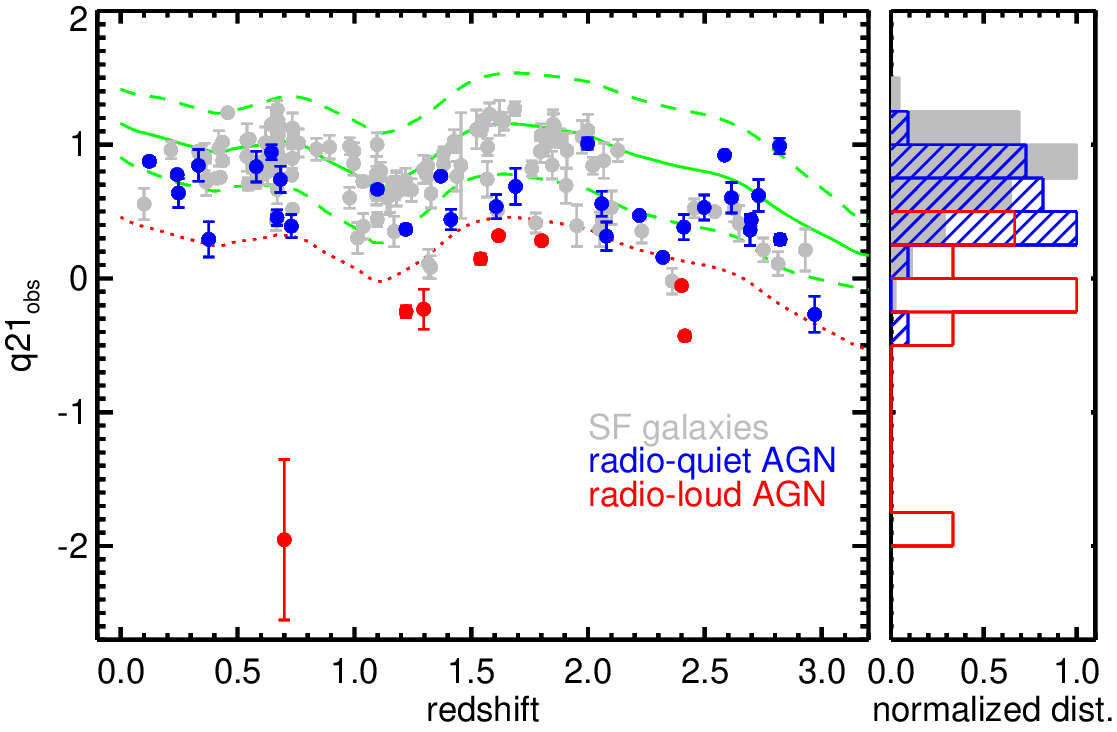}
    \caption{
    The observed IR-to-radio flux $q_{\rm 21, obs}$ as a function of redshift
of all the sources. The red dotted line is 0.5 dex below the mid-point of 
the radio-infrared correlations in the \cite{Rieke2009} SFG templates (the green solid line); we define
the source as a radio-loud AGN if it falls below this line at $\ge$ 2-$\sigma$ significance. We color-code different sources on the main plot and show their relative
distribution of $q_{\rm 21, obs}$ in the right panel.}
  \label{fig:q21_redshift}
    \end{center}
\end{figure}

Since we now have deeper F2100W photometry than the previous {\it Spitzer}/MIPS 24$\mu$m data, it is possible to improve the previous radio-loud AGN selections 
based on the ratio between the radio and mid-IR fluxes (i.e., [rtype\_rl] in \citealt{Lyu2022}). As the F2100W data is deeper than F2550W, we adopt
$q_{\rm 21, obs} = \log(S_{\rm 21 \mu m, obs}/S_{\rm 1.4~GHz, obs})$ to reveal outliers from the radio-infrared correlations for dusty star-forming galaxies, where
$S_{\rm 21 \mu m, obs}$ is the observed MIRI F2100W flux and $S_{\rm 1.4~GHz, obs}$ is the observed radio flux at 1.4 GHz. For the latter, we assume a
power-law index of $\alpha=-0.7$ to estimate the 1.4 GHz flux density from 3 GHz, similar to \citet{Lyu2022}. Following the same strategy in \citealt{Alberts2020},
we determine the locus of SFGs by calculating $q_{\rm 21, obs}$ at $z=$0--3 from the \citet{Rieke2009} SFG templates at $\log(L_{\rm IR}/L_\odot) = [11, 11.5, 12]$.
An outlier from the SFG radio-infrared relation is defined as being 0.5 dex below the midpoint of SFG $q_{\rm 21, obs}$ locus. Using this criterion, we identify 8 
radio-loud AGNs among 167 $z<3.0$ MIRI sources that have been detected at 3 GHz, as shown in Figure~\ref{fig:q21_redshift}. In total, there are 39 AGNs in this radio-detected sample and the radio loud fraction is about 20\%.

\subsubsection{Spectroscopically-identified broad-line AGNs with JWST}

FRESCO has conducted deep NIRCam/grism observations with the F444W filter for the 62 arcmin$^2$ area of GOODS-S that has a considerable 
overlap with our SMILES survey (see Figure~\ref{fig:footprint-sens}), offering the chance to identify AGNs through broad-line emission.
At $z>5$, only one AGN has been reported at z=5.481 in GOODS-S, based on the broad H$\alpha$ emission line \citep{Matthee2023}. Our MIRI SED fitting
also identified this AGN (MIRI ID v0.4.2: 1104). We have also looked for AGN candidates at lower redshifts based on broad emission lines from Pa$\alpha$, Pa$\beta$ 
and He II (Lyu et al, in prep). All the sources we found have been previously identified in \citet{Lyu2022}.

In addition, a few AGNs in GOODS-S have been identified from JADES/NIRSpec spectral observations \citep{Bunker2023}. In terms of broad-line AGNs, there are only two: 
10013704 ($z$=5.92) and 8083 ($z$=4.64) \citep{Maiolino2023}. In contrast to the FRESCO AGN, these objects are much fainter by selection. In our MIRI images, 
they are only marginally detected in F560W and/or F770W without additional measurements in other bands and thus we cannot conduct robust SED fittings
for AGN selection. 

\subsubsection{Mid-IR Variable AGNs with NIRCam and IRAC}

Given the deep IRAC and NIRCam images available across the field, we can also look for long-term mid-IR variable AGNs.
We have degraded the JADES/NIRCam F356W images to match the PSF of {\it Spitzer}/IRAC band-1 mosaics and carried out local difference photometry to look 
for variable objects (Lyu et al., 2023, in prep). Nevertheless, no new AGNs have been revealed.

\subsubsection{AGNs with weak Pa $\alpha$}

Potential identification of AGN candidates from weak Pa $\alpha$ compared with mid-infrared continuum is discussed in Appendix A. We have not yet applied this method; with current data it adds a small number of candidates.

\subsubsection{AGN Selected in the Pre-JWST Era}\label{sec:pre-jwst-agn}

In \citet{Lyu2022}, we have presented an AGN sample selected based on the pre-JWST data from X-ray to radio with eight different methods 
that included X-ray luminosity, X-ray-to-radio luminosity ratio, UV-to-mid-IR SED fitting, mid-IR IRAC color, radio-to-mid-IR ratio (for radio-loud AGN), 
radio slope (for flat spectral radio sources), and time variability in the X-ray and optical bands. We can now replace the AGN sample selected by IRAC color and SED fittings from the \citet{Lyu2022} pre-JWST catalog with the new SED-identified AGNs obtained in this work using the improved JWST photometry. In the MIRI survey footprint,
there are 203 pre-JWST candidate AGNs in total; only five of them do not have a MIRI 
counterpart. Among these pre-JWST AGNs without MIRI source association, four are identified by variability and one is identified by the X-ray to radio flux 
[x2r]. The latter object is also an X-ray source with very low S/N detection that does not have a MIRI counterpart, which therefore may be spurious.

\begin{figure*}[htp]
    \begin{center}
  \includegraphics[width=0.48\hsize]{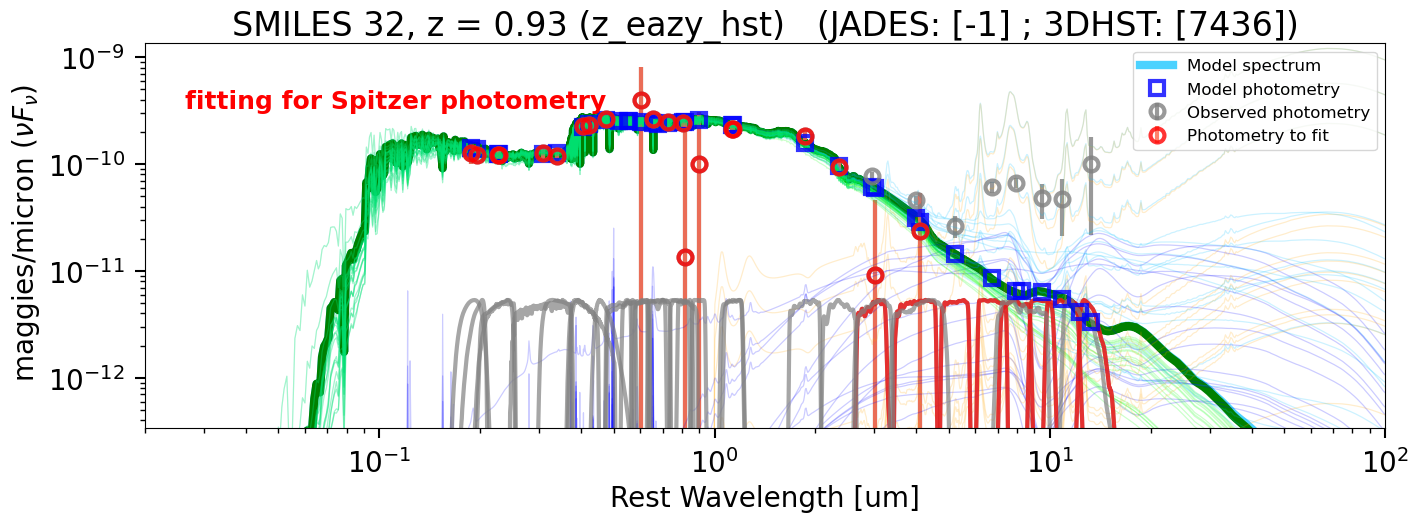}
  \includegraphics[width=0.48\hsize]{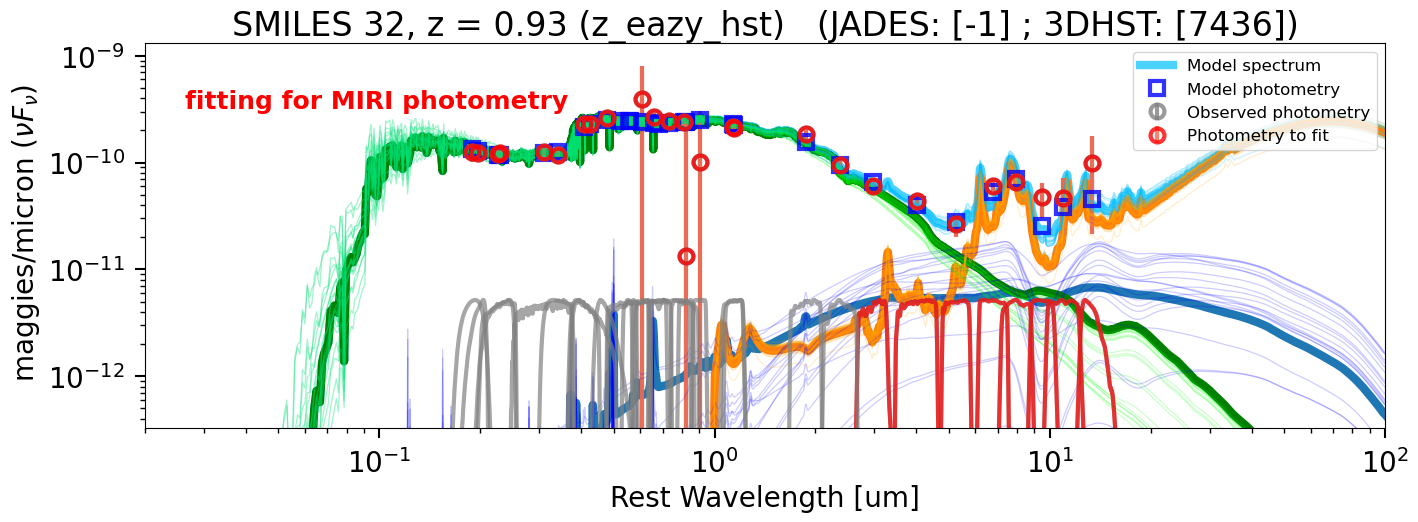}
  \includegraphics[width=0.48\hsize]{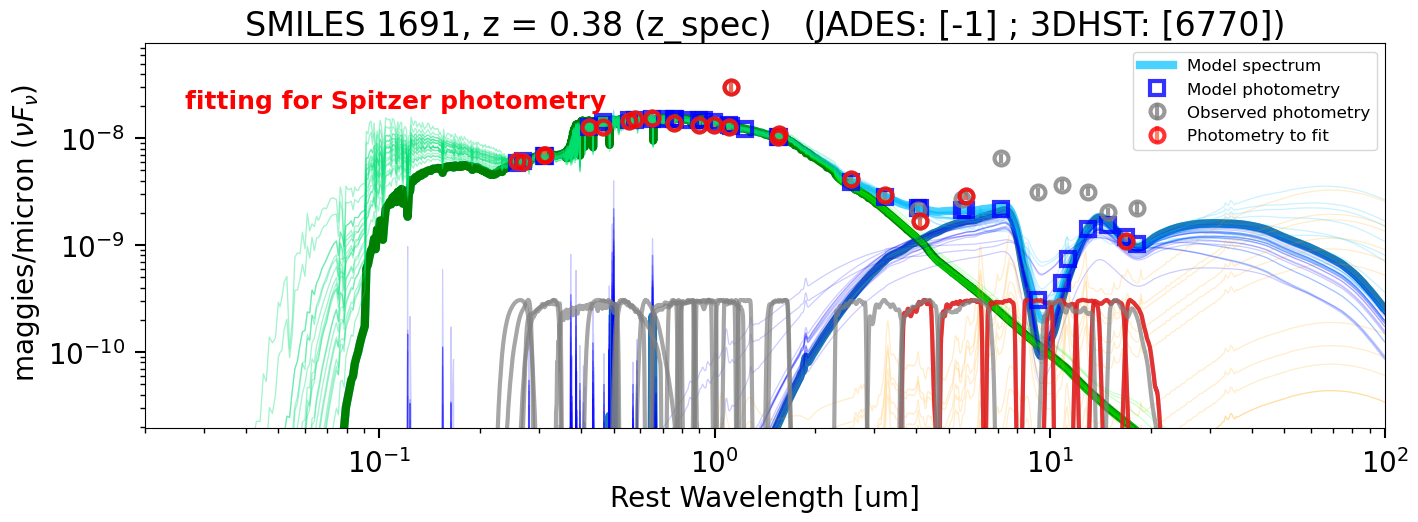}
  \includegraphics[width=0.48\hsize]{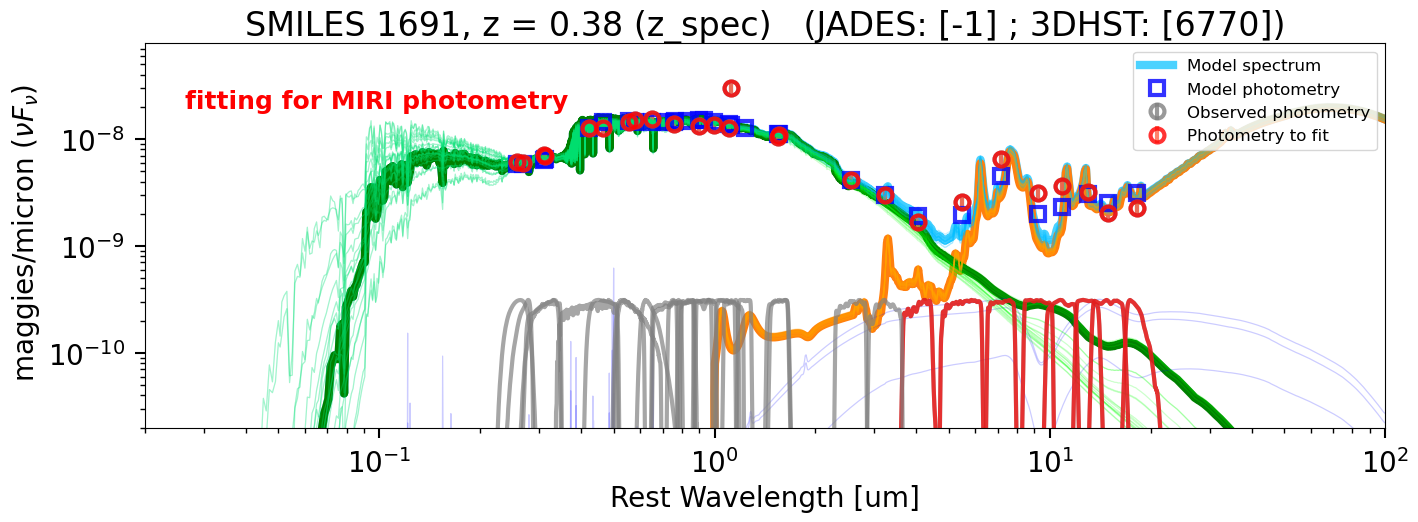}
    \caption{Comparison of the SED fittings with pre-JWST data (left) and JWST data (right). Two examples are shown.}
  \label{fig:sed_comparison}
    \end{center}
\end{figure*}

We can also compare the SED fittings with the {\it Spitzer} and JWST data for the same sources. Two examples are given in Figure~\ref{fig:sed_comparison}. 
The first object (MIRI 32) does not show evidence for an AGN,  since the SED constraints for wavelengths longer than  IRAC band 2 are very limited in the shallower {\it Spitzer} data. 
In contrast, the new MIRI data points
reveal this object is a composite galaxy with an obscured nucleus in the mid-IR. The second object (MIRI 1691) seems to be an obscured AGN based 
on the limited wavelength coverage by the {\it Spitzer} data while the new fittings with MIRI data prefer the star forming galaxy solution. In fact, the AGN solution 
can be ruled out due the extended MIRI source morphology.

For the other selections based on X-ray or radio properties, the new mid-IR JWST data do not have any influence.

\section{AGN Statistics and Comparison of Selection Techniques}\label{sec:result}

With the various selection methods presented above, we can now compile all known AGNs in the MIRI 
footprint and compare the relative performance of different selections. As described above, four variable AGN candidates
and one X-ray-to-radio AGN candidate do not have MIRI counterparts and they are likely to be spurious; thus we will ignore them and focus the discussion on the MIRI-detected sources.

\subsection{MIRI AGN Number Density}\label{sec:num-dens}

Given the survey area of $\sim$34 arcmin$^2$, the total 
number density for the robust MIRI AGNs in the primary sample is 3.4 per arcmin$^2$, or nearly eight AGNs per MIRI pointing. 
In addition, there are 2.9 AGNs per arcmin$^2$ from the extended sample. For each band, the total MIRI-selected AGN (candidates) number densities are
6.4, 6.3, 5.6, 5.0, 4.7, 3.7, 3.2 and 2.1 arcmin$^{-2}$ from F560W to F2550W (3$\sigma$ detections, corresponding to
flux limits at 0.16, 0.13, 0.26, 0.36, 0.41, 1.0, 1.7 and 9.0 $\mu$Jy). These estimates will evolve as we develop more 
sophisticated methods to differentiate the emission by warm dust in extended-sample  galaxies from that of embedded AGNs. 
For now, these values should be treated as lower limits for total numbers of AGNs detected by MIRI since (1) 
due to the reduced sensitivities at longer wavelengths, we are likely missing obscured AGNs at high-$z$; and (2) for the AGNs 
in the extended-sample galaxy population, we are using an extreme SFG template based on Haro 11 and it is possible that some AGNs are missed due 
to the conservative selection;
(3) there are AGNs that have been identified by other selection techniques and that are detected by MIRI but are missed by the SED analysis, 
as shown in the upcoming section. 

\begin{figure*}[htp]
    \begin{center}
  \includegraphics[width=1.0\hsize]{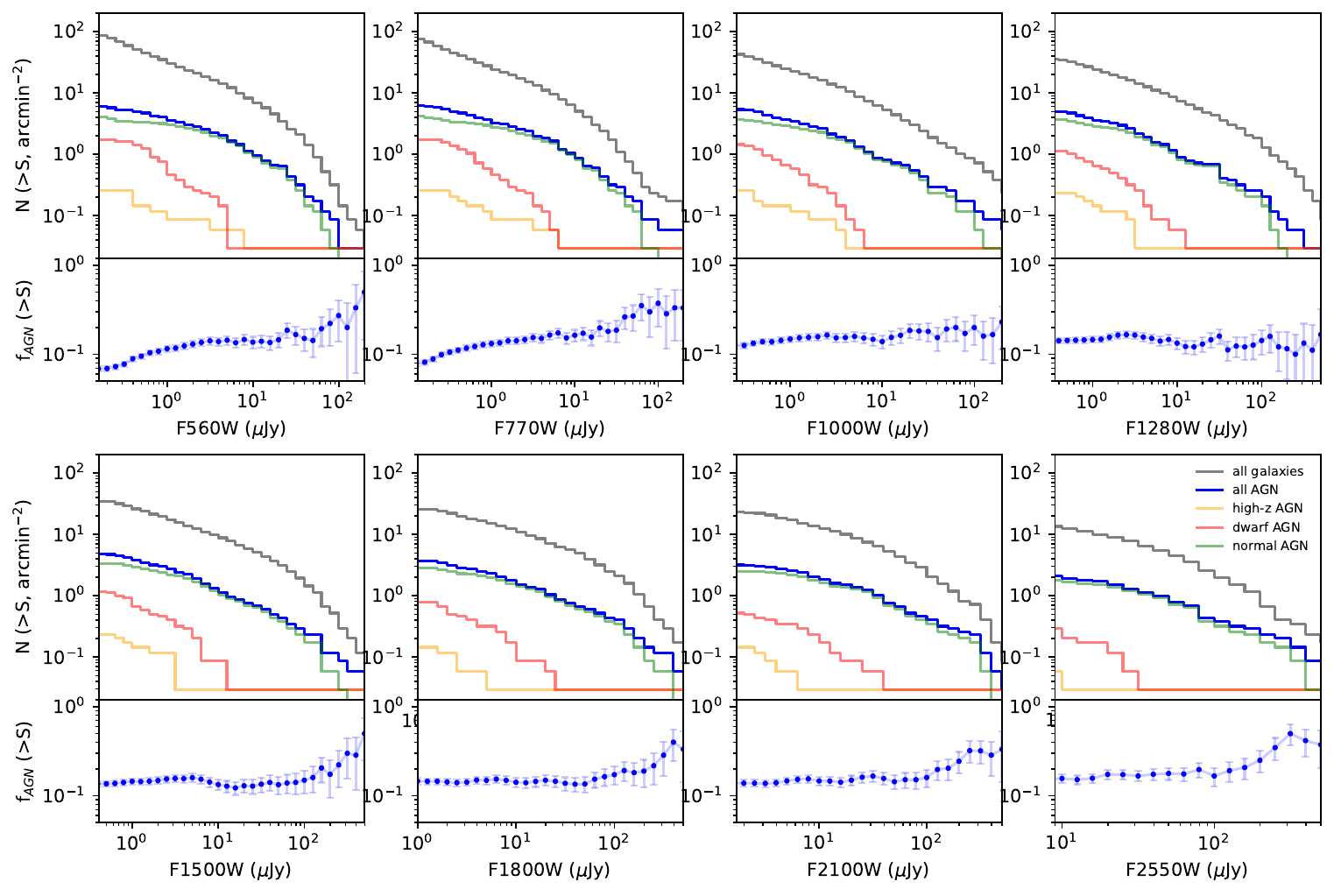}
    \caption{The MIRI source number density and AGN fraction as a function of observed flux in each 
    MIRI band.}
  \label{fig:agn-sky-ct}
    \end{center}
 \label{fig:detectionbands}
\end{figure*}

In Figure~\ref{fig:agn-sky-ct}, we plot the number density of the various MIRI sources as well as the total AGN fraction 
as a function of observed flux for all the MIRI bands. In every band, the AGNs in the primary sample of massive galaxies have dominated
the distribution at the bright end while those in the extended sample of low-mass galaxies and high-$z$ galaxies
contribute mainly at the faint end. The total AGN fraction typically dominates the bright end with a 
fraction $>$40-60\% and gradually drops toward the faint end to about 13--16\%. 
For F560W, F770W and F1000W, the AGN fraction presents an apparent drop at the
faint end, which is expected given our reduced selection efficiency since these sources are not detected at longer
wavelengths. For F1280W, F1500W and F1800W, the AGN fraction seems to present a trend that decreases from the bright end to 
the intermediate flux followed by a gradual upturn towards the faint end. Given that the redshift distribution of our MIRI 
sources is peaked around $z\sim$1--1.5, such trends are likely caused by the strong 7.7$\mu$m PAH features from DSFGs. 

\subsection{Comparison to Previous AGN Samples}\label{sec:compare-sample}

\begin{figure}[htp]
    \begin{center}
    \includegraphics[width=1.0\hsize]{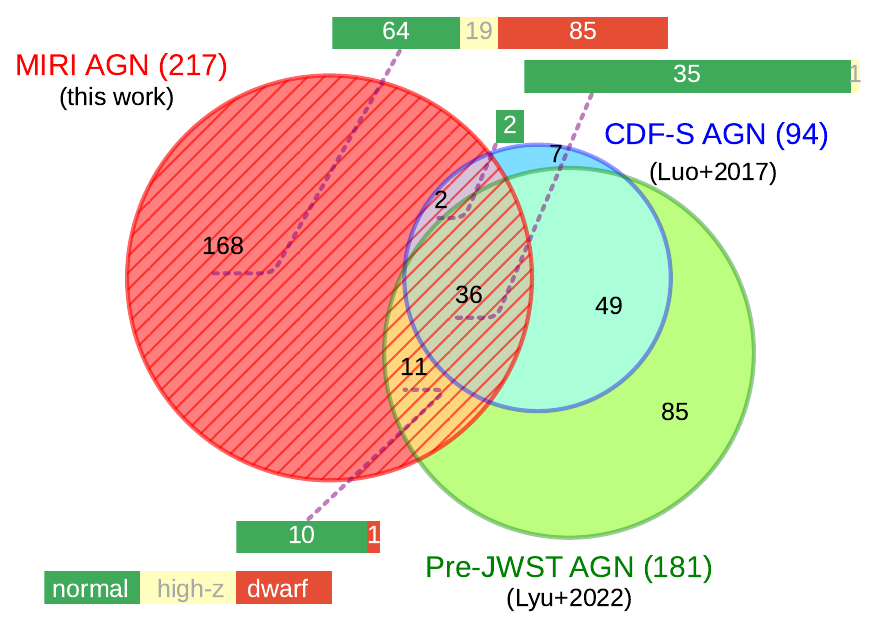}
    \caption{
    Venn diagram of the various AGN samples within the SMILES footprint. We compare the X-ray detected AGN sample in \citet{Luo2017}, the pre-JWST AGN sample compiled in \citet{Lyu2022} and the MIRI SED-identified AGN sample in this work. The numbers in the brackets are the total numbers of AGN from each source. For objects
    identified by MIRI, we also plot a horizontal stacked bar chart to show the number of normal AGNs (green), dwarf AGNs (red) and high-$z$ AGNs (yellow) for each subset.
}
  \label{fig:agn_venn}
    \end{center}
\end{figure}

In Figure~\ref{fig:agn_venn}, we present the Venn diagram to show how the MIRI SED-identified AGN sample 
intersects with the CDF-S AGN catalog \citep{Luo2017} and the pre-JWST AGN catalog \citep{Lyu2022}.  Over 
the same area, there are 94 X-ray detected AGNs reported from the 7Ms CDF-S catalog with an AGN number density 
about 2.8 per arcmin$^2$. Among the 217 MIRI AGNs, only 48 of them (22\%) have X-ray detections in the 7Ms CDF-S catalog 
and 38 of them (18\%) are reported as AGN by \cite{Luo2017}.  In other words, our relatively shallow MIRI survey 
($\sim$2.17 hours per pointing) yields 2.4 times more AGNs than the deepest X-ray survey ($\sim1800$ hr) 
and 80\% are new discoveries.

Within the same footprint, there are 181 sources in the pre-JWST AGN catalog that include not only X-ray detected but also radio- and mid-IR detected AGNs. Still, about 78\% of the MIRI AGN sample does not overlap with the pre-JWST AGN sample, demonstrating 
the huge discovery space offered by JWST. Among the 168 AGNs only identified by MIRI SEDs, there are 64 AGNs 
in massive galaxies at $z<$4.0, 85 AGN candidates in dwarf galaxies and 19 high-$z$ AGN candidates. The latter
two populations are mostly found by MIRI, as there are only 1 dwarf AGN candidate and 1 high-$z$ AGN candidate also identified by other selections.

Meanwhile, there are 56 CDF-S AGNs and 134 pre-JWST AGNs not identified by the MIRI SED analysis, which contribute to $\sim$60\% and $\sim$74\% of 
the corresponding catalog within the MIRI survey footprint.  As pointed out in \citet{Lyu2022}, there is no single method 
or wavelength can identify  all the AGNs. The results here provide very clear support to this argument --- despite
the much improved AGN hunting capability offered by MIRI, there are always some AGN population(s) left out and the 
combination of all possible selections across the electromagnetic spectrum is necessary to complete the AGN census.

\begin{figure}[htp]
    \begin{center}
  \includegraphics[width=1.0\hsize]{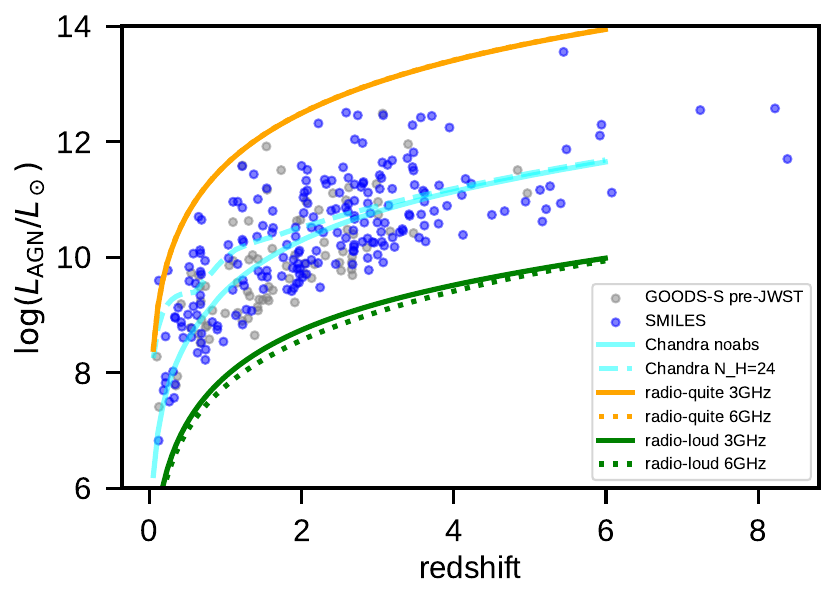}
    \caption{$L_{\rm AGN, bol}$ detection limits as a function of redshift for different bands.}
  \label{fig:agn-lim}
    \end{center}
\end{figure}

To illustrate the new discovery space provided by MIRI in terms of AGN bolometric luminosity, Figure~\ref{fig:agn-lim} shows the detection limits of the Chandra X-ray and JVLA radio bands as a function of redshift. In the X-ray, the Chandra data can only probe the most luminous MIRI AGNs, which is just the tip of the iceberg. This is consistent with the fact that only 33 out of 217 MIRI AGNs (15\%) have reported X-ray luminosity that passes the AGN selection criteria. Most of the MIRI AGN sample are much fainter and the MIRI limit is about one order of magnitude deeper than the Chandra one. 

Regarding the radio emission, there is a wide range of intrinsic variation of its strength from radio-loud to radio-quiet populations and we only show the AGN bolometric luminosity limits from the JVLA data from some average templates (see \citealt{Lyu2022} for details about the template models). It is very clear that the current JVLA data is not deep enough to detect the radio emission from the radio-quiet MIRI AGNs and any detected radio emission is most likely coming from the host
galaxies \citep{Alberts2020}. The JVLA data can probe the radio properties of bright radio-loud MIRI AGNs but would miss relatively faint radio-loud AGNs.

\subsection{The Relative Performance of Different Selection Methods}\label{sec:compare-method}

\begin{figure*}[htp]
    \begin{center}
  \includegraphics[width=1.0\hsize]{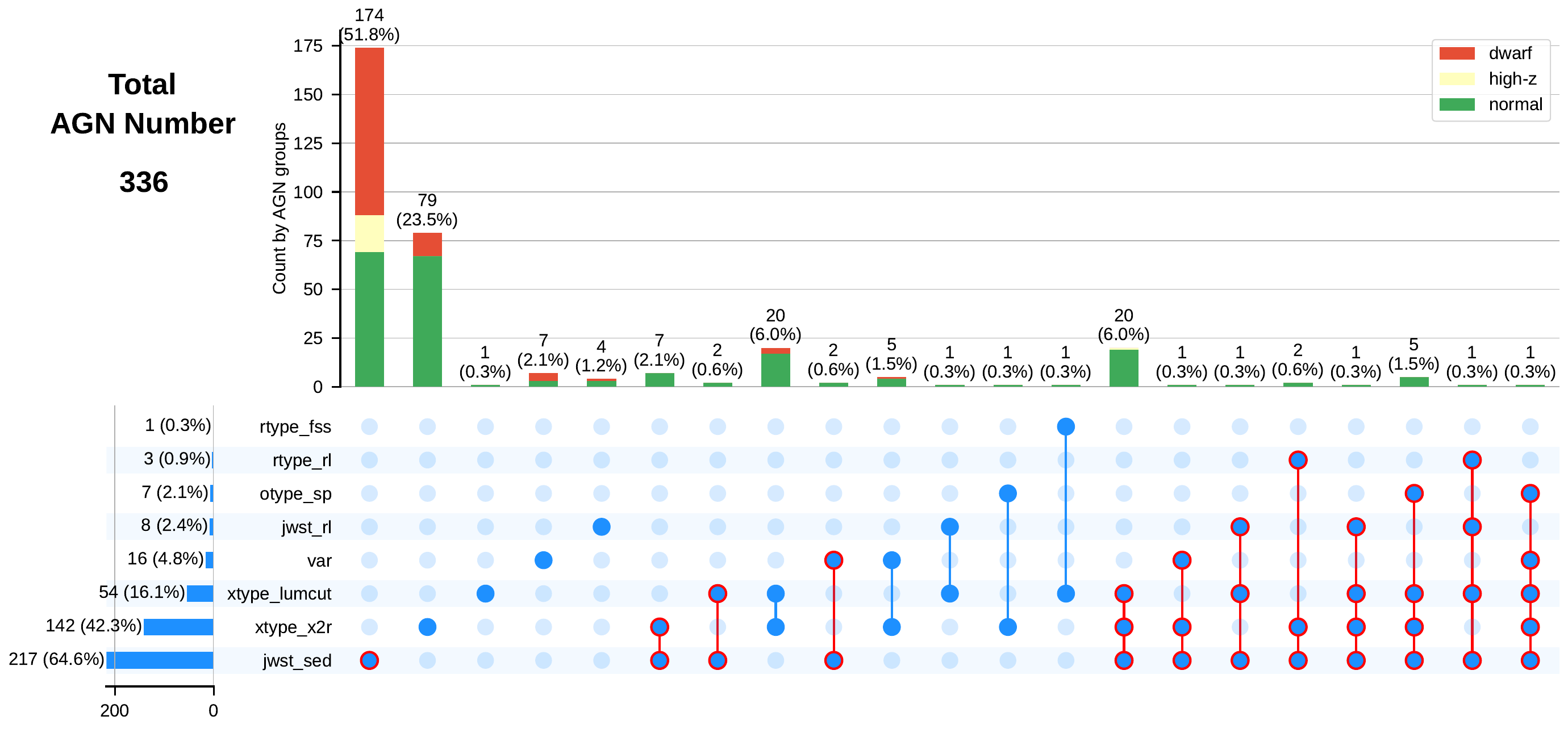}
    \caption{Comparison of AGN sample selected with different methods and their overlaps in the SMILES footprint, 
    visualized with the {\it UpSet} technique.
    The sample intersections of AGNs selected with different methods are plotted as a matrix with bar charts on the left and top to 
    show the corresponding sample sizes. Each row corresponds to one selection method, and the bar charts on the left show the number  
    of AGN identified with a given method. Each column corresponds to a possible intersection: the filled-in cells indicate which 
    selection method is part of the intersection. The lines connecting the filled-in cells show in which direction the plots should be read. 
    The bar charts on the top give the size of the AGN sample identified with the corresponding intersection of various selection 
    techniques. We further break each vertical bar into different colors
    to show the relative fraction of normal, dwarf and high-$z$ AGNs within each subset. We highlight the subsets contributed by JWST SED selections in red. }
  \label{fig:agn-counts}
    \end{center}
\end{figure*}

To compare the yields of different selection methods and the possible overlaps, we used the {\it UpSet} visualization tool \citep{Lex2014} to show
the intersection of different AGN samples in Figure~\ref{fig:agn-counts}. Similar to the conclusions reached in \citet{Lyu2022}, the top three
most efficient selection techniques are SED fittings, X-ray to radio luminosity cut ([X2R]) and X-ray luminosity cut, which identify 97.5\% of the
whole sample. For the remaining AGNs, 1.6\% are identified by variability and 0.9\% by the JWST-JVLA radio-loud AGN selection. 

Despite the substantial improvement of the AGN SED identification with JWST data, about 34\% of
the known AGNs in the field have been missed and most of them have been identified by [X2R] (23\% of the whole AGN sample), { illustrating
the need to combine multi-wavelength selections to reach a more complete AGN census. (We will discuss the nature of these objects in Section~\ref{sec:agn_no_mir}.)} Meanwhile, 53\% of the AGN
are only revealed by the JWST SED analysis. The other selection methods such as [xtype\_lumcut], [var], [jwst\_rl] all include unique AGNs that are solely 
picked out by one method, despite their much smaller contribution compared to the [jwst\_sed] sample. Given the discussion in Section~\ref{sec:compare-sample},
the yields of all the other selection methods are limited by the depth of X-ray or radio data and there should be additional  galaxies detected by current MIRI data that have AGNs missed by the current SED analysis.

Regarding the different AGN groups (primary, extended and high-$z$), the vast majority of extended-sample AGN candidates (89\%) and high-$z$ AGN candidates (100\%) have been identified through the MIRI SED analysis. For all the normal AGNs in massive galaxies at $z<4$, the MIRI SED analysis has identified 114 out of 211 these objects ($\sim$54\%).  Notably, after merging all AGN samples together, 18 out of 20 high-$z$ AGN candidates ($\sim$92\%), 88 out of 109 dwarf AGN candidates (81\%) and 72 out of 211 normal AGNs ($34 \pm 4$\%) are only identified by MIRI SED fittings, where the error is from the sample size. In our previous study \citep{Lyu2022}, which was less effective at identifying AGNs due to less complete data but covered a five times larger area, we found that $\sim$ 20\% of the AGNs were very strongly obscured and 26\% were undetected in the deepest Chandra data, qualitatively in agreement with the new estimate of the fraction missed in previous studies.

\subsection{Comparison to MIRI AGN Studies in other fields}

Pre-JWST efforts to use infrared photometry to identify AGNs are reviewed by \citet{LyuRieke2022}. Although they began with IRAS, they came into their own with the wide and deep surveys obtained with {\it Spitzer} and {\it WISE} \citep[e.g.,][]{Lacy2004, Stern2005,Alonso2006,  Donley2008, Mateos2012, Asmus2020, Hviding2022}. The possibilities are greatly expanded with JWST. 

Based on four MIRI pointings from CEERS,  \citet{Yang2023} reported 102 SF-AGN mixed systems and 25 AGNs from a total of 560 MIRI detections over 
a sky coverage of $\sim$ 9 arcmin$^2$ by fitting the galaxy SEDs with CIGALE. If we combine their SF-AGN mixture and AGN categories, their reported AGN number density is $\sim$14 arcmin$^{-2}$. Our fits include SF-AGN mixtures, so they can be compared directly. We find significantly  lower numbers:  3.4 arcmin$^{-2}$ for the main sample and 2.9  arcmin$^{-2}$ for the extended one, for a total of $\sim$ 6.3 arcmin$^{-2}$. The high AGN numbers in \citet{Yang2023} presumably come with a  caution because of the possibility of significant contamination by low-mass galaxies.  Some other  important distinctions between the two studies are that: (1) we cover a larger field, with five times as many overall MIRI detections and with more ultra-deep ancillary data; (2) we have more accurate redshifts, with $\sim$ 40\% spectroscopic ones and somewhat more accurate photometric ones for the rest; and (3) we separate galaxies assumed to have PAH emission from those without on the basis of mass/metallicity, whereas they leave this difference as a free parameter.\footnote{They reported anomalously weak PAH features in their median star forming SED. However, for the more massive galaxies in our sample at the relevant redshifts, the PAH features have strengths similar to local galaxies, as shown both by our fits in Figure~\ref{fig:sed-galaxy-examples} and by {\it Spitzer} spectra of similar galaxies amplified by lensing, e.g., \citet{Rigby2008}. }

Based on the same CEERS dataset, \citet{Kirkpatrick2023} found $\sim$ 3 AGNs per arcmin$^{-2}$ at F1000W based on mid-IR color selections, compared to our $\sim$5 AGNs
per arcmin$^{-2}$ in the same band.  
There are two likely reasons for this low yield: (1) their template fitting is based on a single set of templates from \citet{Kirkpatrick2015} in which AGN SEDs are represented by power laws with a narrow range of slopes and thus do not represent the full range
of AGN SED variations; (2) color-color plots are less diagnostic than SED fitting, as discussed extensively in \citealt{LyuRieke2022}. Consequently, the AGN numbers identified in this way are expected to be less than those from
SED analysis. Another limitation of the \citet{Kirkpatrick2023} study is that their templates include strong PAH emission for the normal SFGs and do not capture the IR behavior of lower-metallicity galaxies, which may result in misclassifying some dwarf galaxies with strong warm dust emission as obscured AGNs.

\section{The Obscured AGN population}\label{sec:obscured_agn}

Given the AGN sample constructed from this work, we now discuss the obscured AGN population seen at different wavelengths.

\subsection{The Fraction of Obscured AGNs from SED analysis}\label{sec:sed-fobs}

We first characterize the obscured AGN fraction of MIRI-selected AGNs and explore its dependence on the AGN luminosity and redshift. 
Our discussion will focus on the 114 MIRI-selected AGNs at $L_{\rm AGN, bol}>10^{10}~L_\odot$, $M_*\gtrsim10^{9.5}~M_\odot$ and $z$=0--4. 
These criteria ensure a reasonably large sample to carry out statistical studies that cover a large range of source properties while 
minimizing the complications caused by dwarf and high-z AGN candidates where the selections are less robust.

Depending on the observed wavelengths and available  spectral features, the definition of what is an obscured AGN varies in the literature. To first order, the obscuration of an AGN, i.e., whether the AGN can be seen or not, at a given band is caused  by two major effects: (1) how much of 
the AGN emission can be diminished by obscuring materials (e.g., a dusty torus); (2) how strongly the host galaxy emission dilutes the AGN light. Our 
SED fittings provide direct measurements of both effects at rest-frame UV to mid-IR. We define AGN obscuration in three ways:
\begin{itemize}
\item optically obscured: regardless of the AGN intrinsic luminosity and dust attenuation, if the observed AGN light fraction at rest-frame 0.1--1~$\mum$ is less than 20\% of the whole galaxy, we count it as an optically obscured AGN;
\item IR obscured: the near- to mid-IR attenuation level of the AGN component is described by {\it tau\_ir}, which can be transferred to the silicate strength following equation $S10=0.2 - tau_{\rm IR}/5.5$ (negative value for absorption) \citep{Lyu2022}. We define an AGN as IR obscured if $S10<0$, corresponding to {\it tau\_ir}$>$1.1;
\item IR heavily obscured: we define an AGN as IR heavily obscured if {\it tau\_ir}$>$3.85. This corresponds to the silicate strength of $S10<-0.5$, indicating significant dust attenuation \citep[e.g.,][]{Goulding2012}.
\end{itemize}
In a statistical sense, the significance of AGN obscuration increases from optically obscured, IR obscured to IR heavily obscured, though the definition
of the first case is complicated by AGN-host galaxy contrast as well as the SMC extinction law.

\begin{figure*}[htp]
    \begin{center}
    \includegraphics[width=0.495\hsize]{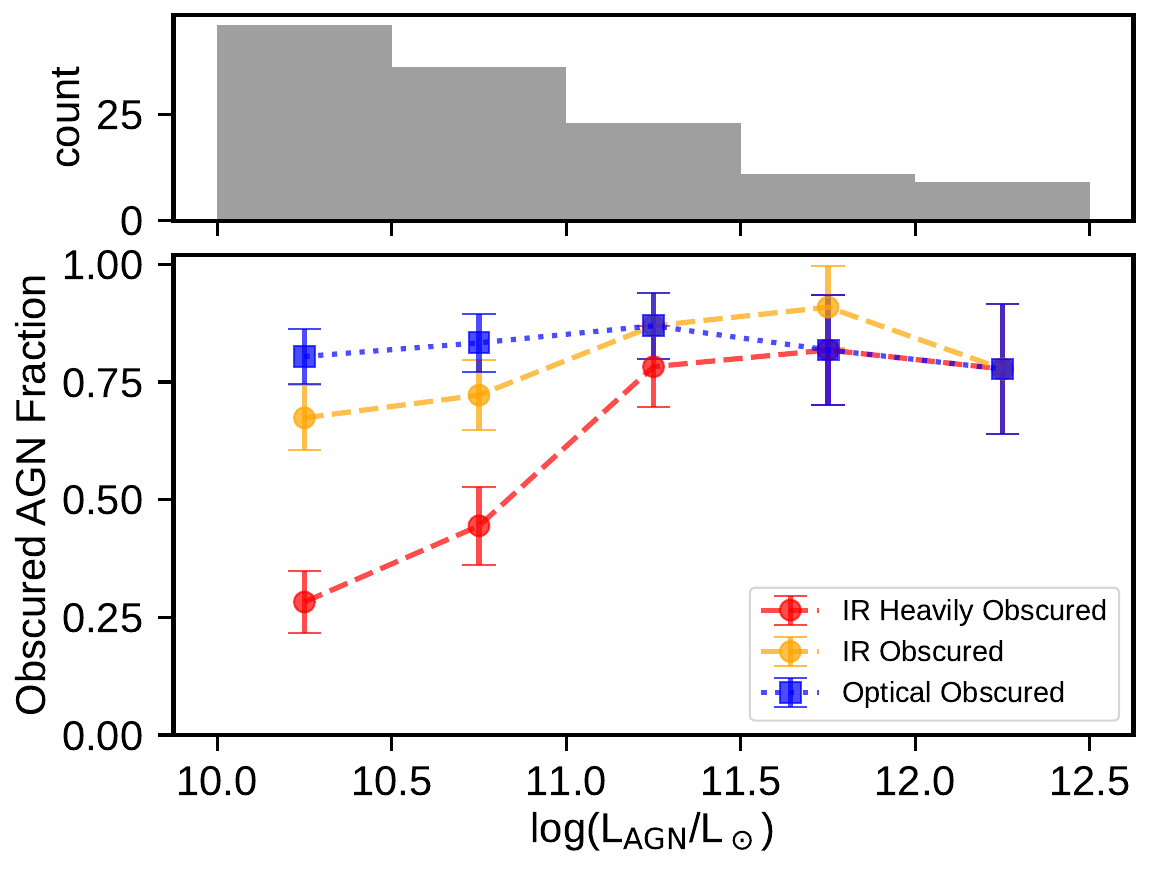}
    \includegraphics[width=0.495\hsize]{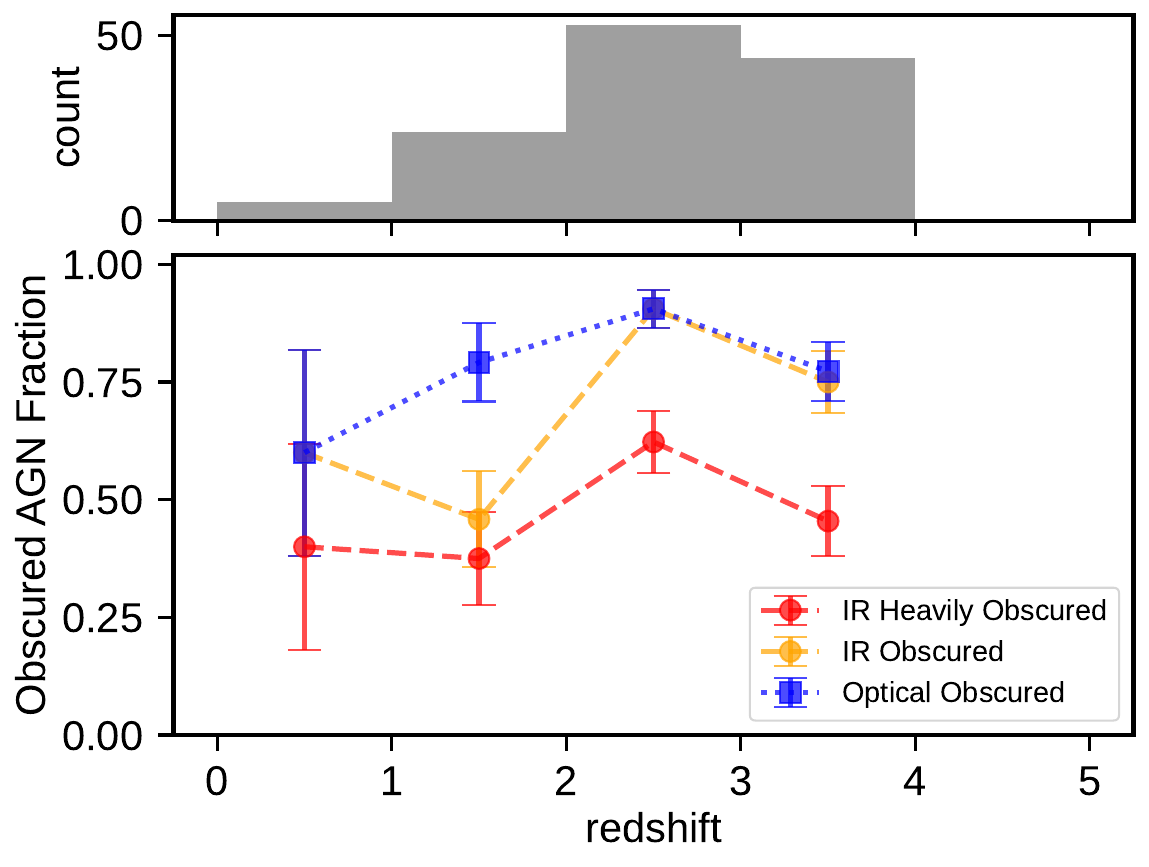}
    \caption{The obscured AGN fraction as a function of AGN luminosity (left) and  object redshift (right). 
    See text for the details.}
  \label{fig:f_obs_evolution}
    \end{center}
\end{figure*}

In Figure~\ref{fig:f_obs_evolution}, we show how the obscured AGN fractions change with different source properties. All three obscuration fractions first increase with AGN bolometric luminosity from
$L_{\rm AGN, bol}\sim10^{10}~L_\odot$ to $L_{\rm AGN, bol}\sim10^{11.5}~L_\odot$, with the slope increasing from optical obscured, IR obscured to IR heavily obscured, and then dropping towards higher luminosity in a consistent way. For redshift evolution, all the obscured AGN fractions increase as a function of 
object redshift. We discuss the implications of these trends below.

Previous studies have repeatedly shown that the obscured AGN fraction decreases with increasing  AGN luminosity \citep[e.g.,][]{Lawrence1991, Simpson2005, Lusso2013}, so the trend in the left panel 
of Figure~\ref{fig:f_obs_evolution} is a surprise at first glance. However, most of these studies have been limited to $L_{\rm AGN, bol}\gtrsim10^{11} L_\odot$. At the low-luminosity end, the obscured fraction has been found to increase with AGN luminosity in the X-ray \citep[e.g.,][]{Burlon2011}. In particular, \citet{Buchner2015} did show a similar
trend with the X-ray obscuration level peaking at $L_{X}\sim10^{43.5}$~erg~s$^{-1}$ and declining both above and below this value around  Cosmic Noon (see also, e.g., \citealt{Brightman2011} for a similar trend reported at low redshifts). Adopting 
the bolometric luminosity correction of $K_{X}\sim11$ \citep{Duras2020}, this corresponds to $L_{\rm AGN, bol}\sim10^{11}~L_\odot$, consistent
with the peak luminosity of our SED-inferred obscuration. 

From a theoretical perspective,  AGN obscuration 
occurs from the gas inflow that feeds the central SMBH, where the disk becomes unstable due to turbulence that produces a torus-like 
structure or dusty wind (e.g., \citealt{Wada2012, Wada2015, Hopkins2012}; see review by \citealt{Netzer2015}). Based on high-resolution hydrodynamic simulations, the obscuring gas column density presents positive correlations with the AGN luminosity \citep[e.g.,][]{Blecha2018, Trebitsch2019}. When the AGN becomes sufficiently bright, strong feedback is expected to blow out the surrounding material, reducing the chance of obscuration \citep[e.g.,][]{Sanders1988, Hopkins2008}.

Based on the analysis of the rest-frame UV to mid-IR SEDs, our study suggests an increasing fraction of obscured AGNs towards higher redshifts. Previous
work in the X-ray based on the gas column density $N_H$ has also revealed 
an increasing fraction of obscured AGNs from the local Universe to Cosmic Noon \citep[e.g.,][]{Ueda2003, LaFranca2005,Treister2006, Buchner2015, Vijarnwannaluk2022}. Such results are possibly associated with the large amount of gas in high redshift galaxies 
\citep[e.g.,][]{Carilli2013} as well as the more vigorous growth of SMBHs in the early Universe \citep{Inayoshi2020}.

Finally, AGN selections in our extended and highz samples (dwarf and  $z>4$ galaxies) are less secure or complete so we only report the integrated numbers of the obscured fractions for reference. For extended sample AGN candidates, the fraction of optically obscured, IR obscured and IR heavily obscured AGN is 86\%, 76\% and 42\%. These numbers are subject, of course, to further confirmation of the AGN nature of individual members of this sample. For the AGN candidates at $z>4$, the fraction of optically obscured, IR obscured and IR heavily obscured AGN is very roughly 54\%,  75\% and 46\%. Given that the MIRI data is shallower than that from NIRCam, the obscured AGN fraction among these galaxies may be underestimated.

\subsection{X-ray Bright but Mid-IR Faint AGNs}\label{sec:agn_no_mir}

As shown in Section~\ref{sec:compare-sample}, our MIRI+NIRCam data is much deeper than other multi-wavelength data 
used for AGN selection in terms of source bolometric luminosity. Naively, one would expect that the superb JWST data can
pick out all the AGNs found at other wavelengths. However, this is not the case. Notable examples are 
23 X-ray bright AGNs not identified by SED fittings. In other words, these systems are X-ray detected bright AGNs but very faint in the optical
to the mid-IR. 

\begin{figure}[htp]
    \begin{center}
  \includegraphics[width=1.0\hsize]{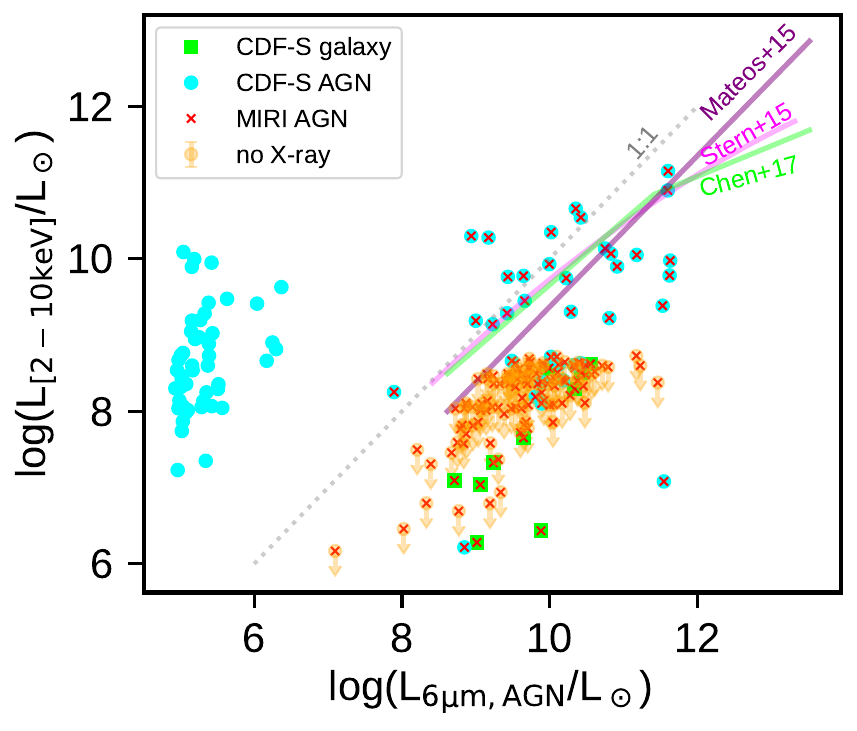}
    \caption{The relation between X-ray luminosity and AGN continuum luminosity at rest-frame 6~$\mu$m for AGN in the SMILES footprint. }
  \label{fig:mir-xray}
    \end{center}
\end{figure}

In Figure~\ref{fig:mir-xray}, we illustrate the locations of such objects on the X-ray luminosity vs the rest-frame 6 $\mum$ AGN luminosity plot. 
Previous studies have established a strong correlation between the absorption-corrected X-ray luminosity and the AGN luminosity at 6 $\mu$m for most
AGNs \citep[e.g.,][]{Asmus2015,Stern2015, Mateos2015, Chen2017}.  For AGNs identified by both MIRI SED fitting and X-rays, such a trend also exists despite
a large scatter. However, there are many X-ray bright AGNs with very weak AGN 6$\mu$m emission (left side of Figure~\ref{fig:mir-xray}). In fact,
even if we ignore the SED decomposition and assume all the observed flux is from the AGN, eleven objects cannot be moved near the typical $L_{X, int}-L_{6\mu m, AGN}$
relation.

\begin{figure*}[htp]
    \begin{center}
        \includegraphics[width=0.495\hsize]{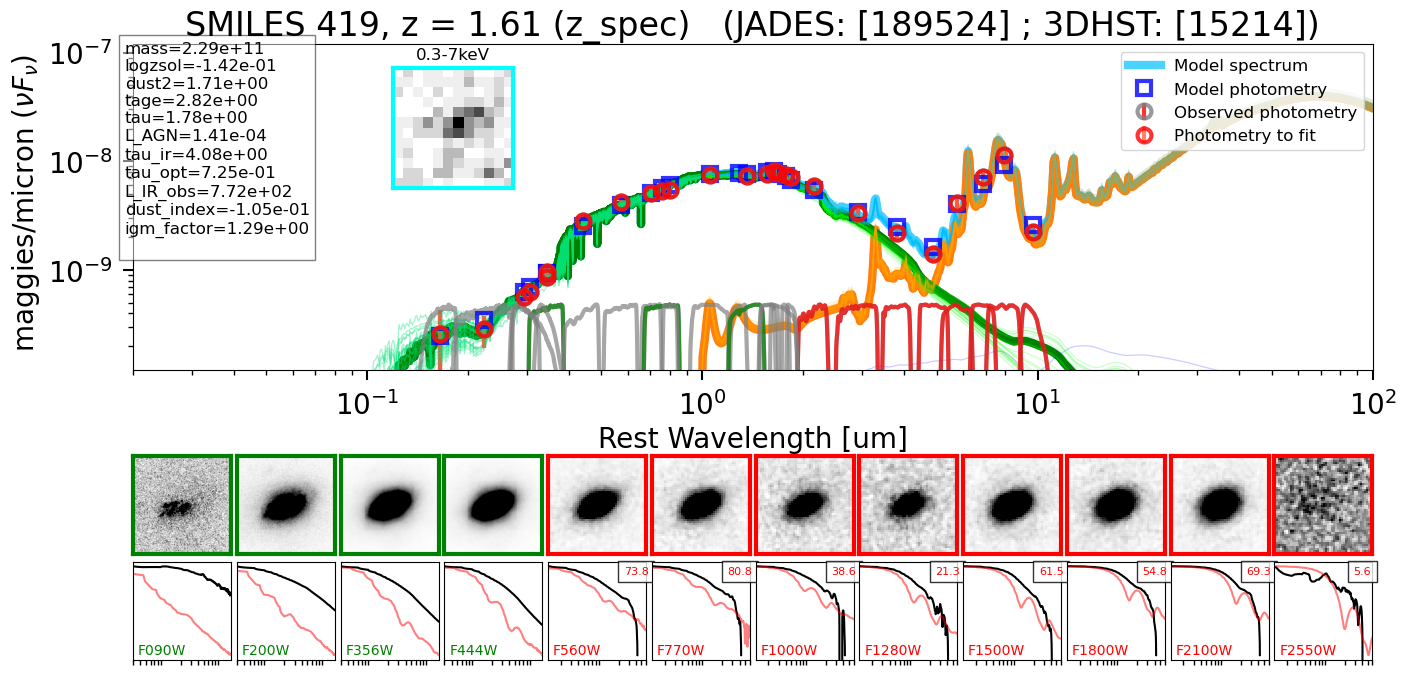}
        \includegraphics[width=0.495\hsize]{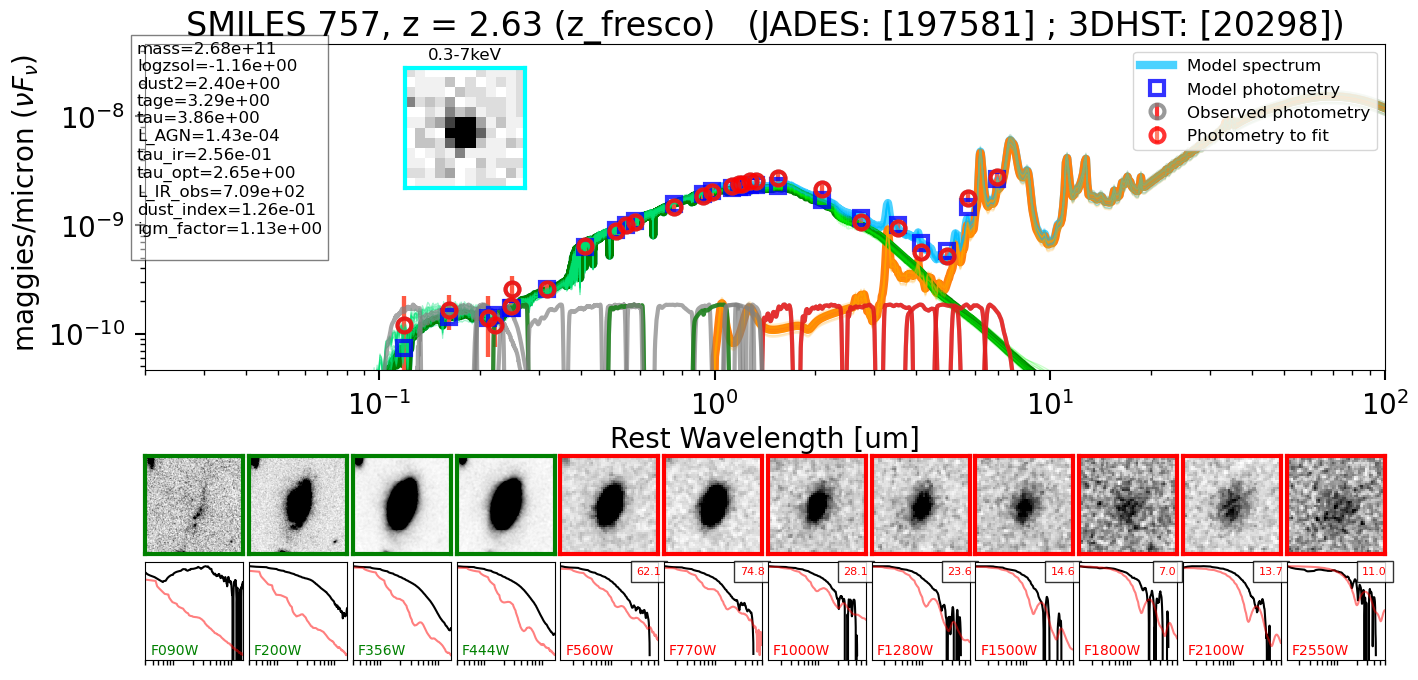}
        \includegraphics[width=0.495\hsize]{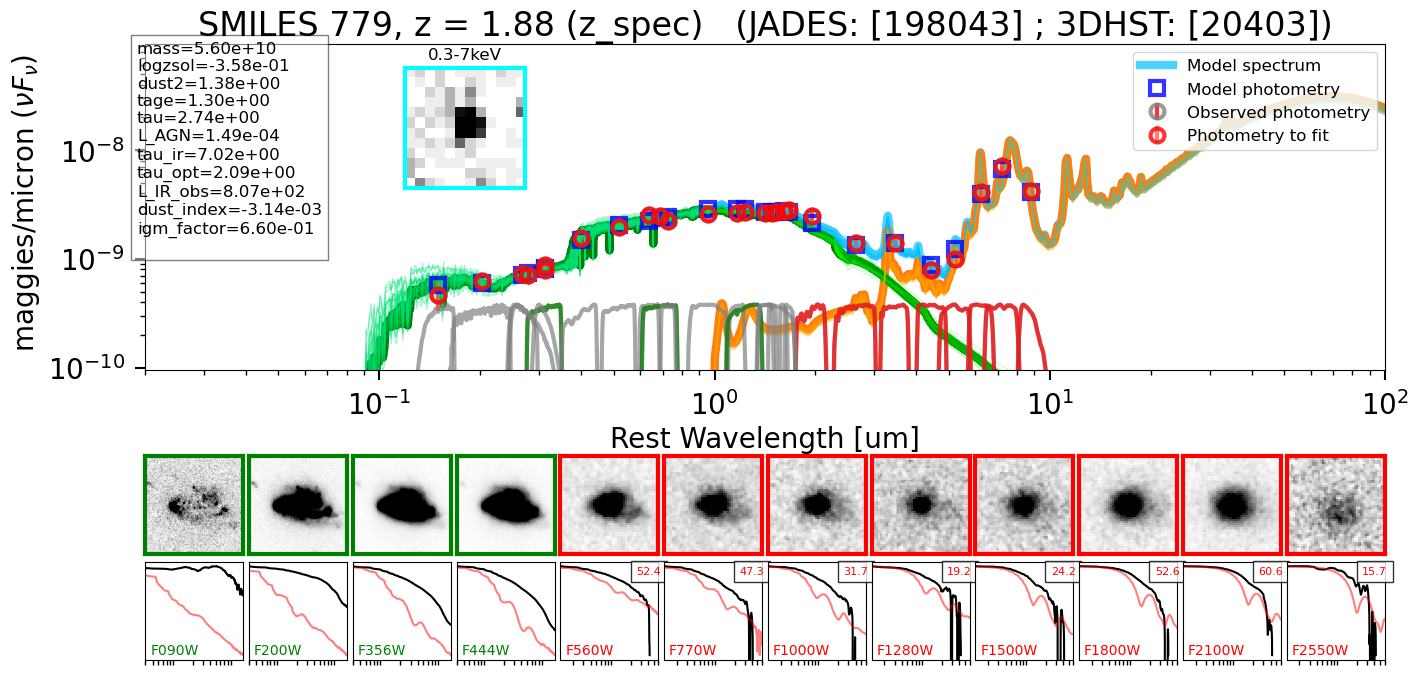}
        \includegraphics[width=0.495\hsize]{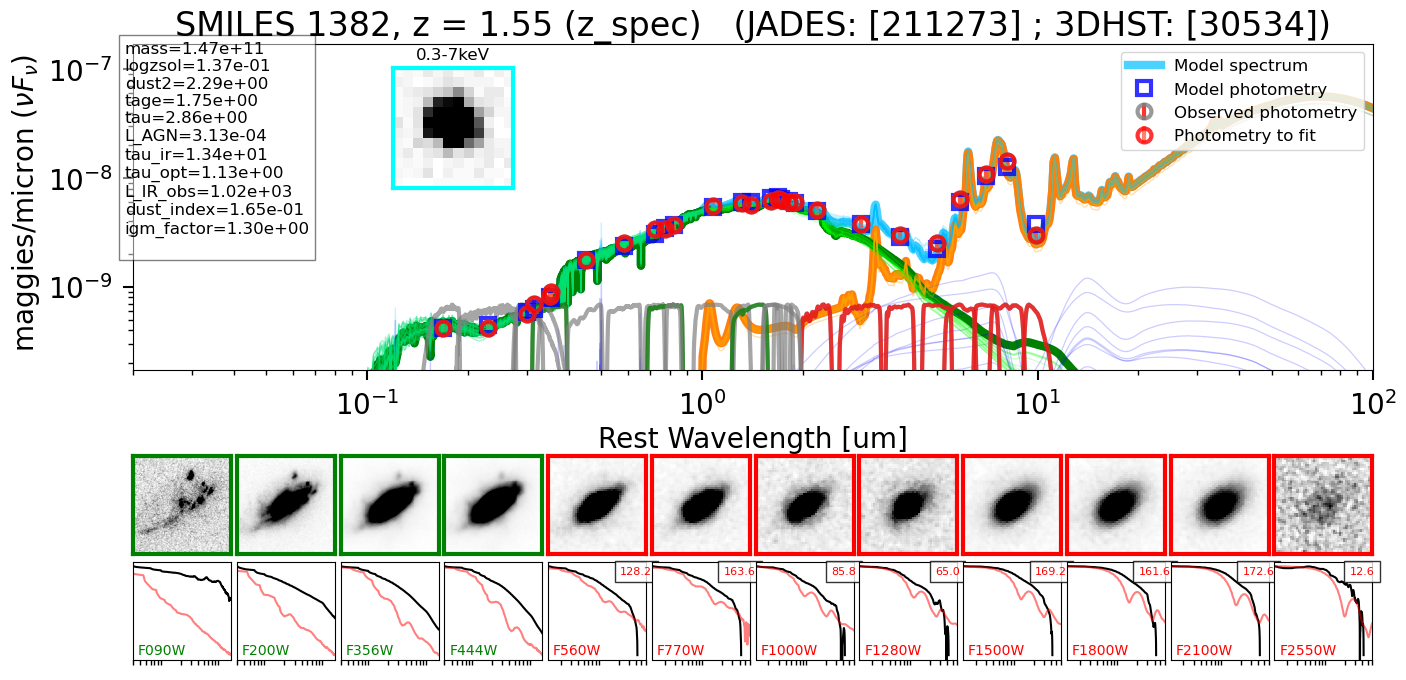}
    \caption{Similar to Figure~\ref{fig:sed-galaxy-examples} but for sample X-ray bright AGNs not confirmed by MIRI SED fittings. We 
    show the {\it Chandra} image cutout of the source with a FOV of 6\arcsec$\times$6\arcsec. According
    to the CDF-S catalog \citep{Luo2017}, the intrinsic 
    X-ray luminosity $\log(L_{\rm X, int}/{\rm erg~s}^{-1})$ and the corresponding gas column density $\log(N_H/{\rm cm}^{-2})$ of each object are: MIRI 419 -- 41.58, 23.47; MIRI 757 -- 42.70, 23.65;  MIRI 779 -- 42.72, 21.84; MIRI 1382 -- 43.12, 22.46.}
  \label{fig:xrayonly-agn-examples}
    \end{center}
\end{figure*}

We have carefully inspected these objects. Besides 8 objects with noisy SEDs or degenerate fitting results, 
15 do not show any AGN evidence in the UV-to-mid-IR SEDs. In Figure~\ref{fig:xrayonly-agn-examples}, we present some example SEDs with the cutouts from NIRCam, MIRI and {\it Chandra}. 
Besides the unresolved SED analysis, all these systems are extended in both NIRCam and MIRI images, consistent with the lack of AGN signatures. Meanwhile, their emission is very strong in the 
X-ray. Similar AGNs
have been reported previously \citep{LaMassa2019, Lyu2022}. 

One possible explanation could be dust-deficient AGNs that lack the hot/warm dust emission \citep{Lyu2017a}. If we limited the sample to X-ray detected AGNs, the fraction of such objects is about 16--25\%, consistent with the typical fraction of dust deficient AGN in an IR-unbiased sample as reported in \citet{Lyu2017a}.
Another possibility is that the X-ray emission of these AGNs is  boosted by, e.g., jets. Typically, the X-ray jet emission is associated with radio loud AGNs. Among
these 23 X-ray bright AGNs, 15 are detected by the JVLA at 3 GHz -- one is a confirmed radio-loud AGN, two are ambiguous cases and 12 are consistent with typical SFGs. 
It is therefore likely that the X-ray bright but mid-IR faint AGNs arise from both the radio-loud and radio-quiet populations.

{ Lastly, as pointed out in Section~\ref{sec:compare-method}, many [X2R] AGNs are not identified via JWST SEDs (see also Figure~\ref{fig:agn-counts}). These AGNs have been picked out by higher X-ray to radio luminosity ratios than  is expected from stellar processes in a galaxy. It is possible that they are similar to the X-ray bright and IR weak AGNs described above but with even lower AGN power.}

\subsection{MIRI AGNs without X-ray Detections}\label{sec:agn_no_xray}

As described in Section~\ref{sec:compare-sample}, a large fraction of our MIRI AGNs do not have X-ray detections (see also Figure~\ref{fig:mir-xray}). 
The intrinsically faint AGNs will not be detected in X-ray even if they are unobscured due to the shallower {\it Chandra} detection limits
compared to MIRI. For relatively bright AGNs, the X-ray non-detected AGNs can be interpreted as being Compton-thick or intrinsically X-ray weak.
We now focus on the first possibility and put some constraints on the Compton-thick AGN fractions with the aid of the MIRI SED results.

\begin{figure*}[htp]
    \begin{center}
  \includegraphics[width=1.0\hsize]{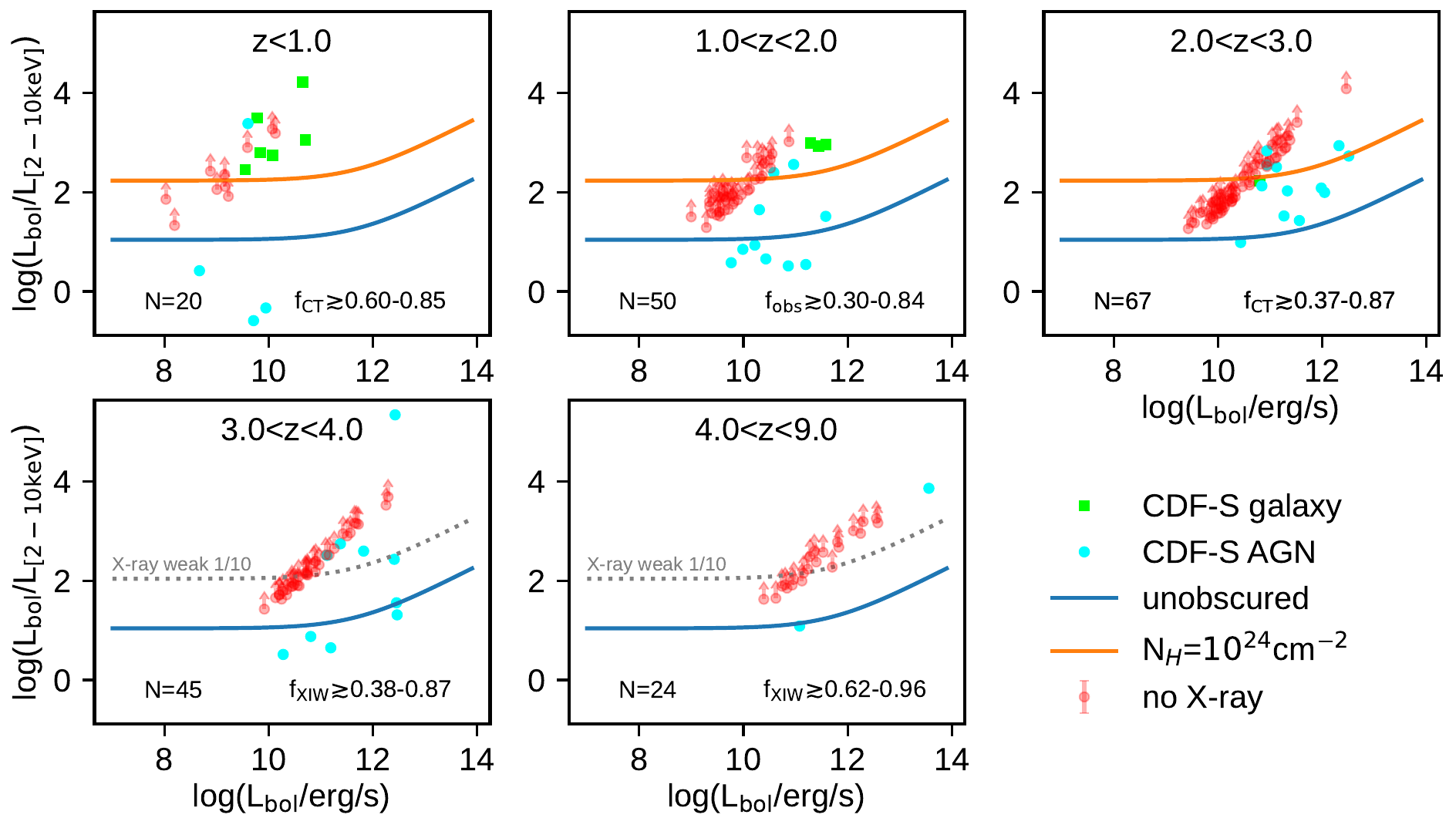}
    \caption{Bolometric to X-ray luminosity fraction as a function of AGN luminosity at different redshifts for MIRI-identified AGNs. On the bottom
    of each panel, we denote the total number of AGNs as $N$ and Compton-thick AGN fraction $f_{\rm CT}$ for $z<3.0$ and X-ray Intrinsic Weak AGN fraction $f_{\rm XIW}$ for $z>3.0$. }
  \label{fig:xray-thick}
    \end{center}
\end{figure*}

Since our SED fittings provide a measurement of the AGN bolometric luminosity, we can convert them to the equivalent X-ray luminosity for different gas column densities
and compare with observations. In Figure~\ref{fig:xray-thick}, we show the ratio between the SED-derived bolometric luminosity and the X-ray luminosity 
at 2--10 keV against the AGN bolometric luminosity for MIRI-identified AGNs at different redshifts. We compute an upper limit to
the 2--10 keV luminosity assuming no extinction for objects without X-ray detections. We also plot
the bolometric correction for the extinction-corrected AGN X-ray luminosity at 2--10 keV from \citet{Duras2020} as well as the expected
correlation if the X-ray emission is Compton-thick with gas column density $N_H\sim10^{24}$~cm$^{-2}$. The relative location of an AGN on this 
diagram is determined by its intrinsic luminosity as well as the gas column density. AGNs above the red curve can be counted as Compton-thick 
AGNs. 

The weak X-ray emission of most AGNs at relatively low redshift is well-explained in terms 
of absorption \citep{saade2022,wang2022}. Above $z\sim$1, we found the obscured AGN fraction gradually increases with redshift, possibly due to an increase in
the number of Compton thick sources.  Our finding of $\sim$ 34\% X-ray-undetected AGNs with SEDs suggestive that they are Compton thick aligns well with the typical
explanation of diffuse X-ray background \citep[e.g.,][]{Akylas2012}. Although there are some variations, a summary of its properties 
is: (1) $92.7 \pm 13.3$ \% of the hard X-ray (2--7 keV) background is resolved into individual sources by the Chandra 7 Msec exposure 
\citep{Luo2017,Cappelluti2017}; (2) models to fit the background are centered on the AGN population at redshifts of 0.5 - 2; and (3) 
many of these models invoke a population of Compton thick AGNs not detected in X-rays to fit the background around 30 keV \citep[e.g.,][]{Gilli2007,
Moretti2009, Treister2014}. It should be possible to advance our understanding of the X-ray background significantly through deep JWST surveys for 
infrared-discovered AGNs, although such studies are beyond the scope of this paper.  

Although the increase of the fraction of X-ray undetected AGN population apparently continues for $z>$3, 
Compton absorption becomes weak at the relevant rest-frame energies\footnote{For example, at z = 5.28, the redshift of MIRI 1104, the hard X-ray 2--7~keV band is at rest energies of 12--45~keV}.  Rather than  obscuration, some of these AGNs may be {\it intrinsically} weak in the X-ray. 
Indeed, the existence of such 
objects has been suggested previously by, e.g.,  \citet{leighly2004,simmonds2016, Lyu2022}. 
Stacking the X-ray observations for all six of the AGNs in Figure~\ref{fig:sed-hzagn-examples} yields non-detections of $3.6 \pm  5.0 \times 10^{-18}$ erg cm$^{-2}$ s$^{-1}$ for the 0.5 $-$ 2 keV band and  $-1.13 \pm 1.28 \times 10^{-17}$ erg cm$^{-2}$ s$^{-1}$ for the 2 $-$ 7 keV band. These stacked values reinforce the result in Figure~\ref{fig:xray-thick} that the high redshift AGNs are in general very X-ray weak. 

We now explore the implications of this result. The X-ray output of high redshift quasars in general does not show any deficiency 
of X-ray emission with an average $\alpha_{OX}$ of $\sim -1.5$ for those with measured values (omitting two outliers), using the 
results from \citet{Li2021}. This does not exclude weak X-rays from the undetected members, which are the majority of the sample, 
although these authors conclude that the relation of the X-ray outputs to the 
other parameters of the quasars is similar to that for lower redshift quasars. The low X-ray fluxes for the lower luminosity high-z AGNs could arise due to Compton thick absorption with $n_H \gtrsim 10^{25}$ cm$^{-2}$, but such high absorbing columns are very rare locally. The infrared AGN-driven SED shapes for the six sources in Figure~\ref{fig:sed-hzagn-examples}, extending down to 1--2 $\mu$m and even shorter wavelengths, are similar to those of typical local Type-2 Seyfert galaxies, rather than extreme Compton thick examples like NGC 1068.  A hint toward an alternative explanation is provided by the very X-ray weak AGNs in a number of low-metallicity dwarf galaxies \citep{Dong2012, simmonds2016, Cann2021, Bohn2021, Latimer2021}, 
often two orders of magnitude below the values expected by analogy with AGNs in massive galaxies. That is, the behavior might be associated 
with low metallicity in modest mass high redshift host galaxies.  

Hard X-rays from AGNs are produced by Compton scattering in the coronal regions above their accretion disks. Therefore, a process that disrupts 
the coronal regions could suppress the X-ray outputs.  The dwarf galaxies have strong AGN-driven outflows that are believed to be effective 
at clearing gas from the galaxies \citep{Bohn2021}; these outflows are also candidates to disrupt the coronal regions. A possible issue with 
this explanation is that a small sample indicates that the outflow and coronal line strengths may be correlated \citep{Bohn2022}. Overall, this 
is a complex situation requiring much larger samples than provided in this paper to test, particularly since AGN variability is likely to influence 
the observational properties of any sample of AGNs.

{ Besides the possibilities discussed above, the lack of X-ray detections for high-$z$ AGNs might be also explained by the possible evolution of the X-ray spectral
shape. For example, \citet{Zappacosta2023} found the X-ray spectra of ten quasars at $z>9$ have steeper average photon index ($\Gamma\approx2.4\pm0.1$) than classical values ($\Gamma\sim$1.8--2). For such systems, even their X-ray luminosity can be high, their redshifted spectrum at the commonly observed X-ray energy band could be a factor up to 4--5 times fainter than standard AGNs, which can greatly reduce the chance of detection. }

\section{The Blind Men and the Elephant: How to Build a Complete AGN Sample?}\label{sec:discussion}

{ As demonstrated in this work, JWST has made a huge step forward in finding AGNs, including those at
very high redshift, highly obscured, or residing in low-mass galaxies. However, 
this does not mean  
the current AGN census is complete. As an epilogue, we discuss the lessons learned so far and give some future outlook.}

The process of selecting AGNs is like the story of the blind men and the elephant. Every study
is limited by the wavelength regime and survey depth, so the results are inevitably biased
and incomplete. In fact, the AGN phenomena is more complicated than elephants and astronomers typically work on different samples built at 
different wavelengths. 
Due to e.g., the intrinsic variation of AGN properties, 
the apparent obscuration caused by dust and gas, the efficiency of the selection method for AGN, 
and the data quality that the selection has to be based on, AGNs
seen in one study can be totally missed in another study (e.g., see Section~\ref{sec:result} on the existence of X-ray bright AGNs without relevant IR signatures in the JWST SEDs and bright mid-IR AGNs without X-ray detections). Thus, a complete picture can be only obtained by combining the results 
across the electromagnetic spectrum together.


In terms of selection techniques, 
the so-called gold standards for AGN identification, such as very broad emission lines, 
bright X-ray emission or clear hot dust signatures, are definitive  but certainly not complete indicators.  
Given how complicated and diverse the  AGN population is, we could miss a substantial fraction of AGNs if the selections rely on 
these most obvious features, as seen in Section~\ref{sec:obscured_agn}. 
When multi-wavelength data is available, it is reasonable to relax the criteria and use multiple (relaxed) selections
to complete the AGN census. From the AGN hints from other wavelengths, selections
at one band can be calibrated and improved \citep[e.g.,][]{Donley2012, Hviding2022}. However, each selection technique has limitations.
For example, we should not adopt the classical AGN color criteria or SED fittings with
normal galaxy templates to look for AGN in the dwarf galaxy population \citep[e.g.,][]{Hainline2016}. As we enter a new regime, relevant 
tools need to be updated or developed. Lastly, besides refining the AGN selection methods and expanding their usage, there must also be 
a balance between robustness and completeness.

Given the various factors that influence the observed AGN behavior, 
whether an AGN sample is complete is actually determined by how we define the term ``AGN'' observationally, which can be very controversial (e.g., the extended discussion about how to classify LINERs, now understood to be a mixed category). 
Our ultimate goal is not to collect all kinds of AGN samples but to understand the physics behind the AGNs themselves and the SMBH-galaxy interaction. 
In this sense, one intriguing future direction is to combine the observed AGN statistics with the predicted observables from cosmological 
simulations that involve reasonable treatments of various selection biases. This approach has more physics behind it and can mitigate the issues of 
obscured AGNs that may not be easily revealed by the data. 
If one day we could build such a physical framework that matched all the key empirical correlations between
SMBHs and galaxies with good predictive power for the observed AGN statistics at different wavelengths, our long journey 
to complete the AGN census might reach its final end.  

\section{Summary}\label{sec:summary}

In this work, we presented the first results on AGN selection and demographics from SMILES, a JWST Cycle 1 GTO program that has targeted the central 
region of GOODS-S with eight MIRI filters from 5.6 to 26~$\mum$. Combining the MIRI data with JADES NIRCam and HST photometry at shorter wavelengths,
we conducted comprehensive SED analyses of 3273 MIRI-sources and reported 217 AGNs over a footprint of 34 arcmin$^{2}$. This 
MIRI-selected AGN sample includes 111 AGNs in the primary sample of normal massive galaxies, 86 AGN candidates in the extended sample of dwarf galaxies and 20 high-z AGN candidates at z $>$ 4. Our major conclusions are:

\begin{itemize}

\item{We reached a total number density of SED-identified AGNs of $\sim$ 6.6 arcmin$^{-2}$, which is over two times larger than that for the confusion-limited {\it Spitzer}/MIPS 24 $\mu$m observations of the same field \citep{Lyu2022}. In addition, compared to the deepest X-ray survey by {\it Chandra} in the same footprint,
our relatively shallow MIRI surveys ($\sim$2.17 hours per pointing) yield 2.4 times more AGNs, of which 80\% are new discoveries.}

\item{AGNs in our primary sample of massive galaxies should be sufficiently luminous for detection by previous searches in other wavebands, assuming they are not obscured. However, 34\% of the AGNs found in our study do not have previous identifications. This result indicates that 34 $\pm$ 6\% of AGNs have been missed due to obscuration, where the error arises from the modest sample size (114); }

\item{For the first time at z $\sim$ 2, JWST/MIRI is detecting significant numbers of AGNs in low-mass galaxies [$\log(M_*/M_\odot)<9.5$] -- from our conservative SED identifications, they are comparable in number with the AGNs we have found in the more massive primary sample;  }

\item{We confirm the existence of X-bright but mid-IR faint AGNs and suggest that they may be dust-deficient AGNs that do not present 
strong hot dust SED features, or AGNs with X-ray emission boosted by e.g. a jet, or a mixture of both populations;}

\item{$\sim$80\% of the MIRI-selected AGNs do not have X-ray detections. Most of them are too faint to be detected by {\it Chandra}, some of which are likely
to be Compton-thick AGNs. For MIRI AGNs without X-ray detections at higher redshifts ($z\sim$4--6), they may be intrinsically weak in the X-ray, possibly due to the lack of a strong corona component due to their lower metallicity environment.}

\item{Based on SED analysis, we find the obscured AGN fraction increases with AGN luminosity towards L$_{AGN} \sim$ 10$^{11}$ L$_\odot$ and then drops at higher luminosity. The obscured AGN fraction also increases with redshift, regardless of the detailed definitions. For AGNs at $z>4$, it is likely half of the population
is heavily obscured.}

\item{Despite the comprehensive MIRI photometric data and the substantially improved SED selections, about 28\% of the known AGN (candidates) in the field have been missed by the SED analysis, indicating a huge diversity of AGN properties that challenges the completion of the AGN census. In other words, no single method or wavelength can identify all the AGNs and a combination of multi-wavelength dataset and selection techniques is always desired;}

\end{itemize}

Our results demonstrate the unique power of JWST MIRI for AGN selection and characterization. This paper summarizes first results from a modest initial survey and we look forward to larger and more ambitious studies.

\begin{acknowledgments}
This work was supported by NASA grants NNX13AD82G and 1255094. We thanks Andras Gaspar for his help on MIRI PSF construction. AJB has received funding from the European Research Council (ERC) under the European Union’s Horizon 2020 Advanced Grant 789056 ``First Galaxies''. PGP-G acknowledges support from grant PID2022-139567NB-I00 funded by Spanish Ministerio de Ciencia e Innovación MCIN/AEI/10.13039/501100011033, FEDER, UE. C.N.A.W. was supported by the NIRCam Development Contract NAS5-02105 from
NASA Goddard Space Flight Center to the University of Arizona. This work is based on observations made with the NASA/ESA/CSA James Webb Space Telescope. The data were obtained 
from the Mikulski Archive for Space Telescopes at the Space Telescope Science Institute, which is operated by the Association of Universities for Research in 
Astronomy, Inc., under NASA contract NAS 5-03127 for JWST. These observations are associated with program \#1207 with the original data can be accessed via \dataset[DOI:10.17909/jmxm-1695]{https://doi.org/10.17909/jmxm-1695}.

\end{acknowledgments}

\appendix{}

\section{AGN Identification from Weak Pa $\alpha$}

\begin{figure}[htp]
   \begin{center}
  \includegraphics[width=.5\hsize]{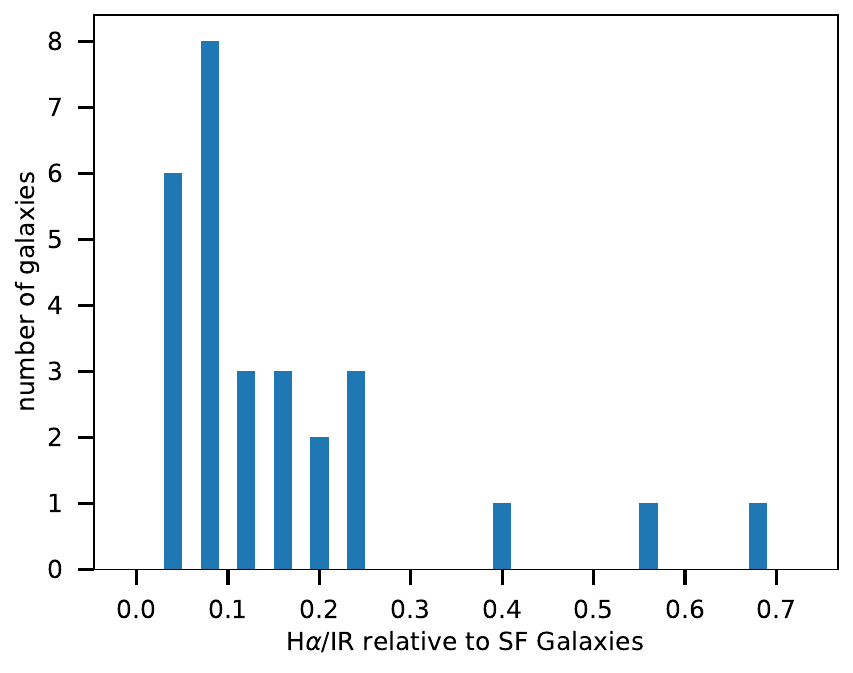}
    \caption{Ratio of H$\alpha$ to 12 $\mu$m luminosity for local Type 2 AGNs, normalized to the average value for star forming galaxies.}
  \label{fig:sy2halpha}
    \end{center}
\end{figure}

Figure~\ref{fig:sy2halpha} shows the distribution of H$\alpha$ to 12 $\mu$m luminosities for Type 2 AGNs, relative to the average value for star forming galaxies. We have selected 12 $\mu$m for this comparison both because of the availability of all-sky data bases, and because the flux at this wavelength is representative of the total luminosity of many galaxies \citep{Spinoglio1989}. The integrated H$\alpha$ fluxes for the star forming galaxies have been taken from \citet{Moustakas2006} and the 12 $\mu$m fluxes from IRAS (which has a sufficiently large beam that the flux is close to incorporating the full galaxy). The luminosity at the latter wavelength is taken to be proportional to $\lambda$F$_\lambda$. The H$\alpha$ fluxes for the AGNs are from \citet{Winter2010,Rose2013} (which provide flux-calibrated line strengths), with 12 $\mu$m fluxes from ALLWISE. The ratio for Type 1 AGNs, obtained from the same references plus \citet{Osterbrock1987}, is nominally just a factor of about two below the average for the star forming galaxies, but Figure~\ref{fig:sy2halpha} shows that the majority of Type 2 AGNs fall more than an order of magnitude low in this ratio.

This behavior is a product of extinction, as is shown by the much higher average ratios for the Type 1 AGNs. If the extinction follows a normal interstellar law and is a foreground screen, the ratio for Pa $\alpha$ would typically still be a factor of 3--4 weaker relative to star forming galaxies. It is likely that the extinction is partially due to optically thick structures (e.g., the circumnuclear torus) and it may be due to relatively large grains due to the harsh environment around the nucleus, and hence be more neutral than the typical interstellar behavior; both effects would increase the difference. 

Since both the Pa $\alpha$ and infrared luminosities are indicators of the star formation rate, this discriminator can be applied by declaring any galaxy where the SFR indicated by Pa $\alpha$ falls significantly below that estimated from the infrared flux is a candidate to host an AGN. This test can be implemented using the FRESCO spectra. At higher redshifts, it may be possible to use FRESCO spectra showing Pa $\beta$.

We have carried out a preliminary analysis using only galaxies with detected Pa $\alpha$ by FRESCO. There are ten galaxies with SFRs calculated from the Pa $\alpha$ luminosity more than three times lower than indicated by the flux density at 18 $\mu$m; we accept this as the indicator of an AGN based on the study of obscuration levels in local LIRGs by \citet{Alonso2006}. They are listed in Table~\ref{tab:paalpha}. Additional candidates can probably be identified among MIRI-detected sources with accurate redshifts in the appropriate range and upper limits on Pa $\alpha$ from the FRESCO spectra.

\begin{deluxetable*}{ccccc}
\tablecaption{AGNs identified from weak Pa $\alpha$}
\tablewidth{50pt}
\tablehead{
\colhead{MIRI ID}&
\colhead {RA} &
\colhead {DEC} &
\colhead {SFR(18 $\mu$m)/SFR(Pa $\alpha$)} &
\colhead {Previous AGN Selections\tablenotemark{a}} 
}
\startdata
 302   &  53.07644&     -27.84873    & 7.5   &  1, 3,4, 5  \\
 509   &  53.10938&	  -27.83495    & 6.1  &    \\
 1183  &  53.14350&	  -27.78327    & 3.2  &    \\
 1425  &  53.16288&     -27.76723    & 8.3  &  1, 3, 4, 6, 7  \\
 1637  &  53.17287&	  -27.74456    & 4.3  &    \\
\enddata
\tablenotetext{a}{1 = from IR SED; 2 = from IR  colors; 3 = from X-ray luminosity; 4 = from X-ray to radio flux ratio; 5 = from radio loudness; 6 = from optical spectrum; 7 = from variability}
\label{tab:paalpha}
\end{deluxetable*}


\begin{thebibliography}{}
\bibitem[National Academies of Sciences(2021)]{astro2020} National Academies of Sciences, E.\ 2021, Pathways to Discovery in Astronomy and Astrophysics for the 2020s, Consenses Study Report. NAtional Academies of Sciences, Engineering, and Medicine. 2021. Washington, DC: The National Academies Press, 2021.. doi:10.17226/26141

\bibitem[Abazajian et al.(2009)]{Abazajian2009} Abazajian, K.~N., Adelman-McCarthy, J.~K., Ag{\"u}eros, M.~A., et al.\ 2009, \apjs, 182, 543. doi:10.1088/0067-0049/182/2/543
\bibitem[Adams(1977)]{Adams1977} Adams, T.~F.\ 1977, \apjs, 33, 19. doi:10.1086/190416
\bibitem[Alberts et al.(2020)]{Alberts2020} Alberts, S., Rujopakarn, W., Rieke, G.~H., et al.\ 2020, \apj, 901, 168. doi:10.3847/1538-4357/abb1a0
\bibitem[Alexander \& Hickox(2012)]{Alexander2012} Alexander, D.~M. \& Hickox, R.~C.\ 2012, \nar, 56, 93. doi:10.1016/j.newar.2011.11.003
\bibitem[Alonso-Herrero et al. (2006)]{Alonso2006} {Alonso-Herrero}, A., {P{\'e}rez-Gonz{\'a}lez}, P.~G.,  {Alexander}, D.~M. et al. 2006,  {\apj}, {640}, 167,  doi:{10.1086/499800}
\bibitem[{\'A}lvarez-M{\'a}rquez et al.(2023)]{Alvarez2023} {\'A}lvarez-M{\'a}rquez, J., Crespo G{\'o}mez, A., Colina, L., et al.\ 2023, \aap, 671, A105. doi:10.1051/0004-6361/202245400
\bibitem[Akylas et al.(2012)]{Akylas2012} Akylas, A., Georgakakis, A., Georgantopoulos, I., et al.\ 2012, \aap, 546, A98. doi:10.1051/0004-6361/201219387
\bibitem[Ananna et al. (2019)]{Ananna2019} {Ananna}, Tonima Tasnim, {Treister}, Ezequiel, {Urry}, C. Megan et al. 2019,  {\apj}, {871}, {240}, doi:{10.3847/1538-4357/aafb77}
\bibitem[Aniano et al. (2020)]{Aniano2020} {Aniano}, G.,  {Draine}, B.~T., {Hunt}, L.~K. et al. 2020,  {\apj}, {889}, {150}, doi:{10.3847/1538-4357/ab5fdb}
\bibitem[Asmus et al.(2015)]{Asmus2015} Asmus, D., Gandhi, P., H{\"o}nig, S.~F., et al.\ 2015, \mnras, 454, 766. doi:10.1093/mnras/stv1950
\bibitem[Asmus et al.(2020)]{Asmus2020} Asmus, D., Greenwell, C.~L., Gandhi, P., et al.\ 2020, \mnras, 494, 1784. doi:10.1093/mnras/staa766
\bibitem[Assef et al.(2013)]{Assef2013} Assef, R.~J., Stern, D., Kochanek, C.~S., et al.\ 2013, \apj, 772, 26. doi:10.1088/0004-637X/772/1/26
\bibitem[Assef et al.(2018)]{Assef2018} Assef, R.~J., Stern, D., Noirot, G., et al.\ 2018, \apjs, 234, 23. doi:10.3847/1538-4365/aaa00a

\bibitem[Bacon et al.(2023)]{Bacon2023} Bacon, R., Brinchmann, J., Conseil, S., et al.\ 2023, \aap, 670, A4. doi:10.1051/0004-6361/202244187
\bibitem[Baldwin et al.(1981)]{Baldwin1981} Baldwin, J.~A., Phillips, M.~M., \& Terlevich, R.\ 1981, \pasp, 93, 5. doi:10.1086/130766

\bibitem[Bunker et al.(2023)]{Bunker2023} Bunker, A.~J., Cameron, A.~J., Curtis-Lake, E., et al.\ 2023, arXiv:2306.02467. doi:10.48550/arXiv.2306.02467

\bibitem[Barro et al.(2011a)]{Barro2011a} Barro, G., P{\'e}rez-Gonz{\'a}lez, P.~G., Gallego, J., et al.\ 2011, \apjs, 193, 13. doi:10.1088/0067-0049/193/1/13
\bibitem[Barro et al.(2011b)]{Barro2011b} Barro, G., P{\'e}rez-Gonz{\'a}lez, P.~G., Gallego, J., et al.\ 2011, \apjs, 193, 30. doi:10.1088/0067-0049/193/2/30
\bibitem[Baskin \& Laor(2018)]{Baskin2018} Baskin, A. \& Laor, A.\ 2018, \mnras, 474, 1970. doi:10.1093/mnras/stx2850
\bibitem[Beckwith et al.(2006)]{Beckwith2006} Beckwith, S.~V.~W., Stiavelli, M., Koekemoer, A.~M., et al.\ 2006, \aj, 132, 1729. doi:10.1086/507302

\bibitem[Blecha et al.(2018)]{Blecha2018} Blecha, L., Snyder, G.~F., Satyapal, S., et al.\ 2018, \mnras, 478, 3056. doi:10.1093/mnras/sty1274

\bibitem[Bohn et al.(2021)]{Bohn2021} Bohn, T., Canalizo, G., Veilleux, S., et al.\ 2021, \apj, 911, 70. doi:10.3847/1538-4357/abe70c

\bibitem[Bohn et al.(2022)]{Bohn2022} Bohn, T., Canalizo, G., Satyapal, S., et al.\ 2022, \apj, 931, 69. doi:10.3847/1538-4357/ac6870



\bibitem[Brammer et al.(2008)]{Brammer2008} Brammer, G.~B., van Dokkum, P.~G., \& Coppi, P.\ 2008, \apj, 686, 1503. doi:10.1086/591786
\bibitem[Brightman \& Nandra(2011)]{Brightman2011} Brightman, M. \& Nandra, K.\ 2011, \mnras, 413, 1206. doi:10.1111/j.1365-2966.2011.18207.x
\bibitem[Buchner et al.(2015)]{Buchner2015} Buchner, J., Georgakakis, A., Nandra, K., et al.\ 2015, \apj, 802, 89. doi:10.1088/0004-637X/802/2/89
\bibitem[Burlon et al.(2011)]{Burlon2011} Burlon, D., Ajello, M., Greiner, J., et al.\ 2011, \apj, 728, 58. doi:10.1088/0004-637X/728/1/58
\bibitem[Bushouse et al.(2022)]{Bushouse2022} Bushouse, H., Eisenhamer, J., Dencheva, N., et al.\ 2022, Zenodo  doi:10.5281/zenodo.7795697

\bibitem[Cackett et al.(2021)]{Cackett2021} Cackett, E.~M., Bentz, M.~C., \& Kara, E.\ 2021, iScience, 24, 102557. doi:10.1016/j.isci.2021.102557
\bibitem[Cann et al.(2021)]{Cann2021} Cann, J.~M., Satyapal, S., Rothberg, B., et al.\ 2021, \apjl, 912, L2. doi:10.3847/2041-8213/abf56d

\bibitem[Carilli \& Walter(2013)]{Carilli2013} Carilli, C.~L. \& Walter, F.\ 2013, \araa, 51, 105. doi:10.1146/annurev-astro-082812-140953

\bibitem[Cappelluti et al.(2017)]{Cappelluti2017} Cappelluti, N., Li, Y., Ricarte, A., et al.\ 2017, \apj, 837, 19. doi:10.3847/1538-4357/aa5ea4
\bibitem[Chen et al.(2017)]{Chen2017} Chen, C.-T.~J., Hickox, R.~C., Goulding, A.~D., et al.\ 2017, \apj, 837, 145. doi:10.3847/1538-4357/837/2/145
\bibitem[Conroy et al.(2009)]{Conroy2009} Conroy, C., Gunn, J.~E., \& White, M.\ 2009, \apj, 699, 486. doi:10.1088/0004-637X/699/1/486
\bibitem[Conroy \& Gunn(2010)]{Conroy2010} Conroy, C. \& Gunn, J.~E.\ 2010, \apj, 712, 833. doi:10.1088/0004-637X/712/2/833
\bibitem[Del Moro et al. (2016)]{Delmoro2016} {Del Moro}, A., {Alexander}, D.~M.,  {Bauer}, F.~E. et al. 2016,  {\mnras}, {456}, 2105,  doi:{10.1093/mnras/stv2748}
\bibitem[De Rossi et al.(2018)]{DeRossi2018} De Rossi, M.~E., Rieke, G.~H., Shivaei, I., et al.\ 2018, \apj, 869, 4. doi:10.3847/1538-4357/aaebf8

\bibitem[Dong et al.(2012)]{Dong2012} Dong, R., Greene, J.~E., \& Ho, L.~C.\ 2012, \apj, 761, 73. doi:10.1088/0004-637X/761/1/73

\bibitem[Donley et al.(2008)]{Donley2008} Donley, J.~L., Rieke, G.~H., P{\'e}rez-Gonz{\'a}lez, P.~G., et al.\ 2008, \apj, 687, 111. doi:10.1086/591510
\bibitem[Donley et al.(2012)]{Donley2012} Donley, J.~L., Koekemoer, A.~M., Brusa, M., et al.\ 2012, \apj, 748, 142. doi:10.1088/0004-637X/748/2/142
\bibitem[Draine et al.(2014)]{Draine2014} Draine, B.~T., Aniano, G., Krause, O., et al.\ 2014, \apj, 780, 172. doi:10.1088/0004-637X/780/2/172
\bibitem[Duras et al.(2020)]{Duras2020} Duras, F., Bongiorno, A., Ricci, F., et al.\ 2020, \aap, 636, A73. doi:10.1051/0004-6361/201936817
\bibitem[Edge et al. (1959)]{Edge1959} {Edge}, D.~O.,  {Shakeshaft}, J.~R., {McAdam}, W.~B., {Baldwin}, J.~E., \& {Archer}, S. 1959, {\memras}, {68}, 37
\bibitem[Eisenstein et al.(2023)]{Eisenstein2023} Eisenstein, D.~J., Willott, C., Alberts, S., et al.\ 2023, arXiv:2306.02465. doi:10.48550/arXiv.2306.02465
\bibitem[Engelbracht et al.(2005)]{Engelbracht2005} Engelbracht, C.~W., Gordon, K.~D., Rieke, G.~H., et al.\ 2005, \apjl, 628, L29. doi:10.1086/432613
\bibitem[Engelbracht et al.(2008)]{Engelbracht2008} Engelbracht, C.~W., Rieke, G.~H., Gordon, K.~D., et al.\ 2008, \apj, 678, 804. doi:10.1086/529513
\bibitem[Fabian(2012)]{Fabian2012} Fabian, A.~C.\ 2012, \araa, 50, 455. doi:10.1146/annurev-astro-081811-125521
\bibitem[Filippenko \& Sargent(1985)]{Filippenko1985} Filippenko, A.~V. \& Sargent, W.~L.~W.\ 1985, \apjs, 57, 503. doi:10.1086/191012
\bibitem[G{\'a}sp{\'a}r et al.(2021)]{Gaspar2021} G{\'a}sp{\'a}r, A., Rieke, G.~H., Guillard, P., et al.\ 2021, \pasp, 133, 014504. doi:10.1088/1538-3873/abcd04
\bibitem[Giacconi et al.(2002)]{Giacconi2002} Giacconi, R., Zirm, A., Wang, J., et al.\ 2002, \apjs, 139, 369. doi:10.1086/338927
\bibitem[Giavalisco et al.(2004)]{Giavalisco2004} Giavalisco, M., Ferguson, H.~C., Koekemoer, A.~M., et al.\ 2004, \apjl, 600, L93. doi:10.1086/379232
\bibitem[Gilli et al.(2007)]{Gilli2007} Gilli, R., Comastri, A., \& Hasinger, G.\ 2007, \aap, 463, 79. doi:10.1051/0004-6361:20066334
\bibitem[Glikman et al.(2006)]{Glikman2006} Glikman, E., Helfand, D.~J., \& White, R.~L.\ 2006, \apj, 640, 579. doi:10.1086/500098
\bibitem[Goulding et al.(2012)]{Goulding2012} Goulding, A.~D., Alexander, D.~M., Bauer, F.~E., et al.\ 2012, \apj, 755, 5. doi:10.1088/0004-637X/755/1/5
\bibitem[Goulding et al.(2023)]{Goulding2023} Goulding, A.~D., Greene, J.~E., Setton, D.~J., et al.\ 2023, arXiv:2308.02750. doi:10.48550/arXiv.2308.02750
\bibitem[Gursky et al. (1971)]{Gursky1971} {Gursky}, H.,  {Kellogg}, E.~M., {Leong}, C., {Tananbaum}, H., \& {Giacconi}, R. 1971,  {\apjl}, {165}, L43, doi:{10.1086/180713}
\bibitem[Hainline et al.(2011)]{Hainline2011} Hainline, K.~N., Shapley, A.~E., Greene, J.~E., et al.\ 2011, \apj, 733, 31. doi:10.1088/0004-637X/733/1/31
\bibitem[Hainline et al.(2016)]{Hainline2016} Hainline, K.~N., Reines, A.~E., Greene, J.~E., et al.\ 2016, \apj, 832, 119. doi:10.3847/0004-637X/832/2/119
\bibitem[Hainline et al.(2023)]{Hainline2023} Hainline, K.~N., Johnson, B.~D., Robertson, B., et al.\ 2023, arXiv:2306.02468. doi:10.48550/arXiv.2306.02468
\bibitem[Hao et al.(2005)]{Hao2005} Hao, L., Strauss, M.~A., Tremonti, C.~A., et al.\ 2005, \aj, 129, 1783. doi:10.1086/428485
\bibitem[Harikane et al.(2023)]{Harikane2023} Harikane, Y., Zhang, Y., Nakajima, K., et al.\ 2023, arXiv:2303.11946. doi:10.48550/arXiv.2303.11946
\bibitem[Heckman \& Best(2014)]{Heckman2014} Heckman, T.~M. \& Best, P.~N.\ 2014, \araa, 52, 589. doi:10.1146/annurev-astro-081913-035722
\bibitem[Hern{\'a}n-Caballero et al.(2016)]{Hernan-Caballero2016} Hern{\'a}n-Caballero, A., Hatziminaoglou, E., Alonso-Herrero, A., et al.\ 2016, \mnras, 463, 2064. doi:10.1093/mnras/stw2107
\bibitem[Hickox \& Alexander(2018)]{Hickox2018} Hickox, R.~C. \& Alexander, D.~M.\ 2018, \araa, 56, 625. doi:10.1146/annurev-astro-081817-051803
\bibitem[Hinshaw et al.(2013)]{WMAP9} Hinshaw, G., Larson, D., Komatsu, E., et al.\ 2013, \apjs, 208, 19. doi:10.1088/0067-0049/208/2/19
\bibitem[Ho et al.(1997)]{Ho1997} Ho, L.~C., Filippenko, A.~V., \& Sargent, W.~L.~W.\ 1997, \apj, 487, 568. doi:10.1086/304638
\bibitem[Hopkins et al.(2008)]{Hopkins2008} Hopkins, P.~F., Hernquist, L., Cox, T.~J., et al.\ 2008, \apjs, 175, 356. doi:10.1086/524362
\bibitem[Hopkins et al.(2012)]{Hopkins2012} Hopkins, P.~F., Hayward, C.~C., Narayanan, D., et al.\ 2012, \mnras, 420, 320. doi:10.1111/j.1365-2966.2011.20035.x
\bibitem[Hopkins \& Quataert(2010)]{Hopkins2010} Hopkins, P.~F. \& Quataert, E.\ 2010, \mnras, 407, 1529. doi:10.1111/j.1365-2966.2010.17064.x
\bibitem[Hviding et al.(2018)]{Hviding2018} {Hviding}, Raphael E.,  {Hickox}, Ryan C.,  {Hainline}, Kevin N. et al. 2018,  {\mnras}, {474}, 1955, 
\bibitem[Hviding et al.(2022)]{Hviding2022} Hviding, R.~E., Hainline, K.~N., Rieke, M., et al.\ 2022, \aj, 163, 224. doi:10.3847/1538-3881/ac5e33
\bibitem[Inayoshi et al.(2020)]{Inayoshi2020} Inayoshi, K., Visbal, E., \& Haiman, Z.\ 2020, \araa, 58, 27. doi:10.1146/annurev-astro-120419-014455
\bibitem[Johnson et al.(2021)]{Johnson2021} Johnson, B.~D., Leja, J., Conroy, C., et al.\ 2021, \apjs, 254, 22. doi:10.3847/1538-4365/abef67
\bibitem[Kauffmann et al.(2003)]{Kauffmann2003} Kauffmann, G., Heckman, T.~M., Tremonti, C., et al.\ 2003, \mnras, 346, 1055. doi:10.1111/j.1365-2966.2003.07154.x
\bibitem[Kewley et al.(2001)]{Kewley2001} Kewley, L.~J., Dopita, M.~A., Sutherland, R.~S., et al.\ 2001, \apj, 556, 121. doi:10.1086/321545
\bibitem[Kirkpatrick et al.(2015)]{Kirkpatrick2015} Kirkpatrick, A., Pope, A., Sajina, A., et al.\ 2015, \apj, 814, 9. doi:10.1088/0004-637X/814/1/9
\bibitem[Kirkpatrick et al.(2023)]{Kirkpatrick2023} Kirkpatrick, A., Yang, G., Le Bail, A., et al.\ 2023, arXiv:2308.09750. doi:10.48550/arXiv.2308.09750
\bibitem[Kocevski et al.(2023)]{Kocevski2023} Kocevski, D.~D., Onoue, M., Inayoshi, K., et al.\ 2023, arXiv:2302.00012. doi:10.48550/arXiv.2302.00012
\bibitem[Kodra et al.(2023)]{Kodra2023} Kodra, D., Andrews, B.~H., Newman, J.~A., et al.\ 2023, \apj, 942, 36. doi:10.3847/1538-4357/ac9f12
\bibitem[Kormendy \& Ho(2013)]{Kormendy2013} Kormendy, J. \& Ho, L.~C.\ 2013, \araa, 51, 511. doi:10.1146/annurev-astro-082708-101811
\bibitem[Kriek \& Conroy(2013)]{Kriek2013} Kriek, M. \& Conroy, C.\ 2013, \apjl, 775, L16. doi:10.1088/2041-8205/775/1/L16
\bibitem[Lacy et al. (2004)]{Lacy2004} {Lacy}, M., {Storrie-Lombardi}, L.~J., {Sajina}, A. et al. 2004, {\apjs}, {154}, 166, doi:{10.1086/422816}
\bibitem[La Franca et al.(2005)]{LaFranca2005} La Franca, F., Fiore, F., Comastri, A., et al.\ 2005, \apj, 635, 864. doi:10.1086/497586
\bibitem[LaMassa et al.(2019)]{LaMassa2019} LaMassa, S.~M., Georgakakis, A., Vivek, M., et al.\ 2019, \apj, 876, 50. doi:10.3847/1538-4357/ab108b
\bibitem[Lanzuisi et al.(2018)]{Lanzuisi2018} Lanzuisi, G., Civano, F., Marchesi, S., et al.\ 2018, \mnras, 480, 2578. doi:10.1093/mnras/sty2025
\bibitem[Larson et al.(2023)]{Larson2023} Larson, R.~L., Finkelstein, S.~L., Kocevski, D.~D., et al.\ 2023, arXiv:2303.08918. doi:10.48550/arXiv.2303.08918

\bibitem[Latimer et al.(2021)]{Latimer2021} Latimer, L.~J., Reines, A.~E., Hainline, K.~N., et al.\ 2021, \apj, 914, 133. doi:10.3847/1538-4357/abfe0c
\bibitem[Lawrence(1991)]{Lawrence1991} Lawrence, A.\ 1991, \mnras, 252, 586. doi:10.1093/mnras/252.4.586
\bibitem[Lee et al.(2006)]{Lee2006} Lee, H., Skillman, E.~D., Cannon, J.~M., et al.\ 2006, \apj, 647, 970. doi:10.1086/505573
\bibitem[Leighly(2004)]{leighly2004} Leighly, K.~M.\ 2004, \apj, 611, 125. doi:10.1086/422089
\bibitem[Lex et al.(2014)]{Lex2014} Lex, A.,, Gehlenborg, B., Strobelt, H., Vuillemot, R., Pfister, H.\ 2014,  IEEE Transactions on Visualization and Computer Graphics (InfoVis ‘14), 20, 12. doi:10.1109/TVCG.2014.2346248

\bibitem[Li et al.(2021)]{Li2021} Li, J.-T., Wang, F., Yang, J., et al.\ 2021, \mnras, 504, 2767. doi:10.1093/mnras/stab1042

\bibitem[Luo et al.(2017)]{Luo2017} Luo, B., Brandt, W.~N., Xue, Y.~Q., et al.\ 2017, \apjs, 228, 2. doi:10.3847/1538-4365/228/1/2
\bibitem[Lusso et al.(2013)]{Lusso2013} Lusso, E., Hennawi, J.~F., Comastri, A., et al.\ 2013, \apj, 777, 86. doi:10.1088/0004-637X/777/2/86
\bibitem[Lyu et al.(2016)]{Lyu2016} Lyu, J., Rieke, G.~H., \& Alberts, S.\ 2016, \apj, 816, 85. doi:10.3847/0004-637X/816/2/85
\bibitem[Lyu et al.(2017)]{Lyu2017a} Lyu, J., Rieke, G.~H., \& Shi, Y.\ 2017, \apj, 835, 257. doi:10.3847/1538-4357/835/2/257
\bibitem[Lyu et al.(2022a)]{Lyu2022} Lyu, J., Alberts, S., Rieke, G.~H., et al.\ 2022, \apj, 941, 191. doi:10.3847/1538-4357/ac9e5d
\bibitem[Lyu \& Rieke(2017)]{Lyu2017b} Lyu, J. \& Rieke, G.~H.\ 2017, \apj, 841, 76. doi:10.3847/1538-4357/aa7051
\bibitem[Lyu \& Rieke(2018)]{Lyu2018} Lyu, J. \& Rieke, G.~H.\ 2018, \apj, 866, 92. doi:10.3847/1538-4357/aae075
\bibitem[Lyu \& Rieke (2022a)]{LyuRieke2022}{Lyu}, Jianwei \& {Rieke}, George 2022,  {Universe}, {8}, {304}, doi:{10.3390/universe8060304}
\bibitem[Lyu \& Rieke (2022b)]{Lyu2022c} {Lyu}, Jianwei, \& {Rieke}, George H. 2022b,  {\apjl}, {940}, {L31}, doi:{10.3847/2041-8213/ac9e5c}
\bibitem[Madden et al.(2006)]{Madden2006} Madden, S.~C., Galliano, F., Jones, A.~P., et al.\ 2006, \aap, 446, 877. doi:10.1051/0004-6361:20053890
\bibitem[Ma et al.(2016)]{Ma2016} Ma, X., Hopkins, P.~F., Faucher-Gigu{\`e}re, C.-A., et al.\ 2016, \mnras, 456, 2140. doi:10.1093/mnras/stv2659

\bibitem[Maiolino \& Rieke (1995)]{Maiolino1995} {Maiolino}, R.\& {Rieke}, G.~H. 1995,  {\apj}, {454}, {95}, doi:{10.1086/176468}

\bibitem[Maiolino et al.(2023)]{Maiolino2023} Maiolino, R., Scholtz, J., Witstok, J., et al.\ 2023, arXiv:2305.12492. doi:10.48550/arXiv.2305.12492
\bibitem[Marble et al. (2010)]{Marble2010} {Marble}, A.~R.,  {Engelbracht}, C.~W., {van Zee}, L. et al. 2010,  {\apj}, {715}, 506, doi:{10.1088/0004-637X/715/1/506}
\bibitem[Markarian (1977)]{Markarian1977} {Markarian}, B.~E. 1977, {\aap}, {58}, 139

\bibitem[Mateos et al.(2012)]{Mateos2012} Mateos, S., Alonso-Herrero, A., Carrera, F.~J., et al.\ 2012, \mnras, 426, 3271. doi:10.1111/j.1365-2966.2012.21843.x
\bibitem[Mateos et al.(2015)]{Mateos2015} Mateos, S., Carrera, F.~J., Alonso-Herrero, A., et al.\ 2015, \mnras, 449, 1422. doi:10.1093/mnras/stv299
\bibitem[Matthee et al.(2023)]{Matthee2023} Matthee, J., Naidu, R.~P., Brammer, G., et al.\ 2023, arXiv:2306.05448. doi:10.48550/arXiv.2306.05448
\bibitem[Mendez et al. (2013)]{Mendez2013} {Mendez}, Alexander J.,  {Coil}, Alison L., {Aird}, James et al. 2013,  {\apj}, {770}, {40}, doi:{10.1088/0004-637X/770/1/40}
\bibitem[Merlin et al.(2021)]{Merlin2021} Merlin, E., Castellano, M., Santini, P., et al.\ 2021, \aap, 649, A22. doi:10.1051/0004-6361/202140310
\bibitem[M. Rieke et al. (2023)]{Rieke2023} Rieke, M. \& the JADES Collaboration\ 2023, arXiv:2306.02466. doi:10.48550/arXiv.2306.02466
\bibitem[Moretti et al.(2009)]{Moretti2009} Moretti, A., Pagani, C., Cusumano, G., et al.\ 2009, \aap, 493, 501. doi:10.1051/0004-6361:200811197
\bibitem[Moustakas \& Kennicutt (2006)]{Moustakas2006} {Moustakas}, John, \& {Kennicutt}, Robert C., Jr. 2006, {\apjs}, {164}, 81,   doi:10.1086/500971
\bibitem[Netzer(2015)]{Netzer2015} Netzer, H.\ 2015, \araa, 53, 365. doi:10.1146/annurev-astro-082214-122302
\bibitem[Oesch et al.(2023)]{Oesch2023} Oesch, P.~A., Brammer, G., Naidu, R.~P., et al.\ 2023, arXiv:2304.02026. doi:10.48550/arXiv.2304.02026
\bibitem[Osterbrock \& Pogge (1987)]{Osterbrock1987} {Osterbrock}, Donald E. \& {Pogge}, Richard W. 1987,  {\apj}, {323}, {108}, doi:10.1086/165810
\bibitem[Padovani(2016)]{Padovani2016} Padovani, P.\ 2016, \aapr, 24, 13. doi:10.1007/s00159-016-0098-6
\bibitem[Padovani et al.(2017)]{Padovani2017} Padovani, P., Alexander, D.~M., Assef, R.~J., et al.\ 2017, \aapr, 25, 2. doi:10.1007/s00159-017-0102-9
\bibitem[Padovani(2017)]{Padovani2017b} Padovani, P.\ 2017, Frontiers in Astronomy and Space Sciences, 4, 35. doi:10.3389/fspas.2017.00035
\bibitem[Papovich et al.(2023)]{Papovich2023} Papovich, C., Cole, J.~W., Yang, G., et al.\ 2023, \apjl, 949, L18. doi:10.3847/2041-8213/acc948
\bibitem[P{\'e}rez-Gonz{\'a}lez et al.(2005)]{PerezGonzalez2005} P{\'e}rez-Gonz{\'a}lez, P.~G., Rieke, G.~H., Egami, E., et al.\ 2005, \apj, 630, 82. doi:10.1086/431894
\bibitem[P{\'e}rez-Gonz{\'a}lez et al.(2008)]{PerezGonzalez2008} P{\'e}rez-Gonz{\'a}lez, P.~G., Rieke, G.~H., Villar, V., et al.\ 2008, \apj, 675, 234. doi:10.1086/523690
\bibitem[Perrin et al.(2014)]{Perrin2014} Perrin, M.~D., Sivaramakrishnan, A., Lajoie, C.-P., et al.\ 2014, \procspie, 9143, 91433X. doi:10.1117/12.2056689
\bibitem[Prestwich et al.(2015)]{Prestwich2015} Prestwich, A.~H., Jackson, F., Kaaret, P., et al.\ 2015, \apj, 812, 166. doi:10.1088/0004-637X/812/2/166
\bibitem[Pouliasis et al.(2019)]{Pouliasis2019} Pouliasis, E., Georgantopoulos, I., Bonanos, A.~Z., et al.\ 2019, \mnras, 487, 4285. doi:10.1093/mnras/stz1483
\bibitem[R{\'e}my-Ruyer et al.(2013)]{Remy-Ruyer2013} R{\'e}my-Ruyer, A., Madden, S.~C., Galliano, F., et al.\ 2013, \aap, 557, A95. doi:10.1051/0004-6361/201321602

\bibitem[R{\'e}my-Ruyer et al.(2015)]{Remy-Ruyer2015} R{\'e}my-Ruyer, A., Madden, S.~C., Galliano, F., et al.\ 2015, \aap, 582, A121. doi:10.1051/0004-6361/201526067

\bibitem[Richards et al.(2002)]{Richards2002} Richards, G.~T., Fan, X., Newberg, H.~J., et al.\ 2002, \aj, 123, 2945. doi:10.1086/340187
\bibitem[Rieke et al.(2009)]{Rieke2009} Rieke, G.~H., Alonso-Herrero, A., Weiner, B.~J., et al.\ 2009, \apj, 692, 556. doi:10.1088/0004-637X/692/1/556
\bibitem[Rieke et al.(2015)]{Rieke2015} Rieke, G.~H., Wright, G.~S., B{\"o}ker, T., et al.\ 2015, \pasp, 127, 584. doi:10.1086/682252
\bibitem[Rieke et al.(2017)]{Rieke-1207} Rieke, G., Alberts, S., Lyu, J., et al.\ 2017, JWST Proposal. Cycle 1, 1207

\bibitem[Rigby et al.(2008)]{Rigby2008} Rigby, J.~R., Marcillac, D., Egami, E., et al.\ 2008, \apj, 675, 262. doi:10.1086/525273


\bibitem[Rose et al. (2013)]{Rose2013} {Rose}, M.,  {Tadhunter}, C.~N., {Holt}, J., \&  {Rodr{\'\i}guez Zaur{\'\i}n}, J. 2013,  {\mnras}, {432}, {2150}, doi:10.1093/mnras/stt564

\bibitem[Saade et al.(2022)]{saade2022} Saade, M.~L., Brightman, M., Stern, D., et al.\ 2022, \apj, 936, 162. doi:10.3847/1538-4357/ac88cf

\bibitem[Sanders et al.(1988)]{Sanders1988} Sanders, D.~B., Soifer, B.~T., Elias, J.~H., et al.\ 1988, \apj, 325, 74. doi:10.1086/165983
\bibitem[Santini et al.(2009)]{Santini2009} Santini, P., Fontana, A., Grazian, A., et al.\ 2009, \aap, 504, 751. doi:10.1051/0004-6361/200811434
\bibitem[Schmidt \& Green(1983)]{Schmidt1983} Schmidt, M. \& Green, R.~F.\ 1983, \apj, 269, 352. doi:10.1086/161048
\bibitem[Seyfert (1943)]{Seyfert1943} {Seyfert}, Carl K.1943, {\apj}, {97}, {28}, doi:{10.1086/144488}
\bibitem[Shivaei et al.(2017)]{Shivaei2017} Shivaei, I., Reddy, N.~A., Shapley, A.~E., et al.\ 2017, \apj, 837, 157. doi:10.3847/1538-4357/aa619c
\bibitem[Shivaei et al.(2020)]{Shivaei2020} Shivaei, I., Darvish, B., Sattari, Z., et al.\ 2020, \apjl, 903, L28. doi:10.3847/2041-8213/abc1ef
\bibitem[Shivaei et al.(2022)]{Shivaei2022} Shivaei, I., Popping, G., Rieke, G., et al.\ 2022, \apj, 928, 68. doi:10.3847/1538-4357/ac54a9

\bibitem[Silverman et al.(2010)]{Silverman2010} Silverman, J.~D., Mainieri, V., Salvato, M., et al.\ 2010, \apjs, 191, 124. doi:10.1088/0067-0049/191/1/124
\bibitem[Simmonds et al.(2016)]{simmonds2016} Simmonds, C., Bauer, F.~E., Thuan, T.~X., et al.\ 2016, \aap, 596, A64. doi:10.1051/0004-6361/201629310
\bibitem[Simpson(2005)]{Simpson2005} Simpson, C.\ 2005, \mnras, 360, 565. doi:10.1111/j.1365-2966.2005.09043.x
\bibitem[Spinoglio \& Malkan(1989)]{Spinoglio1989} Spinoglio, L. \& Malkan, M.~A.\ 1989, \apj, 342, 83. doi:10.1086/167577
\bibitem[Stern et al.(2012)]{Stern2012} Stern, D., Assef, R.~J., Benford, D.~J., et al.\ 2012, \apj, 753, 30. doi:10.1088/0004-637X/753/1/30
\bibitem[Stern(2015)]{Stern2015} Stern, D.\ 2015, \apj, 807, 129. doi:10.1088/0004-637X/807/2/129
\bibitem[Stern et al. (2005)]{Stern2005} {Stern}, Daniel,  {Eisenhardt}, Peter, {Gorjian}, Varoujan et al. 2005,  {\apj}, {631}, 163, doi:{10.1086/432523}
\bibitem[Tadhunter(2016)]{Tadhunter2016} Tadhunter, C.\ 2016, \aapr, 24, 10. doi:10.1007/s00159-016-0094-x
\bibitem[Trebitsch et al.(2019)]{Trebitsch2019} Trebitsch, M., Volonteri, M., \& Dubois, Y.\ 2019, \mnras, 487, 819. doi:10.1093/mnras/stz1280
\bibitem[Treister \& Urry(2006)]{Treister2006} Treister, E. \& Urry, C.~M.\ 2006, \apjl, 652, L79. doi:10.1086/510237
\bibitem[Treister et al.(2014)]{Treister2014} Treister, E., Urry, C.~M., Schawinski, K., et al.\ 2014, Multiwavelength AGN Surveys and Studies, 304, 188. doi:10.1017/S1743921314003731
\bibitem[Tremonti et al.(2004)]{Tremonti2004} Tremonti, C.~A., Heckman, T.~M., Kauffmann, G., et al.\ 2004, \apj, 613, 898. doi:10.1086/423264
\bibitem[{\"U}bler et al.(2023)]{Ubler2023} {\"U}bler, H., Maiolino, R., Curtis-Lake, E., et al.\ 2023, arXiv:2302.06647. doi:10.48550/arXiv.2302.06647
\bibitem[Ueda et al.(2003)]{Ueda2003} Ueda, Y., Akiyama, M., Ohta, K., et al.\ 2003, \apj, 598, 886. doi:10.1086/378940
\bibitem[Vijarnwannaluk et al.(2022)]{Vijarnwannaluk2022} Vijarnwannaluk, B., Akiyama, M., Schramm, M., et al.\ 2022, \apj, 941, 97. doi:10.3847/1538-4357/ac9c07
\bibitem[Wada(2012)]{Wada2012} Wada, K.\ 2012, \apj, 758, 66. doi:10.1088/0004-637X/758/1/66
\bibitem[Wada(2015)]{Wada2015} Wada, K.\ 2015, \apj, 812, 82. doi:10.1088/0004-637X/812/1/82

\bibitem[Wang et al.(2022)]{wang2022} Wang, C., Luo, B., Brandt, W.~N., et al.\ 2022, \apj, 936, 95. doi:10.3847/1538-4357/ac886e


\bibitem[Whitaker et al.(2011)]{Whitaker2011} Whitaker, K.~E., Labb{\'e}, I., van Dokkum, P.~G., et al.\ 2011, \apj, 735, 86. doi:10.1088/0004-637X/735/2/86

\bibitem[Williams et al.(2023)]{Williams2023} Williams, C.~C., Tacchella, S., Maseda, M.~V., et al.\ 2023, \apjs, 268, 64. doi:10.3847/1538-4365/acf130
\bibitem[Winter et al. (2010)]{Winter2010} {Winter}, Lisa M., {Lewis}, Karen T.,  {Koss} et al. 2010,  {\apj}, {710}, 503, doi:{10.1088/0004-637X/710/1/503}
\bibitem[Yang et al.(2023)]{Yang2023} Yang, G., Caputi, K.~I., Papovich, C., et al.\ 2023, \apjl, 950, L5. doi:10.3847/2041-8213/acd639

\bibitem[Wright et al.(2023)]{Wright2023} Wright, G.~S., Rieke, G.~H., Glasse, A., et al.\ 2023, \pasp, 135, 048003. doi:10.1088/1538-3873/acbe66

\bibitem[Zahid et al.(2013)]{Zahid2013} Zahid, H.~J., Geller, M.~J., Kewley, L.~J., et al.\ 2013, \apjl, 771, L19. doi:10.1088/2041-8205/771/2/L19
\bibitem[Zappacosta et al.(2023)]{Zappacosta2023} Zappacosta, L., Piconcelli, E., Fiore, F., et al.\ 2023, \aap, 678, A201. doi:10.1051/0004-6361/202346795
\end{thebibliography}
\end{document}